\newcommand\be {\begin{equation}}
\newcommand\ee {\end{equation}}
\newcommand \bmat {\begin{displaymath}}
\newcommand \emat {\end{displaymath}}
\newcommand \eps {\epsilon}

\newcommand \lan {\langle}
\newcommand \ran {\rangle}

\documentclass[aps]{revtex4}
\usepackage{graphicx}

\begin{document}

\title{Fragility of the Free-Energy Landscape of a
Directed Polymer in Random Media}

\author{Marta Sales$^{1}$ and Hajime Yoshino$^{2}$}
\affiliation{
$^{1}$Departament de F\'{\i}sica Fonamental, \\
Facultat de F\'{\i}sica, Universitat de Barcelona\\
Diagonal 647, 08028 Barcelona, Spain\\
$^{2}$ Department of Earth and Space Science, Faculty of Science,
Osaka University\\
Toyonaka, 560-0043 Osaka , Japan
}

\begin{abstract} 

We examine the sensitiveness of the free-energy landscape of a directed 
polymer in random media with respect to various kinds of infinitesimally 
weak perturbation including the intriguing case of temperature-chaos.
To this end, we combine the replica Bethe ansatz approach outlined
in cond-mat/0112384, the mapping to a modified Sinai model 
and numerically exact calculations by the 
transfer-matrix method. Our results imply that for all 
the perturbations under study there is a slow crossover from a 
weakly perturbed regime where rare events take place to a strongly 
perturbed regime at larger length scales beyond the so called 
overlap length where typical events take place leading to 
chaos, i.e. a complete reshuffling of the free-energy landscape. 
Within the replica space, the evidence for chaos is found 
in the factorization of 
the replicated partition function induced by infinitesimal 
perturbations. This is the reflex of explicit replica symmetry breaking.

\end{abstract}
\maketitle

\titlepage

\newcommand{\eq}[1]{(\ref{#1})}
\newcommand{\kb}{k_{\rm B}}

\section{Introduction}
\label{sec.intro}

A very interesting problem of glassy systems with disorder and frustration
is the possible instability of the glassy frozen states against 
infinitesimally weak perturbations such as an infinitesimal 
change of temperatures and realizations of quenched randomness. 
Such a perturbation does not bring the system 
out of the frozen phase but possibly changes the lugged landscape 
of the free-energy  in a dramatic way. Let us call this intriguing property
as {\it fragility of the free-energy landscape}.
A class of phenomenological scaling theories started first 
in the context of spin-glass by Bray and Moore,
and, Fisher and Huse \cite{BM87,FH88,FH91}
generically implies that equilibrium states of systems with disorder 
and frustration resist against such infinitesimally weak 
perturbations of strength $\delta \ll 1$ 
up to a finite crossover length scale $L_{c}(\delta)$ called
overlap length but change into completely different states at larger 
length scales, resulting in the vanishing of the correlations between 
the two states. The overlap length $L_{c}(\delta)$
diverges as $\delta \to 0$ but remains finite for any non-zero $\delta$.
Such an anomalous response is called as {\it chaos} referring 
to the feature that the distance between the perturbed and
unperturbed systems becomes infinitely large in phase space even
by infinitesimally weak perturbation as the system size $L$ becomes 
macroscopically large $L/L_{c}(\delta) \to \infty$ \cite{BM87}.
Unfortunately, the validity of the prediction has not been proven
explicitly by theoretical studies except for some Migdal Kadanoff 
type real space renormalization-group (MKRG) studies \cite{BB87,NH93}.
Especially, the issue of {\it temperature-chaos}, i.~e. the sensitivity 
of glassy phases with respect to a small change of temperature, 
has been of great interest because of its potential relevance 
for the rejuvenation 
(chaos) effects found in temperature-shift and temperature-cycling
experiments \cite{JVHBN,sg-experiment-review,JYN02}. 

The majority of the previous theoretical and numerical studies
concerning the problem of the fragility of glassy phases 
has been done on Edwards-Anderson (EA) spin-glass models, which has been
considered as prototypical model for glassy systems. 
While a rich amount of numerical evidences for the anomalous response 
to non-thermal perturbations has been accumulated 
\cite{BM87,R94,AFF95,RSBDJ96,N97}, 
the intriguing problem of temperature-chaos remains very controversial. 
For the Sherrington-Kirkpatrick (SK) mean-field spin-glass model, which 
is the EA model embedded in infinite dimensional space. 
It is realized that saddle point solutions both with and without 
temperature-chaos \cite{Sompolinski,Kon89,KV93,FN95,R01} exist
and apparently new theoretical ideas are needed.
On the other hand, numerical studies report conflicting 
results \cite{R94,AFF95,N97,HK97,BM00,MPP00,BM02}. 

Recently we developed an analytical scheme to study the fragility 
of the free-energy landscape of randomly frustrated systems 
against various kind of perturbations \cite{SY}. Especially we proposed 
to prove the onset of chaos in terms of statistical decoupling 
of a set of replicated partition functions, and we applied the method to the directed polymer in random media (DPRM).
The DPRM \cite{HZ95} is a simple model compared to spin-glass models. 
In spite of this, it is believed to possess many of the subtle properties 
of glassy systems, thus, it deserves to be called as  
``baby spin-glass''\cite{M90}. Indeed, the anomalous response of DPRM
towards various kinds of weak perturbations has already been reported 
by many numerical studies  \cite{M90,Z87,FV88,S91,HH93} including a signature
of temperature-chaos \cite{FH91}.
DPRM belongs to the wide class of elastic manifolds in random media 
\cite{MP91,BF93,BBM,GD,PGD00}
which encompasses a variety of physical systems of much
interest such as the domain walls of ferromagnets \cite{HH85,MDW}
with weak bond randomness and the flux lines in type-II superconductors 
with randomly distributed point like pinning centers
\cite{blatt,bolle}, CDW  and vortex lattice systems 
with weak random-periodic pinnings \cite{FL78,Gru88,GD}.

The scope of this paper is to present a unified study on the fragility
of the free-energy landscape of DPRM with respect to various
weak perturbations using the replica Bethe ansatz 
approach outlined in \cite{SY}, mapping to a modified Sinai model 
and numerically exact transfer matrix calculations. 
Our main results are the following. We find that infinitesimally weak 
perturbations amount to replica symmetry breaking terms in the effective
action which lead to the statistical decoupling of two sets of replicas. 
The outcome can be naturally understood as a manifestation of
spontaneous replica symmetry  breaking following the definition 
of Parisi and Virasoro \cite{PV89}. Interestingly enough, the
replica approach turns out to give results quite consistent with
the phenomenological scaling approach \cite{FH91,M90,Z87,FV88,S91,HH93}
and predict the same overlap length $L_{c}(\delta)$. 
Within the replica approach, apparently different
perturbations can be naturally classified into a few universality classes. 
Concerning the well known correspondence between the effective
free-energy landscape of DPRM and the Sinai model, the statistical 
decoupling of replicas (chaos) naturally suggests the emergence of 
statistically independent Sinai valleys for different subsets of replicas.
To examine the anticipated universal aspects of the anomalous response,
we present and discuss the outcome of a detailed numerical analysis
using transfer matrix methods. 

The plan of the paper is the following. In the next sections we 
propose a general framework to define and study the fragility of the
free-energy landscape of randomly frustrated systems.
In  section \ref{sec.model}, we define the DPRM model.
In section \ref{sec.scaling}, we review and summarize 
the previous scaling arguments. In section \ref{sec.replica}
we present details of the replica Bethe ansatz approach outlined in \cite{SY}. 
Then, in section \ref{sec.numerical}, we present the outcome of an  
exhaustive numerical analysis using the transfer matrix method.
Finally, we summarize our results in section \ref{sec.conclusion}.

\section{Statistical Decoupling of Real Replicas}
\label{sec-statistical-decoupling}

In this section we discuss a general strategy to study the sensitivity 
of the free-energy landscape of a generic class of systems. 
The free-energy $F$ of a random system is a random quantity with 
certain mean and variance. Let us denote the deviation of 
the free-energy of a given sample from the mean as,
\be
\Delta F= F-\overline{F}.
\ee
Here and hereafter $\overline{(\cdots)}$ denotes the average over 
different realizations of randomness.

Now let us consider two systems say A and B. Initially they
are prepared as two identical copies with the same randomness, 
temperature and other parameters. Such systems are called
as {\it real replicas}.
We are interested in how the statistical correlation between
A and B changes by introducing a perturbation of strength $\delta$.
Then it is useful to define a disorder-averaged correlation function,
\begin{equation}
C_{F}(L, \delta) = \frac{\overline{\Delta F_{A}(L)\Delta F_{B}(L)}}{
\sqrt{\overline{\Delta F_{A}^{2}(L)}}
\sqrt{\overline{\Delta F_{B}^{2}(L)}}}
\label{eq-c-f}
\end{equation}
If the correlation function vanishes at large length scales 
\be
\lim_{\delta \to 0}\;\lim_{L\to \infty}~C_F(L,\delta)\to 0
\label{eq-cf-vanish}
\ee
it implies the free-energy landscape of A and B decorrelates completely.
If the statistical decoupling between A and B 
happens even with arbitrarily weak perturbation $\delta  \ll 1$, 
we say that there is {\it chaos}.

Now let us consider an equivalent definition of chaos which is 
more suited for analytical approaches based on the replica method. 
Let us suppose that each of the systems A and B are replicated further into
$n$ replicas and consider the disorder-averaged partition 
function of the total system  $Z_{A+B}^{n}(L)$.
As noticed by Kardar \cite{K87}, if an analytical continuation for $n \to 0$ is
possible, the disorder average of such a partition function 
can be identified as the generator of cumulant correlation functions
of sample-to-sample fluctuations of free-energies \cite{K96}.
Thus the complete knowledge of the disorder average of 
the replicated partition function 
allows one to obtain the distribution function of sample-to-sample 
fluctuation of the free-energy \cite{GB}.
In our present context, the disorder-averaged partition function 
$\overline{Z_{A+B}^{n}(L)}$ generates cumulant correlation functions of the 
total free-energy as the following,
\begin{eqnarray}
\lim_{n \rightarrow 0} \ln \overline{Z_{A+B}^{n}(L)}
&=& n \overline{ \ln Z_{A+B}(L)} 
+ \frac{n^{2}}{2} \overline{[\ln Z_{A+B}(L)]_{c}^2}
+ \ldots +\frac{n^{p}}{p!} \overline{[\ln Z_{A+B}(L)]_{c}^p} +\ldots \nonumber \\
&=& n \overline{[-\beta_A F_{A}(L)-\beta_B F_{B}(L)]}
+ \frac{n^{2}}{2} \overline{[-\beta_{A} F_{A}(L)-\beta_{B} F_{B}(L)]_{c}^2}
\nonumber \\
&+& \ldots +\frac{n^{p}}{p!} \overline{[-\beta_A F_{A}-\beta_B F_{B}]_{c}^p} 
+ \ldots
\label{eq5}
\end{eqnarray}
where $[\ldots]^{p}_{c}$ stands for $p$-th cumulant correlation functions of 
the total free-energies $-\beta_A F_{A}-\beta_B F_{B}$ 
with $F_{A}(L)$ and $F_{B}(L)$ being free-energies of
subsystems A and B respectively and  $\beta_{A}$ and $\beta_{B}$ being
inverse temperatures of A and B respectively.

Obviously, the decorrelation of the free-energy fluctuations
between A and B is equivalent to 
the factorization of the replicated partition function,
\begin{equation}
\lim_{\delta\to 0}\;\lim_{L\to \infty}\;\lim_{n\to 0} 
\qquad \overline{Z^{n}_{A+B}(L,\delta)}
=\;\overline{Z^{n}_{A}(L)}\;\times\;\overline{Z^{n}_{B}(L)}
\label{eq-dec}
\end{equation}
Note that if the latter result holds, automatically\eq{eq-cf-vanish} holds too.
An important remark is that the order in which limits 
are taken is crucial to obtain sensible results: 
the limit {\it $n \to 0$  {\it must} be taken  before than the thermodynamic 
limit $L \to \infty$} and finally the limit $\delta\to 0$ must be taken.  
In what follows, we will use \eq{eq-dec}  as our definition of chaos
in the replica approach. We have to stress, though, that this 
definition is general and holds for generic random systems. 

The above definition of chaos implies that it can be regarded 
as a spontaneous symmetry breaking phenomenon. If the perturbation
is absent, A and B are equivalent and one expects the exchange 
symmetry $A \leftrightarrow B$ to be present. One also expects to have permutation symmetry
among the replicas associated to each group  A or B. 
Such an invariance under permutations is usually called in short as
replica symmetry. However, in general, 
it turns out that the disorder-averaged replicated partition function of 
the $2 \times n$ replicas $\overline{Z^{n}_{A+B}}$
without any perturbation has an  even higher symmetry: it is
invariant under any permutation among the $2 \times n$ replicas.
Now, if \eq{eq-dec} holds, this higher  symmetry is reduced: after having introduced a perturbation the permutation symmetry remains at most within 
each subset associated with A and B. Thus in order the this phenomena
happens, the perturbation should show up in the replicated partition
function as a symmetry breaking term which tries to break the full
permutation symmetry. Now, the definition \eq{eq-dec}
tells us that this symmetry breaking happens even with an arbitrary weak perturbation. Therefore chaos defined as \eq{eq-dec} is a spontaneous replica symmetry breaking phenomenon.
We note that such a definition of
replica symmetry breaking was introduced under the name {\it explicit
replica symmetry} first by Parisi and Virasoro
\cite{PV89} who tried to give a sound thermodynamic definition for the
replica symmetry breaking phenomena known in the saddle point solutions of 
mean-field models \cite{P80,MFT} of a class of glassy systems.

\section{Model}
\label{sec.model}

We study DPRM in $1+1$ dimensions 
which is described by the following Hamiltonian
in the continuous limit, 
\begin{equation}
  H_{0}[V, h, \phi]   =   \int_{0}^{L} dz 
 \left [\frac{\kappa}{2}
  \left (\frac{d \phi(z)}{d z} \right)^{2}
+   V_{0}(\phi(z),z) 
\right].
\label{hamiltonian}
\end{equation}
The scalar field $\phi$  represents the displacement of the 
elastic object at point $z$ in a 1-dimensional internal space
of size $L$.  
We assume that the field $\phi$ is a single-valued function of $z$ 
which means that {\it oriented} objects with no overhangs
are considered. In the following, we assume that one end of the string
is fixed as $\phi(0)=0$  while the other end 
$\phi(L)$ is allowed to move freely.
The 1st term in the Hamiltonian is the elastic energy, $\kappa$ 
being the elastic constant. The random pinning media is modeled by 
the quenched random potential $V_{0}(\phi,z)$ with zero mean
and short-ranged spatial correlation, 
\begin{equation}
\overline{V_{0}(\phi,z)}=0 \qquad
\overline{V_{0}(\phi,z)V_{0}(\phi',z')} =
2D\delta(\phi-\phi')\delta(z-z')
\label{random_potential_correlator}
\end{equation}
Many exact properties of this $1+1$ dimensional model are known 
\cite{HZ95}. 
It is in the frozen phase at all finite temperatures 
in the sense that its scaling properties are always governed 
by the $T=0$ glassy fixed point.

We implement the basic strategy explained in the previous section
as the following.
First, we start with a system of two real replicas, say A and B, 
whose configurations  $\phi_{A}(z)$  and $\phi_{B}(z)$ are subjected to
exactly the same random potential and temperature. Second, 
we apply small perturbations to them.
In the present paper we consider 5 different kinds of perturbations:
\begin{itemize}
\item[I)] {\it Tilt field} \cite{M90}: A and B replicas are subjected
to a tilting field of opposite sign
$
-h \phi_A(L) +h \phi_B(L)
$
with $h \ll 1$.
\item[II)]  {\it Explicit short-ranged repulsive 
coupling} \cite{P90,M90}: A and B replicas are subjected to
explicit repulsive short-ranged interaction
$ \epsilon \int_{0}^{L}dz \delta (\phi_{A}(z)-\phi_{B}(z))$
with $0 < \epsilon \ll 1$.
\item[III-i)] {\it Decorrelation of random potential} \cite{Z87,FV88}: the random potential
of B is made from that of A
as $V_{B}=(V_{A}+\delta V')/\sqrt{1+\delta^{2}}$
where $|\delta| \ll 1 $  and $V'$ follows the same Gaussian distribution
as $V$. Then $\overline{V_{G}(\phi,z)}=0$ 
and $\overline{V_{G}\phi,z)V_{G'}(\phi',z')} =
2D_{G G'}\delta(\phi-\phi')\delta(z-z')$
with $D_{AA}=D_{BB}=D$ and $D_{AB}=D/\sqrt{1+\delta^{2}} < D$.
\item[III-ii)] {\it Random Tilt Field}: A and B 
are subjected to statistically independent weak random tilt field.
\item[III-iii)]{\it Temperature difference} \cite{FH91}:
slightly different temperatures $T_{A}=T + \delta T$  $T_{B}=T - \delta T$ 
for A and B respectively with $\delta T/T \ll 1$
\end{itemize}

\section{Droplet Scaling Approach}

\label{sec.scaling}

We first review and discuss the scaling approach picture
\cite{FH91,M90,Z87,FV88,S91,HH93} for the problem of the anomalous
response. Let us consider a simple-minded picture 
consisting in the deepest valley corresponding 
to the ground state configuration
and many branched valleys of low (free-)energy excitations
which for given longitudinal size $L$ differ from the 'ground state' 
over a transverse size $u_{0}(L/L_{0})^{\zeta}$, $\zeta$ being the so called roughness
exponent. Note that we have introduced a characteristic longitudinal length $L_{0}$,
which should be understood as the Larkin length \cite{Larkin}
beyond which pinning becomes important, as well as  its associated
transverse length scale $u_{0}$.
The free-energy gap of these excited states with respect to the 'ground
state'  is expected to scale typically as,
\begin{equation}
\Delta F^{\rm typ}_{L}=U_{0}(L/L_{0})^{\theta}.
\label{eq:typ-gap}
\end{equation}
Here $U_{0}$ is the energy scale associated with the Larkin length
and $\theta$ is the stiffness exponent which is related to 
the roughness exponent $\zeta$ by the exact scaling relation,

\begin{equation}
\theta=2\zeta-1.
\label{eq:theta-zeta}
\end{equation}
In a $1+1$ dimensional system these exponents are believed to be
exactly $\theta=1/3$ and $\zeta=2/3$ \cite{HH85,K87,HZ95}.
The probability distribution function of the free-energy gap $\Delta F_{L}$
is expected to have a natural scaling form,
\begin{equation}
\rho_{L}(\Delta F_{L}) d (\Delta F_{L}) = 
\tilde{\rho}\left( \frac{\Delta F_{L}}{U_{0}(L/L_{0})^{\theta}} \right)
\frac{d (\Delta F_{L})}{U_{0}(L/L_{0})^{\theta}},
\end{equation}
with non-vanishing amplitude at the origin,
\begin{equation}
\tilde{\rho}(0) > 0,
\end{equation}
which allows rare, gap-less excited states  \cite{FH91,M90}.

Let us now consider a generic perturbation 
which triggers an excitation from the 'ground state' with a 
{\it free-energy gain} of order,
\begin{equation}
\delta U \left(\frac{L}{L_{0}} \right)^{\alpha},
\label{eq:dU}
\end{equation}
in the {\it infinitesimally weak perturbation limit},
\begin{equation}
\delta U/U_{0} \to 0.
\end{equation}
In the following we consider perturbations such that $\alpha > \theta$.
Under the influence of such a perturbation, the system in the
deepest valley may jump into other valleys with free-energy gap $\Delta F$
if the possible gain of free-energy due to perturbation \eq{eq:dU}
becomes larger than the original free-energy gap itself.
The probability of such an event  is estimated as,
\begin{equation}
p_{\rm jump}(L, \delta U)=
\int_{0}^{\delta U (L/L_{0})^{\alpha}}
\rho_{L} \left( \Delta F_{L} \right) d\;\Delta \;F
\sim \left ( \frac{L}{L_{c}(\delta U)}\right)^{\alpha-\theta},
\label{eq:prob-jump}
\end{equation}
with a characteristic length scale called {\it overlap length},
\begin{equation}
L_{c}(\delta U) 
\sim L_{0} \left (\frac{\delta U}{U_{0}} \right)^{-1/(\alpha-\theta)}
\qquad \mbox{as} \qquad \delta U/U_{0} \to 0.
\label{lc}
\end{equation}
Let us also define a characteristic transverse length scale which is 
conjugate to $L_{c}(\delta U)$,
\begin{equation}
u_{c}(\delta U)=u_{0} \left (\frac{L_{c}(\delta U)}{L_{0}}\right)^{\zeta}.
\end{equation}

It is important to note that the above expressions make 
sense only for short enough length scales $L \ll L_{c}(\delta U)$. 
In this regime the effect of the jumps on physical quantities 
can be analyzed in a perturbative  
way because the probability of a jump is small enough.
Let us call this regime  {\it weakly perturbed regime}.
However, in {\it strongly perturbed regime} $L \gg L_{c}(\delta U)$,
perturbative treatments will fail because 
jump events will happen with probability one.
The latter implies that after having 
applied the perturbation the free-energy landscape 
is drastically different from the
original on lengthscales larger than the overlap length.
The overlap length \eq{lc}
diverges as $\delta U/U_{0} \to 0$ with exponent $-1/(\alpha-\theta)$
which is sometimes called as chaos exponent 
\footnote{In the spin-glass problems 
the chaos exponent is denoted as $\zeta$. Here we do not use the
convention because it is usually used for the roughness exponent $\zeta$.}
but remains finite for arbitrary small strength of perturbation
$\delta$.

\newcommand{\hu}{h_{\rm uni}}

\subsection{Uniform Tilt Field}
\label{subsec:perturbation-uniform}

We first consider the application of a uniform tilt field $h$ 
to the end-point of the real replica B at $z=L$
by which the statistical rotational symmetry is violated.
In the presence of the tilt field the Hamiltonian becomes,
\begin{equation}
H_{A+B} =  H_{0}[V_{0},\phi_{A}]+H_{0}[V_{0}, \phi_{B}]  
-\hu \int_{0}^{L} dz \frac{d \phi_B(z)}{dz}
\label{unif}
\end{equation}
The unperturbed Hamiltonian $H_{0}$ is given in \eq{hamiltonian}. 
If the string makes a jump responding to the 
uniform tilt field over a distance of order $u_{0}(L/L_{0})^{\zeta}$
into the next valley, it obtains an energy gain of order 
$\hu u_{0}(L/L_{0})^{\zeta}$. Thus the unit for the gain 
in energy \eq{eq:dU} reads as  $\delta U= \hu u_{0}$ with 
characteristic exponent $\alpha=\zeta=2/3$.
Therefore we find the overlap length \eq{lc} to be, 
\begin{equation}
L_{c}(\hu) \sim L_{0}\left(\frac{\hu u_{0}}{U_{0}}\right) ^{-3} 
\label{lc-uniform-tilt}
\end{equation}
This result was previously obtained by M\'{e}zard in \cite{M90}  by using
essentially the same argument and supporting his result by a numerical 
transfer matrix calculation.

\subsection{Explicit Repulsive Coupling}
\label{subsec:perturbation-rsb-1}

The other perturbations that we consider do not break the statistical 
rotational symmetry. First we consider the case of having a short-ranged 
repulsive coupling between the two real replicas by which
the total Hamiltonian becomes,
\begin{equation}
H_{A+B}=  H_{0}[V_{0}, \phi_A]+ H_{0}[V_{0},\phi_B]+ 
\epsilon \int_{0}^{L}dz \delta(\phi_A(z)-\phi_B(z))
\end{equation}
with $\epsilon >0$. 

This type of perturbation was first considered by Parisi and Virasoro
\cite{PV89} in the context of spin-glass models 
in order to give a precise definition of spontaneous replica symmetry
breaking. It explicitly breaks the RS noted in section \ref{sec.intro}.
It was also used in the DPRM problem  by Parisi in \cite{P90} and
was further examined  by M\'{e}zard using the numerical transfer 
matrix method \cite{M90}.\\
If the two replicas jump into different valleys avoiding to touch with
each other, the energy is reduced by an amount of order 
$\epsilon (L/L_{0})$. Thus we read off  $\alpha=1$ and $\delta U= \epsilon$
so that the overlap length \eq{lc} becomes,
\begin{equation}
L_{c}(\epsilon) 
\sim L_{0}  \left( \frac{\epsilon}{U_{0}} \right)^{-3/2} 
\label{lc-rep}
\end{equation}
Again this length scale agrees with the result obtained by  M\'{e}zard for the same quantity
 in \cite{M90}.

\subsection{Potential Change, Random Tilt Field and Temperature Change}
\label{subsec:perturbation-rsb-2}

Now we introduce three other kinds of perturbations which
do not break rotational symmetry.
As we explain in section \ref{sec.replica}, this class of perturbations
also breaks the RS noted 
in section \ref{sec.intro}.
However, the strength of perturbation is sub-extensive 
$\sim L^{1/2}$ $(\alpha=1/2)$
and much weaker than in the case of explicit repulsive coupling which is
extensive $\sim L$  $(\alpha=1)$.

\subsubsection{Potential Change}
\label{subsubsec:perturbation-rsb-2-potential}

We consider three different perturbations of
$\alpha=1/2$. 
The first one is to introduce a  small difference between the  realizations of the
pinning potential for A and B \cite{Z87,FV88,S91,HH93}. Suppose that
A has a certain realization of the pinning potential $V_{0}$. 
Then we can construct  the potential for B as the sum of $V_{0}$ and 
a new statistically independent
random number $V_{1}$. Then the total Hamiltonian becomes,
\begin{equation}
H_{A+B}=  H_{0}[V_{0}, \phi_A]
+H_{0}[(V_{0} +\delta V_{1})/\sqrt{1+\delta^2}, \phi_B].
\label{hamiltonian-potential}
\end{equation}
Here $\delta$ is the strength of the perturbation and $V_1$ has
the same statistical properties as $V_{0}$
given in \eq{random_potential_correlator}.
Namely it has zero mean and short-ranged correlations,
\begin{equation}
\overline{V_{1}(\phi,z)V_{1}(\phi',z)} =
2 D\delta(\phi-\phi')\delta(z-z').
\qquad \overline{V_0(\phi,z)V_1(\phi',z')} =0.
\label{random_potential_correlator_cross}
\end{equation}
Note that the pinning potential for  replica B is normalized by
the factor $1/\sqrt{1+\delta^{2}}$, so that
it has the same amplitude as A replica.

The characteristic fluctuation of the extra energy gain along a configuration due to the
random variation of the potential 
scales typically as $\delta U_{0}\sqrt{L/L_{0}}$ 
since it gives contributions with random signs.
Thus we read off $\alpha=1/2$, $\delta U=\delta U_{0}$ 
and the the overlap length \eq{lc} becomes,
\begin{equation}
L_{c}(\delta) \sim L_{0} \delta^{-6} 
\label{lc-potential}
\end{equation}
This length scale was found by Feigel'man and V. M. Vinokur
by essentially the same argument \cite{FV88}. 
Previous numerical calculations \cite{Z87,HH93} appear
consistent with it but the anticipated crossover phenomena
had remained to be clarified.

\subsubsection{Random Tilt Field}
\label{subsubsec:perturbation-rsb-2-random-tilt}

Similarly, we consider the application of a random tilt field to the end-point of B,
\begin{equation}
H_{A+B} =  H_{0}[V_{0},0, \phi_{A}]+H_{0}[V_{0},0, \phi_{B}]  
- \delta \int_{0}^{L} dz h(z) \frac{d \phi_B(z)}{dz}
\label{eq-random-tilt}
\end{equation}
Here $\delta$ is the strength of the perturbation and 
$h(z)$ is a Gaussian random number with zero mean and short-ranged correlations,
\begin{equation}
\overline{h(z)}=0 \qquad 
\overline{h(z)h(z')} =2 \delta(\phi-\phi')\delta(z-z').
\label{eq-random-tilt-2}
\end{equation}
Within the lattice model we study numerically, the energetic gain of
energy typically scales again as $\delta U_{0} \sqrt{L/L_{0}}$. Thus
we find $\alpha=1/2$ and $\delta U=\delta U_{0}$ which gives
the overlap length,
\begin{equation}
L_{c}(\delta) \sim L_{0}\delta ^{-6} 
\label{lc-random-tilt}
\end{equation}

\subsubsection{Temperature Change}
\label{subsubsec:perturbation-rsb-2-temperature}

All perturbations discussed so far are non-thermal perturbations.  
Finally we consider the introduction of a slight temperature difference between
the two real replicas A and B,
\begin{eqnarray}
T_A &=& T+\delta T_{A} \nonumber \\
T_B &=& T+\delta T_{B}
\end{eqnarray}
where $\delta T_{A} \neq \delta T_{B}$. Although this perturbation 
appears to be rather different from the two other cases above,  
it is also expected to give $\alpha=1/2$ based on the following observation.

Fisher and Huse\cite{FH91} conjectured that
valley-to-valley fluctuations of 
the energy and the entropy are just that of a sum of random variables 
put on a string of length $L$. Thus the amplitude of 
valley-to-valley fluctuation scales as,  
\begin{eqnarray}
 \Delta S(L) & \sim & \kb (L/L_{0})^{1/2} \nonumber \\
 \Delta E(L) & \sim & U_{0} (L/L_{0})^{1/2}
\label{eq:se}
\end{eqnarray}
%
However, it is argued that the free-energy is
optimized so that these wild fluctuations cancel with each other
as much as possible 
in such a way that valley-to-valley fluctuations of the free-energy
is much smaller,
\begin{equation}
 \Delta F(L) = U_{0} (L/L_{0})^{\theta} \qquad \mbox{with} \qquad \theta < 1/2.
\end{equation}
In other words, there is a strong {\it negative} correlation between the fluctuations 
of entropy and energy such that
\begin{equation}
(\Delta S/\kb)  (\delta E/U_{0}) \sim   -(L/L_{0}) < 0
\end{equation}
due to the thermodynamic relation $\Delta F=\Delta E-\kb T \Delta S$.
Note that a similar argument lies at the heart of the droplet theory for 
spin-glasses which suggests temperature-chaos \cite{BM87,FH88}.
Actually the exponent for the free-energy fluctuations is believed to be
exactly $\theta=1/3$ which is definitely smaller than $1/2$.
Furthermore, the stronger fluctuation of entropy 
and energy \eq{eq:se} was confirmed numerically by 
a transfer-matrix calculation
while the smaller fluctuation of free-energy with $\theta=1/3$  was also
observed simultaneously \cite{FH91}\cite{note-energy-entropy}.
Then under a slight temperature-difference between the two replicas A and B, 
it is possible that one of the replicas jumps into a different valley 
taking advantage of 
the large gain in entropy. Such a gain should typically scale as 
$\kb |\delta T_{A}- \delta T_{B}| (L/L_{0})^{1/2}$ and therefore $\alpha=1/2$
and $\delta U= \kb |\delta T_{A}- \delta T_{B}|$.
From \eq{lc}, one then finds the overlap length as, 
\begin{equation}
L_{c}(\delta T) \sim  \left (\frac{\kb |\delta T|}{U_{0}} \right)^{-6} 
\label{lc-temperature}
\end{equation}
with $\delta T=\delta T_{A}-\delta T_{B}$.
This length scale was found by Fisher and Huse in \cite{FH91}.
Indeed their transfer matrix calculation presented in \cite{FH91}
suggests the existence of crossover phenomena. However, details of the 
scaling properties and comparison with the case of
the perturbation on potential have remained to be explored. 
So we try to complete the investigation in section \ref{sec.numerical}.

As we summarized above, what is crucial is the role of entropy. 
In the so called Larkin model \cite{Larkin}, in which the effect
of pinning is modeled by quenched random forces with 
short ranged correlations, entropy plays very little role 
and free-energy is dominated by  energy so that there 
is no temperature-chaos (see \cite{FLN}).

\subsection{Moments of transverse jump distances}  
\label{subsec:b}

In order to characterize the jump events triggered by the perturbations,
it is useful to introduce appropriate correlation functions.
First, let us introduce the disorder average of the q-th moment of the
transverse distance between the end points of the two real replicas,
\begin{equation}
B_{q}(L,\delta U)=\overline{(\Phi_{A}(L)-\Phi_{B}(L))^{q}}.
\end{equation}
It was introduced and studied numerically 
by Zhang in \cite{Z87} and continued
more in \cite{HH93} for the case of perturbation on the random potential.
The following is an extension of the argument by 
Feigel'man and V. M. Vinokur described in \cite{FV88}.

\subsubsection{Weakly Perturbed Regime}
\label{subsubsec:b-weak}

In the weakly perturbed regime $L \ll L_{c}(\delta U)$, a jump event
happens with a probability smaller than $1$ as given in \eq{eq:prob-jump}.
By a single event, a transverse displacement of order $u_{0}(L/L_{0})^{\zeta}$
will take place. Thus we expect 
\begin{equation}
B_{q}(L,\delta U) \sim \left[u_{0} \left (\frac{L}{L_{0}} \right)^{\zeta} \right]^{q}
p_{\rm jump}(L, \delta U)
\sim u_{c}(\delta U)^{q} \left(\frac{L}{L_{c}(\delta U)}
 \right)^{(q-2)\zeta+\alpha+1} \qquad L \ll L_{c}(\delta U),
\label{eq:b-weak}
\end{equation}
where in the last step  we have used the scaling relation \eq{eq:theta-zeta}.

\subsubsection{Strongly Perturbed Regime}
\label{subsubsec:b-strong}

In the  strongly perturbed regime $L \gg L_{c}(\delta U)$, the jump events 
with longitudinal size $L_{c}(\delta U)$ and transverse size 
$u_{c}(\delta U)$ will take place with probability one. 
Let us first consider the behavior of the 1st moment 
$B_{1}(L, \delta U)$ in this regime.

In the strongly perturbed regime, the two replicas A and B
are subjected to very different free-energy landscapes.
In such a situation, we expect that the two replicas A and B 
will make excursions independently. Thus we expect 
a simple scaling form,
\begin{equation}
B_{1}(L,\delta U)= u_{c}(\delta U)
\left (\frac{L}{L_{c}(\delta U)} \right)^{\zeta}
 \qquad L \gg L_{c}(\delta U).
\label{b-strong-1}
\end{equation}

However, the situation is slightly different in the case 
of uniform tilt field considered 
in section \ref{subsec:perturbation-uniform}.
Because the
uniform tilt field continues to increase the separation 
between A and B systematically   as $L \to \infty$. 
After making a transverse jump of order 
$u_{0}(L/L_{0})^{\zeta}$ another jump into a further valley in the
direction of the field can take place if
the strength of the field $h$ is increased further.
The latter happens when the new  increment of the ``Zeeman energy'' 
$\delta h u_{0}(L/L_{0})^{\zeta}$ due to another increment of 
the field $\delta h$ becomes again comparable to the typical 
free-energy gap $\Delta F^{\rm typ}(L)$ given in \eq{eq:typ-gap},
\begin{equation}
\delta h u_{0}(L/L_{0})^{\zeta} \sim \Delta F^{\rm typ}(L).
\end{equation}
The number of times that such a sequence of jumps 
happens by increasing the field from $0$ to $h$
will be typically $h/\delta h$.
Each jump will have a typical transverse size of order $u_{0}(L/L_{0})^{\zeta}$.
Thus the 1st moment grows as,
\begin{equation}
B_{1}(L,h u_{0})= u_{0}\left(\frac{L}{L_{0}}\right)^{\zeta} \frac{h}{\delta h}
=u_{c}(hu_{0})\left (\frac{L}{L_{c}(h u_{0})} \right)
 \qquad L \gg L_{c}(h u_{0})
\label{b-strong-1-uni}
\end{equation}
In the last equation, we used the scaling relation \eq{eq:theta-zeta}.
Note that the first moment $(q=1)$ grows linearly with $L$ not only in the
strongly perturbed regime but also in the
weakly perturbed regime as one can see using $\alpha=\zeta$ in
\eq{eq:b-weak}.  Actually the linear growth of the 1st moment can be proved
rigorously using the statistical rotational (tilt) symmetry of the
system \cite{SVBO,M90}. This is a rather special property of the
1st moment. All other moments are sensitive to the crossover from
weak to strong perturbation regimes.

Let us now consider higher moments $q>1$.
Since jump events are {\it typical} in the strongly perturbed regime,
we generically expect a simple relation between different moments, 
\begin{equation}
B_{q}(L,\delta U)=u_{c}^{q}(\delta U) 
\left (\frac{B_{1}(L,\delta U)}{u_{c}(\delta U) } \right)^{q}
 \qquad L \gg L_{c}(\delta U).
\label{b-strong-q}
\end{equation}
where note that the natural unit for 
the q-th moment is now $u_{c}^{q}(\delta U)$.
Note that in the weakly perturbed regime such a simple relation
between different moments does not hold because of the rareness of
the jump events. The 1st moment obeys a scaling law such that
$B_{1}(L,\delta U)/u_{c}(\delta U)$ is a function of 
$L/L_{c}(\delta U)$ also in the strongly perturbed regime 
as we mentioned above. This implies
that the higher moments $(q>1)$ obey a scaling law such 
that $B_{q}(L,\delta U)/u^{q}_{c}(\delta U)$ 
becomes a function of $L/L_{c}(\delta U)(\gg 1)$ in the strongly
perturbed regime.

\subsubsection{Summary}

To summarize, we expect a generic scaling form for 
the behavior of the q-th moment including both 
weakly and strongly perturbed regimes as,
\begin{equation}
B_{q}(L,\delta U)=u^{q}_{c}(\delta U)
\tilde{B}_{q}\left(  \frac{L}{L_{c}(\delta U)} \right).
\end{equation}
Here the scaling function presents the asymptotic forms in
weakly perturbed regime $L \ll L_{c}(\delta U)$ and
strongly perturbed regime $L \gg L_{c}(\delta U)$
which we discussed above.

\subsection{Overlap function}
\label{subsec:q}

Another useful quantity to probe the jump events 
is the overlap function defined as \cite{P90,M90},
\begin{equation}
q(L, \delta U) = \frac{1}{L}\int_{0}^{L} dz 
\delta(x_{0}(z)-x_{\delta U}(z)).
\label{eq.def-q}
\end{equation}
We expect it to scale as,
\begin{equation}
q(L,\Delta U)= \tilde{q} \left ( \frac{L}{L_{c}(\delta U)}\right).
\label{eq-scaling-q}
\end{equation}
Note that $1-q$ is essentially the probability that the string 
jumps to a different valley. Thus in the weakly perturbed regime 
$L \ll L_{c}(\delta U)$, 
we expect that it behaves as,
\begin{equation}
1-q(L,\delta U) \sim  p^{\rm jump}(L,\delta U) \sim
\left ( \frac{L}{L_{c}(\delta U}\right)^{\alpha-\theta}.
\end{equation}
In the strongly perturbed regime $L \gg L_{c}(\delta U)$, 
we expect that $q$ (i.~e. the probability of staying in the same valley) 
decays faster down to $0$ as $L/L_{c}(\delta)\rightarrow \infty$ because the 
free-energy landscapes of the two replicas are increasingly 
different there.

\subsection{Correlation of the free-energy fluctuation}
\label{subsec:cf}

In order to probe the difference of free-energy landscapes 
between the perturbed and unperturbed systems, we study
the correlation function \eq{eq-c-f} of the sample-to-sample fluctuations of
the free-energy between the two systems  which reads as,
\begin{equation}
C_{F}(L, \delta U) = \frac{\overline{\Delta F(L,0)\Delta F(L,\delta U)}}{
\sqrt{\overline{\Delta F^{2}(L,0)}}\sqrt{\overline{\Delta F^{2}(L,\delta U)}}}.
\end{equation}
here $\Delta F(L,0)$ and $\Delta F(L,\delta U)$ are deviations from 
the mean free-energy of the unperturbed and perturbed systems.
A similar correlation function was studied numerically for the
case of perturbation of temperature-shift in \cite{FH91}.
We expect it to scale as,
\begin{equation}
C_{F}(L,\Delta U)= \tilde{C}_{F} \left ( \frac{L}{L_{c}(\delta U)}\right),
\end{equation}
and decay down to $0$ as $L \rightarrow \infty$.

A possible functional form of the correlation function 
in the weakly perturbed regime $L \ll L_{c}(\delta)$ can be guessed
by a simple argument proposed by Bray and Moore \cite{BM87} 
for the equivalent problem in a spin-glass model. 
First, we are considering 
perturbations such that perturbed and unperturbed
system have the same statistical properties. Thus we must have,
\begin{equation}
\sqrt{\overline{\Delta^{2} F(L,\delta U)}}
=\sqrt{\overline{\Delta^{2} F(L,0)}}
\sim U_{0}\left(\frac{L}{L_{0}}\right)^{\theta}.
\label{eq:same-amp}
\end{equation}
Suppose that we introduce a perturbation which scales as $\delta U (L/L_{0})^{\alpha}$
as given in \eq{eq:dU}. Then the fluctuations of the free-energy of 
the perturbed system have two contributions: the original fluctuation $\Delta F(L,0)$ plus the change due
to the perturbation, 
\begin{equation}
\Delta F(L,\delta U)=\frac{1}{\cal N} 
\left(\Delta F(L,0)
+\delta U \left(\frac{L}{L_{0}}\right)^{\alpha}\right)
\label{eq:pert-amp}
\end{equation}
Here ${\cal N}$ is a normalization factor which assures that the
statistics of the perturbed and unperturbed systems remain the same
 as in\eq{eq:same-amp}.  It is assumed that the two terms between brackets in expression  \ref{eq:pert-amp} 
 are uncorrelated.
When performing the average over the disorder, 
the cross-terms due to the two terms in \eq{eq:pert-amp} 
cancel out to give  the following scaling function 
for the correlation function,
\begin{equation}
C_{F}(L,\Delta U) \sim \frac{1}{\cal N} 
\sim \left[ 1+ \left(\frac{\delta U}{U_{0}}
\frac{L}{L_{0}}\right)^{2(\alpha-\theta)} \right]^{-1/2}.
\label{eq-cf-scaling-form}
\end{equation}
For strongly perturbed regime $L \gg L_{c}(\delta)$, the correlation
function may decay faster.

\section{Replica Bethe Ansatz Approach}
\label{sec.replica}

Now let us take the replica approach introduced
in section \ref{sec-statistical-decoupling} to study chaos.
We start from the partition function of $2\times n$ replicas:
A and B and their $n$ copies.
It can be expressed by a path integral
over all possible configurations of $2 \times n$ replicas labeled 
by two indices $G=A,B$ and $\alpha=1,\ldots,n$,
\begin{eqnarray}
&& \overline{Z_{A+B}^{n}(L)} =
\int  \prod_{G=A,B}\prod_{\alpha=1}^{n}{\cal D}\phi_{G,\alpha}
\exp \left(-  S_{A+B}[\phi_{G,\alpha}] \right).
\label{partition-function-start}
\end{eqnarray}
where we have introduced the dimension-less effective action,
\begin{eqnarray}
 S_{A+B}[\phi_{G,\alpha}] &=&
\int_{0}^{L}dz \left[ \sum_{G,\alpha} 
\frac{\kappa}{2 \kb T} \left ( \frac{d \phi_{G,\alpha}(z)}{d z} \right)^{2} 
-
\frac{D}{(\kb T)^{2}} \sum_{G,G',\alpha,\beta}
\delta (\phi_{G,\alpha}(z)-\phi_{G',\beta}(z))
\right]
\label{action}
\end{eqnarray}
To obtain the last equation we have used \eq{random_potential_correlator}.
Here one end of each replica
is fixed as $\phi_{G,\alpha}(0)=0$  while the other end 
$\phi_{G,\alpha}(L)$ is allowed to move freely as we noted above.

The effective action \eq{action} has several important symmetries. 
First, it has a symmetry under global rotation 
in the $(z,\phi)$ plane. Second, it is symmetric under 
all possible permutations among the $2 \times n$ replicas.
Let us call the latter as 'RS' (replica symmetric) for simplicity. 
As we explained in section \ref{sec-statistical-decoupling}
our primary interest is how the RS is broken by infinitesimally
weak perturbations.

%

Now we focus on the study of the disorder averaged partition function
$\overline{Z_{A+B}^{n}(L)}$. 
To this respect we will use the well known mapping 
to an $n$-body imaginary time 
quantum mechanical problem in 1-dimensional space, 
which was also firstly noted by Kardar \cite{K87,K96,EK00}. 
The advantages of this approach is that
one can make use of the Bethe ansatz which provides us with the  exact  ground state of the quantum problem. 
Moreover, from the latter one gets many hints about how to construct the relevant excited states.
In what follows the main steps in this procedure are outlined to
emphasize several points which will become relevant in the analysis 
of the perturbation. 
The path-integral of the partition function 
defined in \eq{partition-function-start} through the action in \eq{action} 
can be reinterpreted as that of a quantum system in imaginary time.
In the absence of temperature difference between A and B, 
the Schr\"odinger equation reads as 
\begin{equation}
-\frac{d}{dt}\overline{Z^{n}_{A+B}(\{x_{G,\alpha}\},t)}=
{\cal H}_0\overline{Z^{n}_{A+B}(\{x_{G,\alpha}\},t)}.
\end{equation}
with the following Schr\"odinger operator for $2 \times n$-bosons,
\begin{equation}
{\cal H}_0=-\sum_{G,\alpha} \frac{\kb T}{2\kappa} \frac{\partial^{2}}{\partial x_{G,\alpha}^{2}}
-\frac{D}{(\kb T)^{2}}\sum_{((G,\alpha),(G',\beta))}\delta (x_{G,\alpha}-x_{G',\beta})~~~.
\label{eq2}
\end{equation}
  The 1st term  represents the kinetic energy.
The 2nd term stands for attractive short-ranged interactions between
the bosons where the sum is taken over all possible pairs of bosons (excluding
unphysical self-interactions which are absent in lattice models). 

Let us note that here we have two kinds of
``bosons''. The bosons of A can be distinguished
from those of B and vice versa while the bosons cannot be distinguished
from each other within the subgroups. However, the Schr\"odinger operator
has an even higher bosonic symmetry: 
it is symmetric under permutations of all the
$2 \times n$ replicas. This is nothing but the RS
we mentioned above.

By integrating out the coordinates of the free ends of the string
$(\{x_{\alpha},G\},L)$
while keeping the other ends fixed at $(0,0)$,
we formally obtain
the disorder-averaged partition function of the replicated system as,
\begin{eqnarray}
\overline{Z^{n}_{A+B}(L)} &=&
\int \prod_{G=A,B}\prod_{\alpha=1}^{n} dx_{G,\alpha}
\overline{Z^{n}_{A+B}(\{x_{G,\alpha}\},L)} \nonumber\\
&=&\sum_{\mu} e^{-L E_{\mu}} 
\int \prod_{G=A,B}\prod_{\alpha=1}^{n} dx_{G,\alpha}
<\{x_{G,\alpha}\}
|\psi_{\mu}><\psi_{\mu}|\{0\}>
\label{eq3}
\end{eqnarray}
where $|\psi_{\mu}>$ and $E_{\mu}$ are the eigenstates and
eigenvalues of the Schr\"{o}dinger operator ${\cal H}_0$ defined in (\ref{eq2}).
In the large $L$ (large time) limit, the partition function will be 
dominated by the eigen states of  the Schr\"{o}dinger operator  with
lowest eigen-values ('energies') including the ground state.

The ground-state wavefunction is well-known 
to satisfy the Bethe ansatz reading,

\begin{equation}
<\Psi_{\rm RS}|\{x_{G,\alpha}\}> 
\sim \exp \left(-\lambda\sum_{((G,\alpha),(G',\beta))}
|x_{\alpha,G}-x_{\beta,G'}|\right)~~~{\rm with}\;\:\:
\lambda=\kappa D/(\kb T)^3~~~,
\label{eqf}
\end{equation}
where the sum is taken over all possible pairs among the $2\times n$ replicas
labeled as $(G(=A,B),\alpha(=1,\ldots,n))$.
The index RS  stands for the fact that this wave-function has
the RS, i.~e. permutation symmetry 
among all $2 \times n$-replicas. In the following we label this state as replica symmetric (RS).

In general, the ground state of  one-dimensional $n$-body problems with
contact interaction is constructed in the following way: the $2 \times n$
particles are ordered and occupy a certain segment within which they
are free. The global wavefunction consists on the product of $2 \times n$ plane waves
whose moments $\lambda_{m}$ have to 
fulfill certain matching and boundary conditions which in our case result in
$\lambda_{m}=(2n+1-2m)\:\lambda$ 
with $m=1,...,2n$. 
The ground-state energy is then the sum of 
the kinetic energy of the $2 \times n$ 'free-particles',
\begin{equation}
E_{\rm g}=-\frac{\kb T}{2\kappa}\sum_{m=1}^{2n}\lambda_{m}^2
=- \frac{\kb T}{2\kappa} \frac{1}{3}\: \lambda^2\: 2n(4n^2-1) ~~~~.
\label{eq4}
\end{equation}
%
%
%

Although the ground state makes most important contribution to the partition function, it may
not be the only one. If one {\it only} takes into account the contribution of 
the ground state neglecting all other excited states, 
one would wrongly conclude from \eq{eq4} and the relation \eq{eq5}
that only the 1st and 3rd cumulants of the correlation functions 
of free-energy fluctuations exist. 
This conclusion is definitely unphysical because the  
2nd cumulant cannot be zero. Such a pathology implies existence of
continuum of  gap-less excited states which give important contributions 
to the partition function.

Orland and Bouchaud \cite{BO90} pointed out that
the translational symmetry of the Schr\"{o}dinger operator
allows to construct a continuous spectrum of excited states 
by considering center of mass (CM) motion. Such an excited state 
with wave-vector $k$ has the form,
\begin{equation}
<\Psi_{\rm{RS}, k}|\{x_{G,\alpha}\}> 
=\exp (ik \sum_{G,\alpha}x_{G,\alpha})
<\Psi_{\rm RS} | \{x_{G,\alpha} \} >
\end{equation}
with eigenvalue
\begin{equation}
E_{\rm CM}(k)=E_{\rm g}+2n \frac{\kb T}{2\kappa} k^{2}.
\end{equation}
The resultant partition function obtained by integrating out
the continuous spectrum
can be put into the following scaling form \cite{K96},
\begin{equation}
\ln\overline{Z_{A+B}^{n}}= -2n \beta \overline{f} L +g(2nL^{1/3})~~~.
\label{eq4b}
\end{equation}
where $\overline{f}$ in the 1st term represents the average free-energy
density. The function $g(x)$ in the 2nd term is analytic 
for small $x$, implying that the q-th cumulant of 
the correlation function of free-energy fluctuations scales as
$L^{q/3}$. Thus the characteristic exponent for the free-energy
fluctuation, which is called stiffness exponent \eq{eq:typ-gap}, 
is obtained as $\theta=1/3$, being consistent with extensive numerical
results of transfer matrix calculations \cite{HZ95} and other analytical
approaches such as the mapping to the noisy Burgers equation \cite{HHF85}.

Parisi \cite{P90} pointed out another important  spectrum of excited states 
in which replicas are grouped into {\it clusters} of bound states. 
Each  cluster  is supposed to be described by a 
Bethe ansatz type wavefunction so that there is replica (permutation) 
symmetry  within each cluster. An important assumption
is that these clusters are located far enough
from each other so that their mutual overlap is negligible. The latter
is allowed if the transverse size of the system is infinitely large.

In our present context, we have two kinds of bosons corresponding 
to the two real replicas A and B which can be distinguished from each
other. Thus it is natural to consider an
excited state which consists of separate Bethe type clusters
$<\Psi^{A}_{\rm RS}|$  for A and $<\Psi^{B}_{\rm RS}|$ for B,
with no mutual overlaps,
\begin{equation}
<\Psi_{\rm RSB}| = <\Psi^{A}_{\rm RS}| \times <\Psi^{B}_{\rm RS}|
\qquad <\Psi^{B}_{\rm RS}|\Psi^{A}_{\rm RS}>=0.
\label{zero-overlap}
\end{equation}
It's associated energy is readily obtained as,
\begin{equation}
E_{\rm RSB}=- \frac{\kb T}{2\kappa} \frac{1}{3}\: \lambda^2\: n
\left(n^{2}-1\right) \times 2.
\label{rsb-ene}
\end{equation}
This wavefunction has the reduced replica symmetry mentioned
in section \ref{sec.model}, i .e. it is symmetric under permutations
among A and B groups and the exchange operation $A \leftrightarrow B$.
We will call this state as replica symmetry broken (RSB) 
state in the following.

A very important feature is that the gap of the RSB excited state
with respect to the RS ground state energy, which is of order $O(n^3)$,
become vanishingly small in the $n \rightarrow 0$ limit. 
Thus such an excited state should be also taken into account
since we must take  $n \rightarrow 0$  before 
$L \rightarrow \infty$ in the
evaluation of the replicated partition function.
Presumably each cluster of bound states can have its own center 
of mass motion. Therefore the RSB excited 
state should have the continuum of excited states of CM motion similar
to that associated with the RS ground state mentioned
above. Then the resultant partition function $\overline{Z^{n}_{A+B}}$
which will be obtained 
integrating out these RSB excited states and the associated 
continuum due to CM motions may be put again into the scaling form  \eq{eq4b}.
The latter will again yield $\theta=1/3$. 

To summarize, the replica symmetry is not broken but 
only in a {\it marginal} way. As suggested by Parisi \cite{P90}, 
the role of these RSB excited states will become important if 
perturbations are considered. In the following we generalize the 
approach of \cite{P90} and exploit its implications to study the 
stability of the frozen phase against various perturbations we 
considered in section \ref{sec.scaling}.

\subsection{A Perturbative Approach by Replica Scaling Ansatz}			
\label{sec.perturbation-rsb}

Now we address the situation in which the two real 
replicas $A$ and $B$ are under infinitesimally weak perturbations.
The partition function of the system 
under such a perturbation can be formally written as,
\begin{eqnarray}
&& \overline{Z_{A+B}^{n}(L)} =
\int  \prod_{G=A,B}\prod_{\alpha=1}^{n}{\cal D}\phi_{G,\alpha}
\exp \left(-  S_{A+B}[\phi_{G,\alpha}] - \delta S_{A+B}[\phi_{G,\alpha}]
\right).
\label{eq6}
\end{eqnarray}
where the action $S_{A+B}[\phi_{G,\alpha}]$ is the original one
\eq{action} which is fully replica symmetric 
and the 2nd one $\delta S_{A+B}[\phi_{G,\alpha}]$ is
the perturbation term.
Suppose that we can  map the problem onto the quantum 
mechanical one such that the corresponding Schr\"odinger operator 
becomes,
\begin{equation}
{\cal H}_{A+B}={\cal H}_{0}+  \delta {\cal H}
\label{eq9}
\end{equation}
where ${\cal H}_{0}$ is the original fully replica symmetric
$2 \times n$-boson operator given in \eq{eq2} and $\delta {\cal H}$ 
corresponds to the $\delta S$ in the path-integral.
As we will see in the following, these perturbations try to
break the RS
present in the original system down to the reduced symmetry:
replica symmetric only within A and B subgroups.
At this stage, the whole quantum problem can not be solved exactly. 
However, we can obtain a useful insight into our problem  by
perturbation analysis proposed by Parisi \cite{P90}.

Here let us note a problem in the case of perturbation by random tilt field
considered in section \ref{subsubsec:perturbation-rsb-2-random-tilt}.
If one tries to obtain a continuous model starting from a lattice model
as considered by Kardar \cite{K87}, one can find that 
inter-replica coupling terms due to the random tilt field emerge
at 2nd order in the transverse hopping rate of the lattice string
(denoted as $\gamma$ in \cite{K87}). This  implies that the mapping in the  continuous limit to a 
Schr\"{o}dinger equation 
is invalid in this case, because the Schr\"odinger equation contains only first order time derivatives. Thus we do not consider this case 
in this section.

From standard perturbation theory we can evaluate the first order 
corrections to the original ground-state energy as,
\begin{equation}
E^{\rm RS}_\Delta=-\frac{1}{3}\frac{\kb T}{2\kappa}
\:\lambda^2\: 2n(4n^2-1)+\frac{\lan \Psi_{\rm RS}|\delta {\cal H}|\Psi_{\rm RS}\ran}{\lan \Psi_{\rm RS}|\Psi_{\rm RS}\ran}~~~,
\label{eq10}
\end{equation}
where the label $\Delta$ stands for the perturbation strength
and $\lan \Psi_{\rm RS}|$ is the ground-state wavefunction
given in \eq{eqf}. The 1st term corresponds to  the ground-state energy 
given in  \eq{eq4}.

Following Parisi, we will consider the RSB excited state
\eq{zero-overlap}.
 \begin{equation}
<\Psi_{\rm RSB}| \{x_{G,\alpha }\}>
\propto \exp (-\lambda\sum_{\alpha <
 \beta}|x_{A,\alpha}-x_{A,\beta}|)
\exp(-\lambda\sum_{\alpha <
 \beta}|x_{B,\alpha}-x_{B,\beta}|)
\label{eqf-rsb}
\end{equation}
with $\lambda=\kappa D/(\kb T)^3$. This wavefunction
has the reduced replica symmetry.  At 1st order in perturbation theory, we
can compute the energy of the RSB excited states as follows,
\begin{equation}
E^{\rm RSB}_\Delta=-\frac{1}{3}\frac{\kb T}{2\kappa}
\:\lambda^2\: 2n(n^2-1)+\frac{\lan \Psi_{\rm RSB}|\delta {\cal H}|\Psi_{\rm RSB}\ran}{\lan \Psi_{\rm RSB}|\Psi_{\rm RSB}\ran}~~~,
\label{eq10a}
\end{equation}
where the 1st term is the energy of the unperturbed system 
given by \eq{rsb-ene}.

Let us introduce the ratio of 
the contributions to the partition function
$\overline{Z^{n}_{A+B}}$ 
due to the RS ground state and the RSB excited state,
\begin{equation}
D(n,L)\equiv(E^{\rm RSB}_{\Delta}-E^{\rm RS}_{\Delta})L =
D_{0}(n,L)
 -\delta D(n,L)
\label{gap}
\end{equation}
where
\begin{equation}
D_{0}(n,L)=\frac{\kb T}{\kappa}\:\lambda^2\: n^3L \qquad > 0
\label{gap-original}
\end{equation}
is the original 'energy gap' and the correction 
is due to the 1st order perturbation. 
\begin{equation}
\delta D(n,L)= L \left (
\frac{\lan \Psi_{\rm RSB}|\delta {\cal H}|\Psi_{\rm RSB}\ran}{\lan
\Psi_{\rm RSB}|\Psi_{\rm RSB}\ran}
-\frac{\lan \Psi_{\rm RS}|\delta {\cal H}|\Psi_{\rm RS}\ran}{\lan
\Psi_{\rm RS}|\Psi_{\rm RS}\ran}
\right).
\label{gap-with-correction}
\end{equation}
In the following we call $D(n,L)$ as ``gap''. If it is large enough,
the contribution of the RSB excited state to the partition
function becomes negligible. 
We will find that, in general, the correction term of the gap has the form,  
\begin{equation}
-\delta D(n,L)=-\Delta n^p  L < 0 .
\label{gap-correction}
\end{equation}
Here the symbol $\Delta$ stands for the strength of the perturbation.
Most importantly the correction term 
$-\delta D(n,L)/L=-n^p \Delta$
will turn out to be {\it negative} for all the perturbations under  consideration.
In what follows we will refer to $p$ as the {\em order of the perturbation}
which will play a central role. 
More precisely the correction to the gap $-\delta D(n,L)/L$ 
will contain several terms of different powers of $n$. Here $p$ is the exponent
of the term with {\it smallest} exponent, which becomes most relevant in
the $n \to 0$ limit.

If the 1st order correction turns out to give a null contribution, we have to 
proceed to higher order perturbation calculations which is 
obviously impossible without the complete knowledge of the whole spectrum of excited
states. Fortunately for all the cases except for the case of the perturbation
by uniform tilt field we will find non-zero first order corrections.
Higher order correction terms will be higher order 
in $\Delta$ which will be unimportant since we are
interested in the scaling properties in  
infinitesimally weak perturbation limit $\Delta \to 0$.
Furthermore it is unlikely that the higher order terms are
lower orders of $n$. Thus they will be irrelevant in the $n \to 0$ limit.
For the case of uniform tilt field, we will fortunately find 
exact RS and RSB bound states of the system
which will allow the  evaluation of the gap $D(n,L)$ also in this situation.

Now using \eq{gap-correction} and \eq{gap-original} in \eq{gap} we find,
\begin{equation}
D(n,L)=
D_{0}(n,L) \left [1- \left (\frac{n}{n^{*}(\Delta)} \right)^{-(3-p)}\right]
\label{eq-gap-crossing}
\end{equation}
with
\begin{equation}
n^{*}(\Delta)= \left(\frac{\Delta}{\lambda^2 \kb T/\kappa}\right)^{1/(3-p)}~~.
\label{eq14}
\end{equation}
From the above result we can generalize the an argument used by Parisi for the explicit repulsive case (p=1) to extract the following conclusions.
As far as $n$ is integer and the strength of the perturbation 
$\Delta$ is small,
the contribution of the RSB state becomes negligible in the
thermodynamic limit $L \to \infty$. However, we have to consider
the other limiting case: the $n \to 0$ limit should 
be taken before $L \to \infty$. 
Now if $p<3$, which will turn out to be the case for all 
the perturbations under study,
an arbitrarily small perturbation $\Delta$ will induce
a level crossing at $n^{*}(\Delta)$ below which the contribution of 
RSB excited state becomes larger than that of the 
original ground state (RS).
The result \eq{eq-gap-crossing} matches perfectly with 
our definition of chaos \eq{eq-dec}
since  it suggests  that the partition function of the total
system factorizes in the $n \to 0$ limit as
\begin{equation}
\lim_{n/n* \to
0}\overline{Z^{n}_{A+B}}=\overline{Z^{n}_{A}}\;\times\;\overline{Z^{n}_{B}})\qquad
\mbox{if $p < 3$}.
\end{equation}
implying a
complete change of the free-energy landscape.

Now let us further exploit from the above result to find a more physical picture. 
In the absence of perturbations, the logarithm of the replicated partition
function has a functional form \eq{eq4b} which reads as,
$
\ln\overline{Z_{A+B}^{n}}= -\beta \overline{f} L (2n) +g(2nL^{1/3}).
$
On the other hand, \eq{eq-gap-crossing} implies $n/n^{*}$ is another
natural variable of the replicated partition function.\footnote{
To treat the replica number $n$ as a scaling variable has been proposed
by several authors \cite{HZ95,K96,EK00}.}
Combining the two, we conjecture the following scaling ansatz,
\begin{equation}
\ln\overline{ Z_{A+B}^{n}}+\beta \overline{f} L (2n) 
=\tilde{g}(2nL^{1/3}, n/n^*)=\tilde{g}(2nL^{1/3}, L/L^*)~~~.
\label{ansatz}
\end{equation}
where we introduced a characteristic length $L^*$ defined as
\begin{equation}
L^*\sim(n^*)^{-3} \sim \Delta^{-3/(3-p)}
\label{eq17}
\end{equation} 
An interesting observation is that $n \to 0$ limit induces the  thermodynamic
limit $L \to $ if the variable $nL^{1/3}=x$ is fixed.
Then for fixed $x$ we expect,
\begin{eqnarray}
\tilde{g}(x, L/L^{*}\to 0) & \simeq & g(2x) \qquad~~~~ L/L^* \ll 1  \qquad
(n/n^* \gg 1) \qquad \mbox{'weak perturbation regime'} \nonumber \\
\tilde{g}(x, L/L^{*}\to \infty) & \simeq & 2 \times g(x)  \qquad L/L^* \gg 1 \qquad
(n/n^* \ll 1) \qquad \mbox{'strong perturbation regime'} .
\end{eqnarray}
The 1st equation means that for small enough lengthscales,
the effect of perturbation is small and the partition function
is essentially the same as that of the unperturbed system of 
$2 \times n$-replicas
given in \eq{eq4b}. The 2nd equation is the consequence of having to two statistically independent systems in the limit $L\to\infty$.

From the above scaling ansatz, it follows
that the correlation function of the free-energy
fluctuations $C_{F}(L)$ considered in section \ref{subsec:cf}
should have the scaling form $\tilde{C}_{F}(L/L^*)$ which goes to $0$ 
as $L/L^* \to \infty$. Similarly the overlap function $q(L,\delta U)$
considered in section \ref{subsec:q} should also have the scaling 
form $\tilde{q}(L/L^*)$ which goes to $0$ as $L/L^* \to \infty$.
Thus the crossover length $L^*$ should be identified with the 
overlap length $L_{c}(\delta U)$. 
The above ansatz implies that the de-correlation of the
free-energy landscape between A and B takes place as a universal phenomenon 
whose features are classified according to the order of the perturbation $p$.
In the following, we consider the perturbations considered in 
previous real-space scaling argument section \ref{sec.scaling}
specifically one by one based on the replica approach 
and evaluate the correction to the gap \eq{gap-correction} explicitly and
extract the strength of perturbation $\Delta$ and the order of
perturbation $p$.
Interestingly enough, we will find that the two approaches give
the same overlap length.

Finally let us comment on how to choose detailed forms of 
perturbations which we discuss in the following. 
We consider perturbations such that the original symmetry 
is preserved as much as possible : 
the $2 \times n$-replica system remains invariant at least under permutation 
among n-replicas belonging to the same subset A and B and 
exchange  $A \leftrightarrow B$, i.~e. the reduced replica symmetry.
  
\subsubsection{Short-Ranged Repulsive Coupling}
\label{subsubsec:replica-rep}

Let us begin with the perturbation which 
introduces an explicit repulsion term  between  
strings $A$ and $B$ as given by the Hamiltonian  in \eq{unif}.
The corresponding Schr\"odinger operator for the replicated
system can be obviously put into the form of \eq{eq9}
 - fully replica symmetric term + perturbation -  to obtain,
\begin{equation}
{\cal H}_{A+B}= H_{0}+\delta {\cal H}\qquad
\mbox{with}\qquad
\delta {\cal H}=\frac{\eps}{\kb T} 
\sum_{\alpha=1,\ldots,n}\delta(x_{A,\alpha}-x_{B,\alpha}) \qquad \eps >0.
\label{eqr1}
\end{equation}
Clearly the repulsive perturbing term breaks the
original RS \cite{P90}.

Computing explicitly the expectation value of the a delta-interaction term
with respect to the Bethe ground state one obtains \cite{P90},
\be
\frac{\lan \Psi_{\rm
RS}|\delta(x_{A,\alpha}-x_{B,\alpha})|\Psi_{\rm RS}\ran}{\lan \Psi_{\rm
RS}|\Psi_{\rm RS}\ran}=  \frac{\lambda}{6}\:(2n+1).
\label{eqr2}
\end{equation}
while
\begin{equation}
\frac{\lan \Psi_{\rm
RSB}|\delta(x_{A,\alpha}-x_{B,\alpha})|\Psi_{\rm RSB}\ran}{\lan \Psi_{\rm
RSB}|\Psi_{\rm RSB}\ran}= 0.
\label{eqr2b}
\end{equation}
because the bound states of A and B subsets have no overlap
$<\Psi_{RSB}|\Psi_{RSB}>=0$ \eq{zero-overlap}. 

Thus the correction term to the gap \eq{gap-correction} is obtained as,
\begin{equation}
-\frac{\delta D(n,L)}{L}=-\frac{\lambda}{6}\:(2n+1)
\times \left(\frac{\eps}{\kb T}\right) \times n.
\end{equation}
Note that the reduced replica symmetry: permutation 
symmetry among n-replicas belonging to the same subset plus
the exchange symmetry $A \leftrightarrow B$ is still preserved.
Thus the leading order of the perturbation (smallest power of $n$,
which becomes most relevant in the $n \to 0$ limit)
is read off as  $p=1$ and  the strength of perturbation 
 as $\Delta\sim \eps$. 
Finally, using the  relation (\ref{eq17}) we obtain the crossover 
length $L^*\sim \eps^{-\frac{3}{2}}$ .
Remarkably the latter turns out to be the same 
as the overlap length \eq{lc-rep} found 
in the real space scaling argument.

\subsubsection{Potential Change}
\label{subsubsec:replica-potential}

If a slight difference of the random potential is introduced
as described in \eq{hamiltonian-potential}, the corresponding 
Schr\"odinger operator of the $2 \times n$  replica system reads,
\begin{eqnarray}
 {\cal H}&=&-\sum_{G,\alpha} \frac{\kb T}{2\kappa} 
\frac{\partial^{2}}{\partial x_{G,\alpha}^{2}}
-\frac{D}{(\kb T)^{2}}\sum_{(\alpha,\beta)}
\delta (x_{A,\alpha}-x_{A,\beta})
-\frac{D}{(\kb T)^{2}}\sum_{(\alpha,\beta)}
\delta (x_{B,\alpha}-x_{B,\beta})\nonumber \\
&& -\frac{1}{\sqrt{1+\delta^2}}\frac{D}{(\kb T)^{2}}\sum_{(\alpha,\beta)}
\delta (x_{A,\alpha}-x_{B,\beta}) \nonumber \\
&=& {\cal  H}_{0}+\delta {\cal H}
\label{eqb1}
\end{eqnarray}
with the symmetry breaking term
\begin{equation}
\delta {\cal H}=\frac{\delta^2}{2}\sum_{(\alpha,\beta)}\delta\left(x_{A,\alpha}(z)-x_{B,\beta}(z)\right).
\end{equation}
Here we are in  the  infinitesimally
weak perturbation limit, $\delta \to 0$ so higher order terms can be ignored.

A remarkable feature is that the 2nd term of the last equation, 
which is the perturbation term $ \delta {\cal H}$,
is again {\it repulsive}. Note that 
the sum is taken over $n^{2}$ rather than $n(n-1)/2$ pairs.
The expectation value of the delta-function
with respect to the RS ground state and the RSB excited state
has already been computed in \eq{eqr2} and \eq{eqr2b}, hence 
we immediately find the correction to the gap as,
\begin{equation}
-\frac{\delta D(n,L)}{L}=-
\frac{\lambda }{6} (2n+1) \times \frac{\delta^{2}}{2} \times n^{2}
\end{equation}
Note that this perturbation contains the reduced replica symmetry.
The latter was made possible by a specific choice of
the perturbation by introducing the rescaling factor $1/\sqrt{1+\delta^2}$
used in \eq{hamiltonian-potential}.
We can now read off the order of the perturbation as 
$p=2$ and the strength of the perturbation as 
$\Delta\sim \delta^2$. Now using
\eq{eq17}, we obtain the overlap length 
$L^*\sim\delta^{-6}$. Indeed, the latter turns
out to be the one obtained by the real space scaling argument
given in \eq{lc-potential}.

\subsubsection{Temperature Change}
\label{subsubsec:replica-temp}

Now two real replicas in the same quenched random potential $V(\phi(z),z)$ are subjected to
a small temperature difference.
 
The Schr\"odinger operator for the $2 \times n$-replica system with
A at temperature $T_{A}$ and B at temperature $T_{B}$ is the following,
\begin{equation}
{\cal H}=-\sum_{\alpha} \frac{\kb T_{A}}{2\kappa} 
\frac{\partial^{2}}{\partial x_{A,\alpha}^{2}}
-\sum_{\alpha} \frac{\kb T_{B}}{2\kappa} 
\frac{\partial^{2}}{\partial x_{B,\alpha}^{2}}
-\sum_{((G,\alpha),(G',\beta))}\frac{D}{(\kb T_{G})(\kb T_{G'})}
\delta (x_{G,\alpha}-x_{G',\beta})~~~.
\label{operator-TA-TB}
\end{equation}
The RS is apparently lost in the operator. 
Let us  choose the following specific parameters of the perturbation,
\begin{eqnarray}
T_A & \rightarrow & T+\delta T \nonumber \\
T_B &\rightarrow  & T-\delta T \nonumber \\
D & \rightarrow & D \left[ 1-3 \left (\frac{\delta T}{T} \right)^2 \right]~~~.
\label{eqt1}
\end{eqnarray}
Then we can put the operator in the form,
\begin{equation}
{\cal H} = {\cal H}_{0} +\delta {\cal H},
\end{equation}
with the symmetry breaking terms,
\begin{eqnarray}
\delta {\cal H}= && -\sum_{\alpha} \frac{\kb \delta T}{2\kappa} 
\frac{\partial^{2}}{\partial x_{A,\alpha}^{2}}
+2\frac{\delta T}{T}\sum_{\alpha,\beta}\frac{D}{(\kb T)^{2}}
\delta (x_{A,\alpha}-x_{A,\beta})
\nonumber \\
&& +\sum_{\alpha} \frac{\kb \delta T}{2\kappa} 
\frac{\partial^{2}}{\partial x_{B,\alpha}^{2}}
-2\frac{\delta T}{T}\sum_{\alpha,\beta}\frac{D}{(\kb T)^{2}}
\delta (x_{B,\alpha}-x_{B,\beta}) \nonumber \\
&& +2 \left (\frac{\delta T}{T}\right)^{2}
\sum_{\alpha,\beta}\frac{D}{(\kb T)^{2}}
\delta (x_{A,\alpha}-x_{B,\beta})
+O\left(\frac{\delta T}{T}\right)^{3}.
\label{operator-dT}
\end{eqnarray}
In this last equation, we are considering the limit of an infinitesimally
weak perturbation $\delta T/T  \to 0$ to neglect higher order terms.
The expectation value of the perturbing operator with respect
to the RS ground state is obtained as,
\begin{eqnarray}
<\Psi_{\rm RS}| \delta {\cal H} |\Psi_{\rm RS}>
&& =2 \left (\frac{\delta T}{T}\right)^{2}\frac{D}{(\kb T)^{2}}
\sum_{\alpha,\beta}
<\Psi_{\rm RS}| \delta (x_{A,\alpha}-x_{B,\beta}) |\Psi_{\rm RS}> \nonumber \\
&& =\frac{\lambda}{6} \frac{D}{(\kb T)^{2}} 
(2n +1)\:  \times 2 \left (\frac{\delta T}{T}\right)^{2} \times n^{2}.
\end{eqnarray}
Here we have used the fact that the ground-state wavefunction is 
symmetric with respect to  the exchange $A \leftrightarrow B$ plus \eq{eqr2}.
Due to the latter, the terms of order $O(\delta T)$ cancel out
and we are left with the $O(\delta T^{2})$ term.
Note also that the sum is again taken over $n \times n$ pairs of
replicas rather than $n(n-1)/2$.

On the other hand, the expectation value of the perturbing term with respect the RSB
excited state is obtained immediately as
$<\Psi_{\rm RSB}| \delta {\cal H} |\Psi_{\rm RSB}>=0$
using \eq{eqr2b} and the fact that RSB wavefunction is 
symmetric with respect to the exchange $A \leftrightarrow B$ and \eq{zero-overlap}.
Using the above results we find the correction to the
gap as, \footnote{The choice of ``renormalized'' $D$ 
in \eq{eqt1} allowed to simplify the calculation. 
One can work also with original
$D$ and will find the same {\it positive}  $\delta D(n,L)/L$ of 
order $O(n^{2})$ up to some irrelevant differences.}
\begin{equation}
-\frac{\delta D(n,L)}{L}=-
\frac{\lambda}{6} 
\frac{D}{(\kb T)^{2}}(2n +1)
\times 2 \left(\frac{\delta T}{T}\right)^{2} \times n^{2} 
\end{equation}
Note that the resultant gap 
is invariant under the exchange $A \leftrightarrow B$
which was made possible by the anti-symmetric direction of
the change of temperature \eq{eqt1}. 
From the above results, we read off the order of the 
perturbation as $p=2$ and  the strength of the perturbation as
$\Delta\sim (\delta T)^2$. 
Quite remarkably the latter used in \eq{eq17} again yields
the crossover length 
$L^*\sim(\delta T)^{-6}$ which is the same as the one 
found by the real space scaling argument \eq{lc-temperature}.

\subsubsection{Uniform Tilt Field}
\label{subsubsec:tilt}

Finally we consider to apply a uniform tilt $h$ to 
one real replica and $-h$ to the other.
The effective action describing the uniform field perturbation 
\eq{unif}  is the following,
\begin{eqnarray}
 S_{A+B}[\phi_{G,\alpha}] &=&
\int_{0}^{L}dz \left[ \sum_{G,\alpha} 
\frac{\kappa}{2 \kb T} \left ( \frac{d \phi_{G,\alpha}(z)}{d z} \right)^{2} 
-
\frac{D}{(\kb T)^{2}} \sum_{G , G',\alpha ,\beta}
\delta (\phi_{G,\alpha}(z)-\phi_{G',\beta}(z))
\right. \nonumber \\
&-& \left.
\frac{h}{\kb T}  \sum_{\alpha} \frac{d \phi_{A,\alpha}(z)}{d z} 
+ \frac{h}{\kb T}  \sum_{\alpha} \frac{d \phi_{B,\alpha}(z)}{d z} 
 \right]~~~.
\label{eqs3}
\end{eqnarray}
Here not only the full permutation symmetry among the $2 \times n$ replicas 
but also the global rotational symmetry is lost due to the field.
Thus the universality of this perturbation should be 
very different from the ones  discussed so far.
The corresponding Schr\"odinger operator of the quantum mechanical 
problem reads as,
\begin{equation}
{\cal H}=-\sum_{G,\alpha} 
\frac{\kb T}{2\kappa} \frac{\partial^{2}}{\partial x_{G,\alpha}^{2}}
-\frac{D}{(\kb T)^{2}}
\sum_{G,G',\alpha,\beta}
\delta (x_{\alpha}-x_{\beta})-\frac{h}{\kappa}\sum_{\alpha}
\frac{\partial}{\partial x_{A,\alpha}}
+\frac{h}{\kappa}\sum_{\alpha}\frac{\partial}{\partial x_{B,\alpha}}.
\label{eqs4}
\end{equation}
Note that the first two terms are the original operator ${\cal H}_{0}$ given in \eq{eq2}.

Now let us analyze the change of the RS state \eq{eqf}.
One can easily see that the 1st order perturbation vanishes simply because 
the total ``momentum'' of the ground state is zero. 
On the other hand, one can also easily note that when a field is applied, 
the original wavefunction is no longer an eigenstate.
Fortunately, the exact eigenstate can be found 
in this odd situation in which particles belonging to different
subsets (A and B) are driven into opposite directions.
The former Schr\"odinger operator\eq{eqs4} can be rewritten 
into the fully symmetric form of the original problem \eq{eq2} by shifting 
the momenta,
\begin{equation}
\frac{\partial}{\partial x'_{A,\alpha}}=
\frac{\partial}{\partial x_{A,\alpha}} 
- \frac{h}{\kb T} 
\qquad \frac{\partial}{\partial x'_{B,\alpha}}
=\frac{\partial}{\partial x_{B,\alpha}} + \frac{h}{\kb T}~~~.
\label{eqs5}
\end{equation}
Notice that this transformation  preserves 
the commutation relations  between conjugated coordinates and moments 
\nolinebreak{(i.e. $[\frac{\partial}{\partial x'_{G,\alpha}}$ ,~$x_{G,\alpha}]=[\frac{\partial}{\partial x_{G,\alpha}}$ ,~$x_{G,\alpha}]$)}.
In terms of these new coordinates the RS ground state again 
takes the form of the Bethe Ansatz solution of \eq{eqf}. 
And, therefore, the final groundstate can be obtained 
from Bethe's wave-function by undoing the previous shifting of moments,
\be
\Psi\sim \Psi_{\rm RS}(\{x_{G,\alpha}\}) 
\exp \left(\frac{h}{\kb T}\sum_{\alpha}x_{A,\alpha}\right)
\exp \left(-\frac{h}{\kb T}\sum_{\alpha}x_{B,\alpha}\right)~~~,
\label{eqs6}
\end{equation} 
where $\Psi_{\rm RS}(\{x_{G,\alpha}\})$ is the original Bethe ansatz
wavefunction for $2 \times n$ replicas given in \eq{eqf}.
The eigenvalue $E_h$ corresponding to this wave function is obtained as
\begin{equation}
E_h=E_0+\frac{n h^2}{\kappa \kb T},
\label{rs-h}
\end{equation}
which does not depend on the ordering of the particles.
Here $E_{0}$ is the original ground-state energy $E_{g}$ given in \eq{eq4}.
Although the original full permutation symmetry 
is lost in the wave function \eq{eqs6}, it still described a sort
of bound state on $2 \times n$-particles. So we may refer to it as RS state.
In the next subsection, we will discuss the mapping onto the Sinai model and the physical meaning will become clearer.
The 2nd term of \eq{rs-h} gives the change of the eigenvalue of the RS state 
due to the perturbation $\Delta E_{RS}=n h^2/(\kappa \kb T)$.

Next let us consider the change of the eigenvalue corresponding to the 
RSB excited state which again is formed by two separate bound states for
A and B subsets. Here it is useful to note that if {\it all} the
particles are subjected to the common field, 
the unperturbed single-bound-state wavefunction is still 
an eigenstate  of the operator.
Based on this observation, one immediately finds that the
unperturbed RSB wavefunction is still valid eigenstate 
under the field because of the two fold reasons: 
i) there is no overlap between A and B and 
ii) rotational and replica symmetries are preserved within the same subsets.  
Thus the eigenvalue of the RSB state does not change by the
perturbation
$\Delta E_{RSB}=0$.

Using the above values of $\Delta E_{RS}$ and $\Delta E_{RSB}$
we obtain 
\begin{equation}
-\frac{D(n,L)}{L}=-\frac{n h^2}{\kappa\kb T}.
\end{equation}
We can now read off $p=1$ and $\Delta \sim h^{2}$,
which yields the overlap  length $L_c\sim h^{-3}$. 
Then using \eq{eq17} we 
find the same overlap length $L^{*} \sim h^{-3}$
being consistent with the result \eq{lc-uniform-tilt}
of the real-space scaling argument.

\section{Mapping to a modified Sinai Model}
\label{sec.sinai}

In the previous section, we found de-correlation the free-energy landscapes 
of perturbed and unperturbed systems. Here we analyze the problem further 
for the case of uniform tilt field based 
on the connection between the $1+1$ dimensional DPRM 
and the statistical mechanics of the 
Sinai model  \cite{Sinai,P90,M90,BO90}.
With this mapping, effective 1-dimensional 
energy landscape for the free end $x(L)$ of the 1+1 DPRM
is obtained as a Sinai potential which is generated by
a simple random walk in a 1-dimensional space. 

Here we consider this mapping onto the Sinai model in the presence
of the uniform tilt field by evaluating the partition function
\eq{eq3}. First, we evaluate the partition function assuming 
the RS and using the ground-state 
wavefunction given in \eq{eqs6}. Second, we perform another evaluation
assuming replica symmetry breaking (RSB) which only allows 
the reduced replica symmetry and using the 'clustered' 
wave function \eq{eqf-rsb}. The former is supposed good
for the weakly perturbed regime $L \ll L_{c}(h)$ while the
latter is good for the strongly perturbed regime $L \gg L_{c}(h)$.
In order to interpolate the two limits, we propose 
a phenomenological model using a bounded Sinai potential.

\subsubsection{Replica Symmetric case}
\label{sec.sinai-rs}

We start by considering the fully replica symmetric (RS) ansatz
following Bouchaud and Orland \cite{BO90}.
The ground-state wavefunction under a uniform tilt
is given by \eq{eqs6}. In order to take into account the motion of
the center of mass (CM) of the $2 \times n$ replicas, we consider the spectrum
of excited states whose wavefunctions are given by,
\begin{equation}
\Psi_{\rm RS}(h,k:\{x_{G,\alpha}\})\sim \Psi_{\rm RS}(\{x_{G,\alpha}\}) 
\exp \left[\frac{h}{\kb T}\left(\sum_{\alpha}x_{A,\alpha}
-\sum_{\alpha}x_{B,\alpha}\right)\right]
\exp \left( ik\sum_{G,\alpha} x_{G,\alpha} \right).
\end{equation} 
The 1st factor is the Bethe wavefunction given in \eq{eqf}
which describes the unperturbed bound state of $2 \times n$ replicas.
Now we use the Gaussian transformation introduced by Parisi in \cite{P90} to represent Bethe's wavefunction as follows,
\begin{equation}
 \Psi_{\rm RS}(\{x_{G,\alpha}\}) \sim
e^{-\lambda\sum_{((G,\alpha),(G',\beta))}
|x_{\alpha,G}-x_{\beta,G'}|}=\int {\cal D V} 
\exp \left[-\int dx \frac{1}{4\lambda}\left(\frac{d{\cal V} }{dx}\right)^{2}\right]
\exp \left(\sum_{G,\alpha} {\cal V} (x_{G,\alpha})\right).
\end{equation}
The 2nd factor arises from the uniform tilt perturbation: $h$ to subset $A$ and $-h$ to subset $B$.
The last factor is the plane wave of wavevector $k$ which accounts for
the free CM motion. Here the ground state is included as
the $k=0$ case. One can easily find the eigenvalues to be,
\begin{equation}
E_{\rm RS}(h,k)=E_{0}+\frac{nh^{2}}{\kappa \kb T}
+n \frac{\kb T}{\kappa}k^{2}.
\end{equation}
The 1st term is the original ground-state energy $E_{g}$ of the unperturbed
system given in \eq{eq4}, the 2nd term is due to the perturbation and
the last term is due to the CM motion.

Let us now suppose that replicas have both ends fixed: one at $(0,0)$ and the other at $x_{A}$ in the case of
 subset $A$ and $x_{B}$ in the case of subset $B$.  
Then the partition function \eq{eq3} is evaluated by integrating out
the spectrum of excited states as,
\begin{eqnarray}
\overline{Z_{\rm RS}(0,0|x_{A},x_{B})} 
&\sim & e^{-(E_{0}+nh^{2}/(2\kappa \kb T))L}
\int {\cal D}{\cal V} \exp \left[
-\int dx \frac{1}{4\lambda}\left(\frac{d{\cal V}}{dx}\right)^{2}\right]
\exp \left(n {\cal V}(x_{A})+n {\cal V}(x_{B})\right)  \nonumber \\
&\times & \exp ((h/\kb T)(n x_{A}+n x_{B})) 
\int dk \sqrt{Ln\frac{\kb T}{\kappa\pi}}
\exp \left(-Ln\frac{\kb T}{\kappa}k^{2}
+ ik (nx_{A}+nx_{B}) \right) \nonumber\\
&\sim & e^{-(E_{0}+nh^{2}/(\kappa \kb T))L}
\overline{\left[\exp \left(-nL^{1/3}E_{\rm RS-Sinai}(T,h,{\cal V}; y_{A},y_{B})\right)\right]}_{\tilde{\cal V}}.
\label{partition-func-RS}
\end{eqnarray}
where $\overline{[\cdots]}_{\tilde{\cal V}}$ means the average 
over the effective potential $\tilde{\cal V}$,
\begin{equation}
\overline{[\cdots]}_{\tilde{\cal V}} = \int {\cal D}\tilde{\cal V} e^{
-\int dy (1/4\lambda)(\frac{d\tilde{\cal V}}{dy})^{2}}
\cdots
\end{equation}
and $E_{\rm RS-Sinai}(T,h,\; y_{A},y_{B})$ is the effective Hamiltonian, 
\begin{equation}
E_{\rm RS-Sinai}(T,h,{\cal V}; y_{A},y_{B})=
\frac{\kappa}{2\kb T}\frac{(y_{A}+y_{B})^{2}}{2}
+\tilde{\cal V}(y_{A})+\tilde{\cal V}(y_{B})
-\frac{\tilde{h}}{\kb T}y_{A}
+\frac{\tilde{h}}{\kb T}y_{B}
\label{effh}
\end{equation}
in terms of the scaled variables,
\begin{equation}
 x=L^{2/3}y  \qquad {\cal V}=-L^{1/3}\tilde{\cal V} \qquad
 h=L^{-1/3}\tilde{h}.
\label{eq:rescale}
\end{equation}

By increasing $nL^{1/3}$, the partition function will be 
dominated by the minimum of the effective Hamiltonian 
$E_{\rm RS-Sinai}(T,h,{\cal V}; y_{A},y_{B})$. 
Then the following physical interpretation can be made: 
the end point of the strings $A$ and $B$ are subjected 
to the {\it same  effective quenched random potential}
which displays the long-ranged correlations in transverse 
space just as the Sinai model,
\begin{equation}
\overline{[(\tilde{\cal V}(y)-\tilde{\cal V}(y'))^2]}_{\tilde{V}}\propto |y-y'|~~~.
\label{eqs2}
\end{equation} 
Furthermore, 
the CM of the total system is subjected to an effective Hookian 
spring which tries to bind together the two real replicas.
The effect of the uniform tilt field amounts to an effective transverse 
force $\tilde{h}$ applied at the endpoints of A and B replicas which tries to
drive them into the opposite directions. From \eq{eq:rescale},
it can be seen that the effective force $\tilde{h}$ increases
by increasing the system size $L$ (with fixed $h$).

\subsubsection{Replica Symmetry Broken Case}
\label{sec.sinai-rsb}

In section \ref{subsubsec:tilt}, we found out that replica symmetry breaking
becomes important at $L \gg L_{c}(h)$ with $L_c(h)\sim h^{-3}$
given in \eq{lc-uniform-tilt}.  
Here we perform the evaluation of the partition function \eq{eq3}
based on the replica symmetry breaking (RSB) ansatz. 
In this case, we consider a spectrum of excited states whose wavefunctions 
are given by,
\begin{equation}
\Psi_{\rm RSB}(h,k_A,k_B:\{x_{G,\alpha}\})\sim \Psi_{\rm RS}(\{x_{A,\alpha}\}) 
\Psi_{\rm RS}(\{x_{B,\alpha}\}) 
\exp \left( ik_{A}\sum_{\alpha} x_{A,\alpha} \right)
\exp \left( ik_{B}\sum_{\alpha} x_{B,\alpha} \right).
\end{equation} 
The first two factors are due to the original wave function of 
the RSB state \eq{eqf-rsb} which consists in two clusters 
of bound states. As we noted in \ref{subsubsec:tilt}, it remains
as an eigenstate even under the uniform tilt field
since it is assumed that these clusters have zero overlap.
Moreover, this absence of overlap also allows
independent CM motions of A and B subsets. 
The later two factors account for such 
separate CM motions. The eigenvalues are the following,
\begin{equation}
E_{\rm RSB}(h,k_A,_{B})=E_{\rm RSB}
+n\frac{h}{\kappa}(i k_{A})+n\frac{h}{\kappa}(i k_{B})
+n \frac{\kb T}{2\kappa}k_{A}^{2}
+n \frac{\kb T}{2\kappa}k_{B}^{2}.
\end{equation}
The 1st term is the original ground-state energy $E_{\rm RSB}$ 
of the unperturbed RSB state given in \eq{rsb-ene}.
The 2nd and 3rd terms come from the perturbation.
The last two terms are due to the separate CM motions.
The partition function \eq{eq3} is evaluated by integrating out
the spectrum of excited states as,
\begin{eqnarray}
&& \overline{Z_{\rm RSB}(0,0|x_{A},x_{B})}   \nonumber \\
&&\sim  e^{-E_{0}L}
\int {\cal D}V_{A} \exp \left[
-\int dx \frac{1}{4\lambda}\left(\frac{dV_{A}}{dx}\right)^{2}\right]
\int {\cal D}V_{B} \exp \left[
-\int dx \frac{1}{4\lambda}\left(\frac{dV_{B}}{dx}\right)^{2}\right]
 \exp \left (n V_{A}(x_{A})+nV_{B}(x_{B}) \right)  \nonumber \\
&&\times  
\int dk_{A} \sqrt{Ln \frac{n\kb T}{2\kappa\pi}}
\int dk_{B} \sqrt{Ln \frac{\kb T}{2\kappa\pi}} 
 \exp \left(-Ln\frac{2\kb T}{\kappa}k_{A}^{2}
-Ln\frac{2\kb T}{\kappa}k_{B}^{2}
-Ln \frac{h}{\kappa}(ik_{A})+Ln \frac{h}{\kappa}(ik_{B})
+ ik (nx_{A}+nx_{B})
 \right) \nonumber \\
&\sim & e^{-(E_{0}+(2n)h^{2}/(2\kappa \kb T))L}
\overline{\left[\exp \left(-nL^{1/3}
E_{\rm RSB-Sinai}(T,h,\tilde{\cal V}_{A},
\tilde{\cal V}_{B}; y_{A},y_{B})\right)\right]}_{\tilde{\cal V}_{A},
\tilde{\cal V}_{B}},
\label{partition-func-RSB}
\end{eqnarray}
where $E_{\rm RSB-Sinai}(T,h,\tilde{\cal V}_{A},\tilde{\cal V}_{B}; y_{A},y_{B})$ is the effective Hamiltonian, again in terms of the scaled variables, 
\begin{eqnarray}
E_{\rm RS-Sinai}(T,h,\tilde{\cal V}_{A},\tilde{\cal V}_{B}; y_{A},y_{B}) &=&
\frac{\kappa}{2\kb T}y_{A}^{2}
+\tilde{\cal V}_{A}(y_{A})
-\frac{\tilde{h}}{\kb T}y_{A}
+ \frac{\kappa}{2\kb T}y_{B}^{2}
+\tilde{\cal V}_{B}(y_{B})
+\frac{\tilde{h}}{\kb T}y_{B}.
\end{eqnarray}

It is interesting to compare the last result with
the replica symmetric (RS) one given in\eq{effh}. 
Here the two subsets $A$ and $B$ are 
now subjected to {\it independent Hookian springs} which try to confine 
the CM of each subset while the {\it total} CM was confined in the RS case. 
Moreover, the two replicas are 
now subjected to {\it completely independent Sinai potentials} 
$\tilde{\cal V}_{A}$ and $\tilde{\cal V}_{B}$. The effect of 
the uniform tilt field again amounts to an effective transverse 
force $\tilde{h}$ applied at the endpoints of A and B replicas which tries to
drive them into opposite directions just as in the replica symmetric case. 

\subsubsection{Discussion}

In section \ref{sec.replica}, we conjectured a possible scaling
form \eq{ansatz} of the crossover from the weakly perturbed regime 
at length scales shorter than the overlap  length $L_{c}$
where the RS holds, to 
the strongly perturbed regime where replica symmetry breaking
becomes relevant,
\begin{equation}
\ln\overline{Z_{A+B}^{n}}+\beta \overline{f} L (2n) 
=\tilde{g}(2nL^{1/3}, L/L_{c})~~~.
\end{equation}
Indeed, the partition function based on the RS
and RSB  ansatz given in \eq{partition-func-RS} and 
\eq{partition-func-RSB} have the expected form; 
the $O(n)$ term which provides the average free-energy 
$\beta \overline{f} L (2n)$ plus a function
which contains the two scaling variables
$nL^{1/3}$ and $\tilde{h}=L^{1/3}h=(L/L_{c}(h))^{1/3}$. In the last
equation we used the relation $L_{c}(h) \sim h^{-3}$ 
given in \eq{lc-uniform-tilt}. 

Here we have only discussed the two limiting ans\"atze: RS and RSB. 
The crossover between the two limits remains an open problem.
Here let us propose a modified Sinai model which interpolates 
the limits.
We define effective Hamiltonian for
the endpoints' positions of replicas $A$ and $B$  at a given length 
reads as follows,
\begin{eqnarray}
H= H_A+H_B \qquad 
H_A=\frac{\kappa}{ 2\kb T}y_A^2+\tilde{\cal V}(y_A)+\tilde{h}\;y_A \qquad
H_B=\frac{\kappa}{ 2\kb T}y_B^2+\tilde{\cal V}(y_B)-\tilde{h}\;y_B
\label{eq-modified-sinai-1}
\end{eqnarray} 
where $\tilde{\cal V}(x)$ is a {\it bounded} Sinai potential with correlations,
\begin{equation}
\overline{(\tilde{\cal V}(x)-\tilde{\cal V}(y))^2}\propto C(|x-y|)
\qquad \mbox{with} \qquad
C(u)=y+(1-u)\theta(u-1)
\label{eq-modified-sinai-2}
\end{equation}
Here the correlation grows as $C(u)=u$ for $u \leq 1$ 
and saturates  $C(u)=1$ for larger separations $u >1$.
The latter saturation (confined random walk) allows to 
realize statistically independent Sinai valleys at large separations (RSB).
Actually such a saturation of the effective energy landscape was
observed numerically in the DPRM by M\'{e}zard \cite{M90}.

In section \ref{subsec:tm-tilted-dprm} we analyze the crossover
phenomena in detail by a transfer matrix method.
Subsequently, in  section \ref{subsec:numerics-tilted-sinai}, we analyze 
the phenomena numerically using the modified Sinai model defined 
above and compare the result with that obtained in the original DPRM.

\section{Numerical Analysis}
\label{sec.numerical}

Now we examine numerically in detail the properties
of anomalous response of the DPRM 
towards various perturbations discussed in the previous sections
by transfer matrix calculations.
We focus on the anticipated universal 
scaling properties of the crossover from the weakly to strongly 
perturbed regime across the overlap length which has not been 
clarified in previous numerical studies (see however \cite{M90}).

\begin{figure}[h]
\begin{center}
\includegraphics*[scale=0.6]{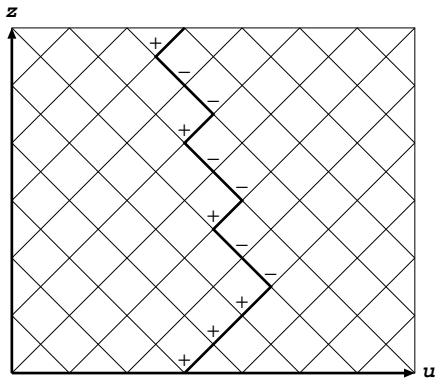}
\end{center}
\caption{The lattice $1+1$ dimensional DPRM model.
This example has longitudinal size $L=12$ and transverse size $U=14$.
The thick zig-zag line is an example of the configuration.
The string is directed in the direction of the $z$-axis with transverse
displacements in the direction of $u$-axis. 
}
\label{lattice_string.fig}
\end{figure}

Specifically, we consider  a lattice model on a two-dimensional lattice of 
size  $L \times U$  as shown in Fig. \ref{lattice_string.fig}.
The string of length $L$ is directed along the $z$ axis
with  transverse displacements in the direction of the $u$-axis.
The configuration of the string is represented by the
positions of the vertices ``X'' which the configuration goes through,
i.~e. $(u(z),z)$ with $z=1,\ldots,L$.
The ``gradient'' $\sigma(z)=u(z+1)-u(z)$
is constrained to take only the values $+1$ or $-1$.
Note that elasticity is realized entropically within this lattice model.
The random potential $V(u,z)$ is defined on each vertex $(u,z)$ on
which it takes a random value drawn from a uniform distribution between
$-V_{0}$ and $V_{0}$. 
The energy of a configuration $\{u(z)\}$ is given by,
\begin{eqnarray}
E[V,u] =  \sum_{z=1}^{L}  V(u(z),z) 
\end{eqnarray}

One end of the configuration is fixed at $(0,0)$ and the other
end is allowed to move freely.
 On the transverse direction we have imposed periodic boundary conditions such that $V(u+U,z)=V(u,z)$.
The natural unit for the temperature is the scaled thermal energy 
$\kb T/V_{0}$ where $V_{0}$ is the unit for the random potential. 
In the following, the Boltzmann's constant is set to $\kb=1$ 
and the unit for the random potential to $V_{0}=1$, so that we will often
denote the scaled thermal energy simply as $T$.

First we prepare two real replicas A and B identically
except for small perturbations which we will describe
in detail. Depending on the type of the problem, 
we use either zero temperature 
\cite{HH85} or finite temperature versions \cite{K85}
of the transfer matrix method to compute correlation functions.
Here and in the following
$\overline{X}$ denotes the average of a quantity $X$ 
over different realizations of the random potential and $<X>$ denotes 
the thermal average of $X$ (or simply the value of $X$ at ground state
in the case of zero temperature). We have examined various system
sizes up to $L=10^{4}$ and have averaged over $N_{s}=10^{4}$ different
realizations of the random potential except for the
explicit repulsive coupling case for which we used
system sizes up to $L=10^{3}$ and $N_{s}=10^{4}$.
The limitation of the system size used for the latter case is that
we have to take into account explicitly the inter-real-replica
coupling in the transfer matrix which requires 
one to keep track of trajectories of two strings
simultaneously and thus much larger computational effort \cite{M90}.

First we examine the mean-squared transverse displacement 
of the end point due to the perturbation,
\begin{equation}
B_{2}(L)=\overline{<u_A(L)-u_B(L)>^{2}}.
\label{eq-b-numerical}
\end{equation}
Here $u_A(L)$ and $u_B(L)$ stand for the position of the end point of 
$A$ and $B$ replicas respectively. 
Second we compute the exact free-energies (or ground-state energies at
zero temperature) of both replicas by the transfer
matrix method and examine the correlation of the free-energies,
\begin{equation}
C_{F}(L) = \frac{\overline{\Delta F_A(L)\Delta F_B(L)}}{
\sqrt{\overline{\Delta F_A^{2}(L)}}\sqrt{\overline{\Delta F_B^{2}(L)}}}.
\label{eq-c-f-numerical}
\end{equation}
where $\Delta F$ is the deviation from the mean free-energy,
\begin{equation}
\Delta F_A(L)=F_{A}(L)-\overline{F_{A}(L)} \qquad
\Delta F_B(L)=F_{B}(L)-\overline{F_{B}(L)}.
\end{equation}

We also computed the overlap function $q(L,\delta U)$ 
defined in \eq{eq.def-q} using the method of \cite{M90}. 
However it requires much computational effort because
one has to keep track of trajectories of two strings
simultaneously and computation 
was limited to smaller system sizes $L \sim 500$. 
So we do not display the result in the following. We only 
note that anticipated scaling \eq{eq-scaling-q} was checked
within the limited system sizes.

\subsection{Uniform Tilt Field}
\label{subsec:tm-tilted-dprm}

First we examine the case of the perturbation by a uniform 
tilt field. For simplify, the temperature is set to zero $T=0$.
The two replicas have exactly the same 
random potential. The difference is that replica B
is subjected to a uniform tilt field $h$ which amounts
to a force acting just on its end,
\begin{equation}
E_{A}[V,h_{A}=0,u_A] =  
\sum_{z=1}^{L} V(u_A(z),z) 
\qquad E_{B}[V,h_{B}=h,u_B] =  
\sum_{z=1}^{L} \left [ 
 V(u_B(z),z) \right] - h u_{B}(L).
\end{equation}
We have used the $T=0$ transfer matrix method and obtained
the ground states with various perturbation strengths:
$h=0,0.05,0.1,0.2,0.3,0.4$ for each realization of random potential.

Let us begin with the mean-squared transverse displacement 
of the end point due to perturbation $B_{2}(L)$ defined 
in \eq{eq-b-numerical}. 
$B_{2}(L)$ is expected to grow with increasing size $L$ 
as $L^{1+2/3=5/3}$ in the weakly 
perturbed regime (see \eq{eq:b-weak}) and  as $L^2$ 
in the strongly perturbed regime (see \eq{b-strong-q} 
and \eq{b-strong-1-uni}).
Here we used the exponent associated to this perturbation 
$\alpha=\zeta=2/3$ (see section 
\ref{subsec:perturbation-uniform}).
The crossover between the two is expected to take place at
the overlap length $L_{c} \sim h^{-3}$ given in \eq{lc-uniform-tilt}.

\begin{figure}[h]
\begin{center}
\includegraphics*[scale=0.45]{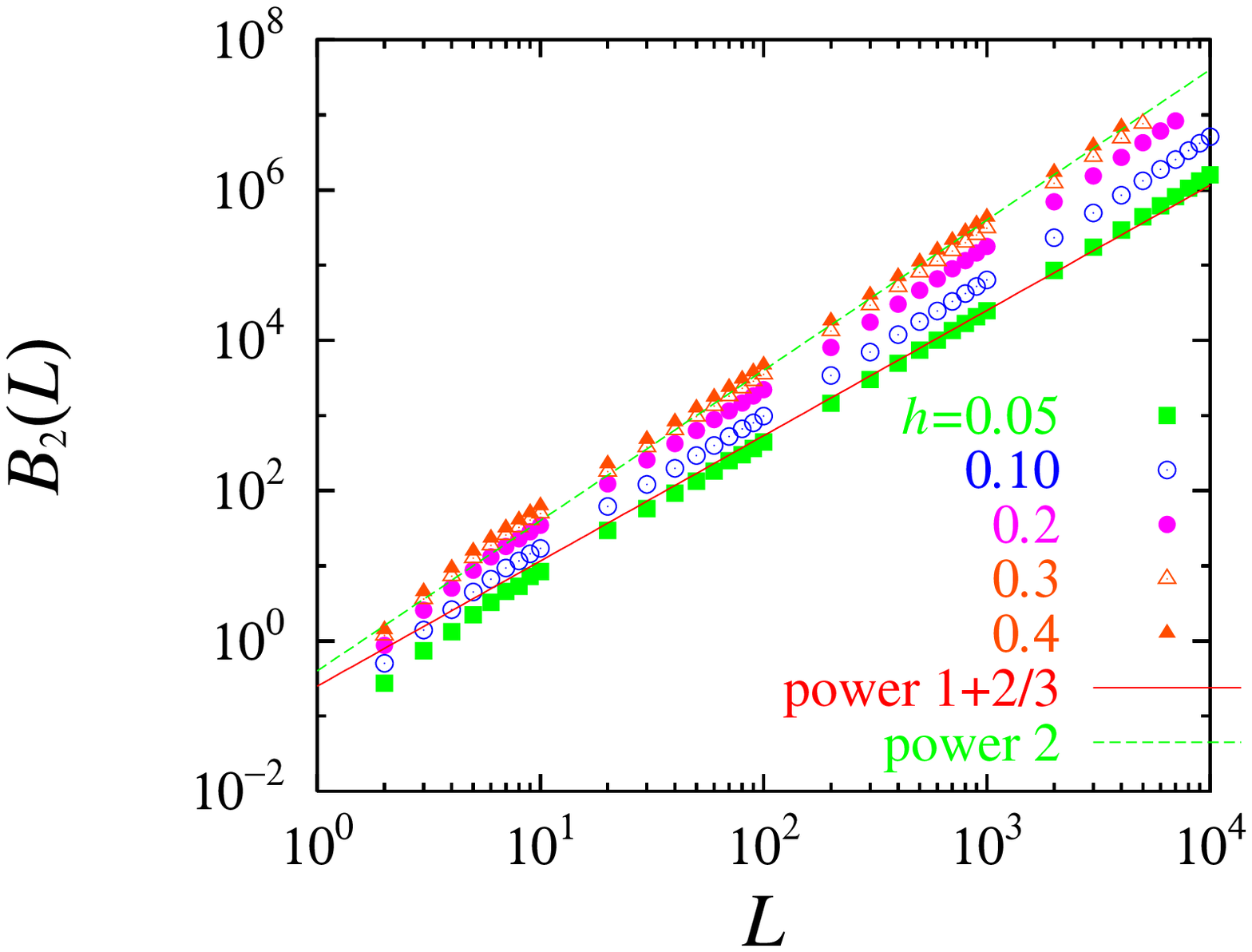}
\includegraphics*[scale=0.45]{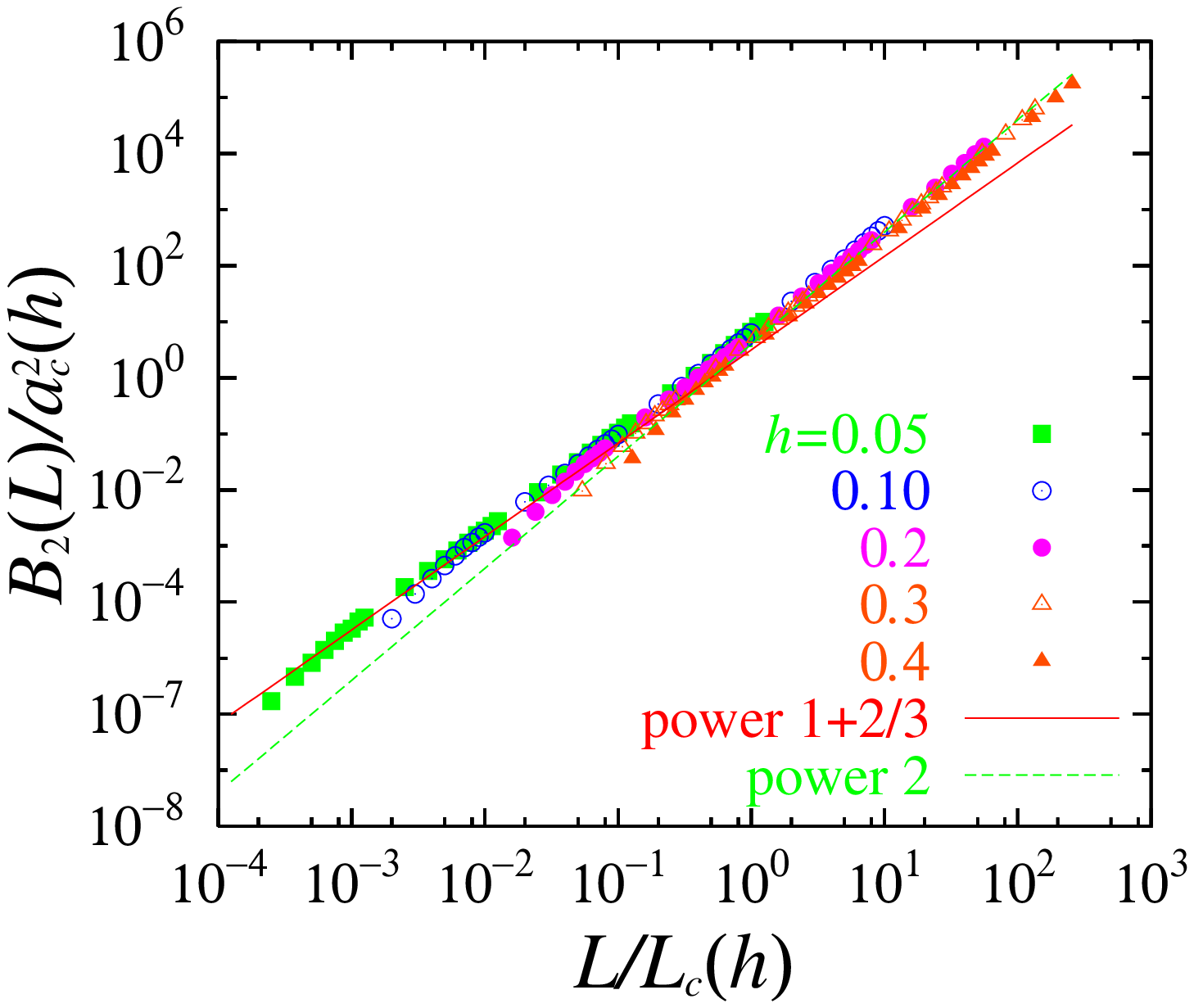}
\end{center}
\caption{$B_{2}(L)$ of uniform tilt field case.
The data is shown on the left
and its scaling plot is shown on the right. 
Here the scaling parameters are $L_{c}(h)=h^{-3}$ and
$u_{c}(h)=L_{c}(h)^{\zeta=2/3}$.
}
\label{b-uni.fig}
\end{figure}

In Fig. \ref{b-uni.fig}, the data of $B_{2}(L)$ and 
its scaling plot is shown. For very weak perturbations $h=0.05$,
the data grows almost entirely as $L^{1+2/3}$ except for a
short length transient. On the contrary,
the data corresponding to a strongest perturbation $h=0.4$ 
grows almost entirely as $L^{2}$, again except for
a short length transient. The data for the intermediate 
range of $h$ displays a crossover between the two. Indeed, 
the scaling plot confirms the expected crossover scaling between 
the two regimes with no adjustable parameters.

\begin{figure}[h]
\begin{center}
\includegraphics*[scale=0.45]{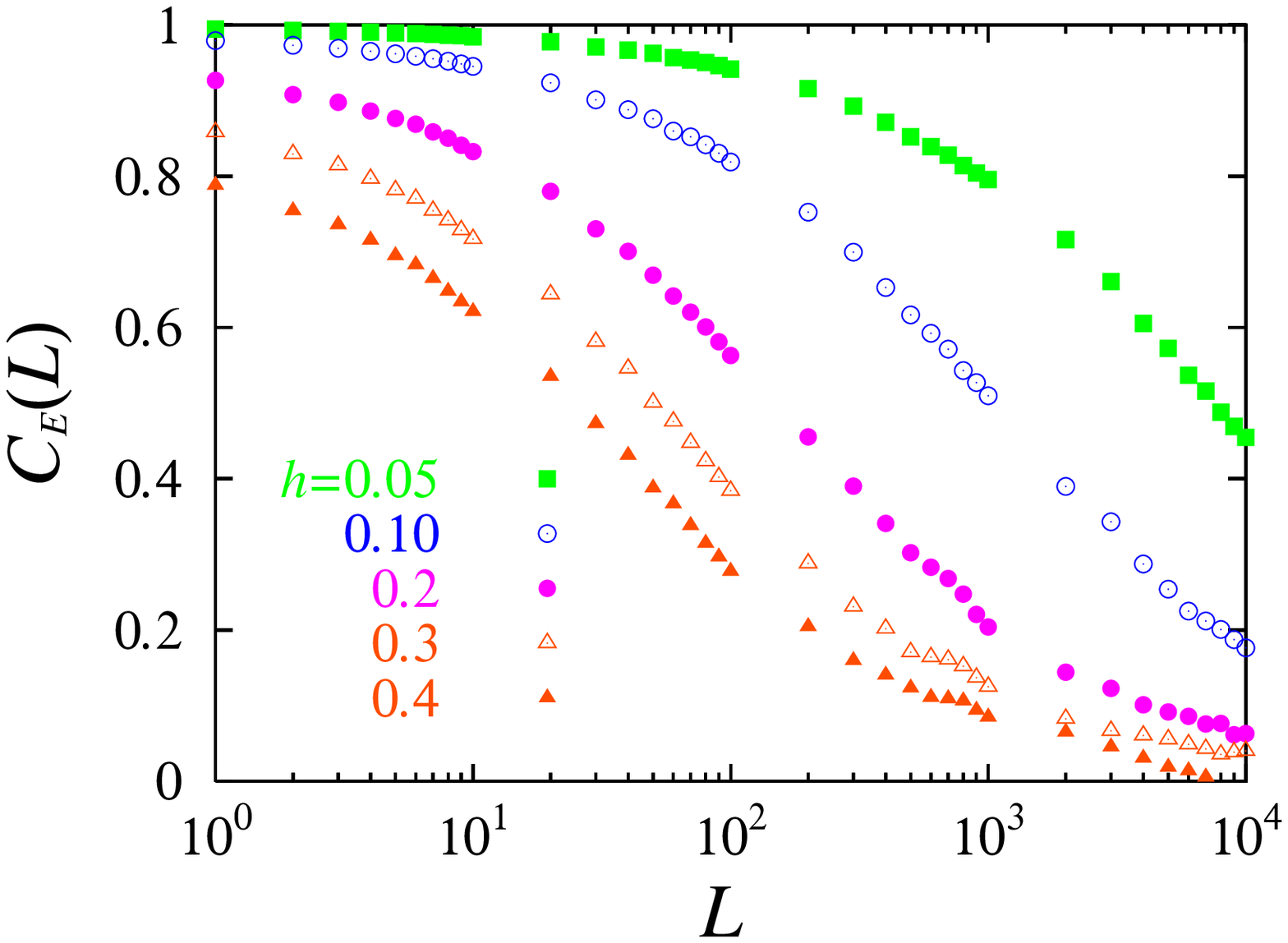}
\includegraphics*[scale=0.45]{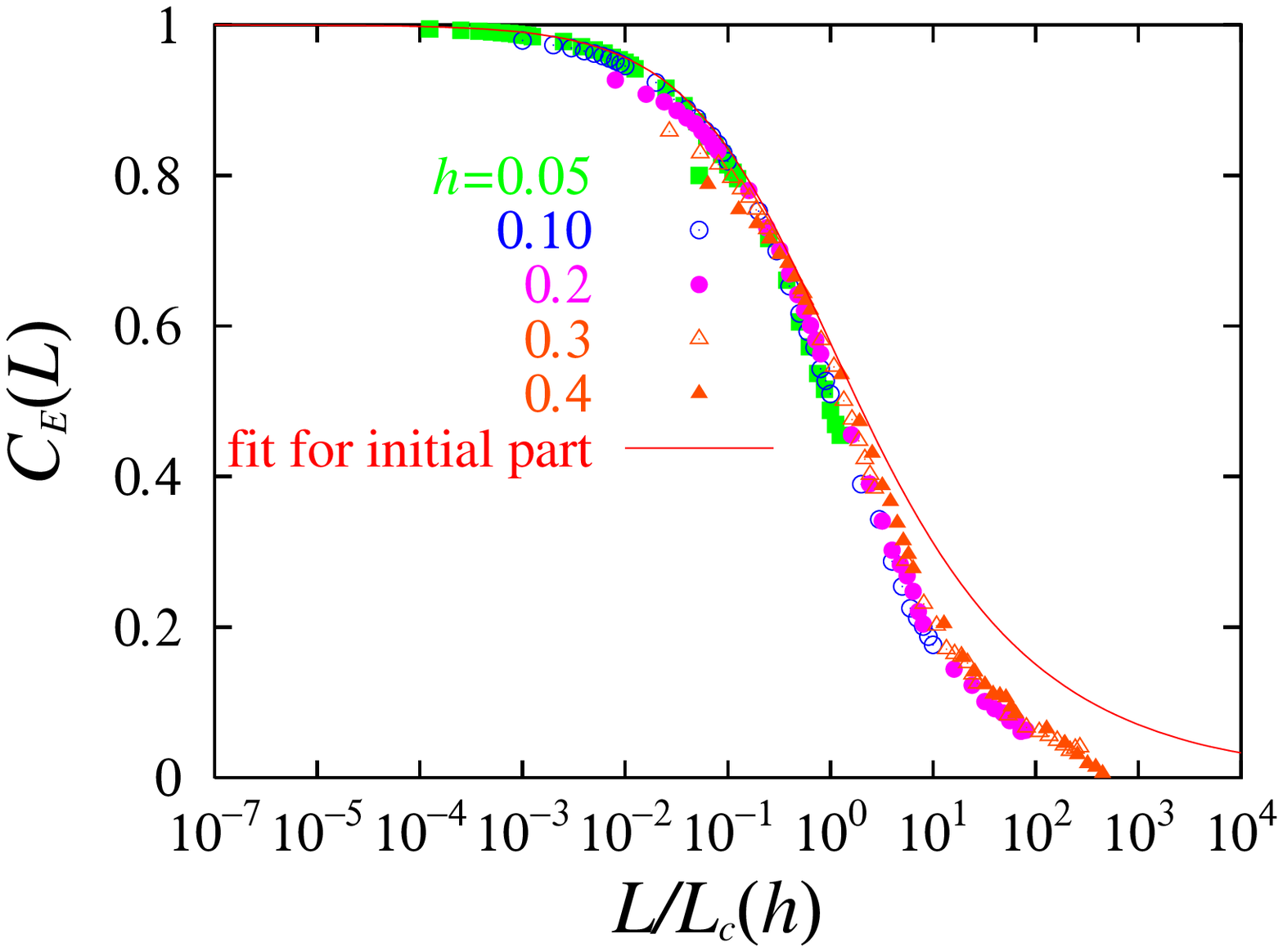}
\end{center}
\caption{$C_{E}(L)$ of the uniform tilt field case
and its scaling plot with $L_{c}(h)=h^{-3}$.
The fit is $C_{F}(L)=1/(1+A(L/L_{c}(h))^{2(2/3-1/3)})$ with $A\sim 2.0$.
}
\label{corr-e-uni.fig}
\end{figure}

Next let us examine the correlation of
the ground-state energies of the perturbed and unperturbed systems
through \eq{eq-c-f-numerical}. In Fig. \ref{corr-e-uni.fig},
the data of the correlation function and its scaling plot
is shown.The data shows a  de-correlation of the (free-) energy landscape
of the two systems as expected. The scaling plot is obtained
again without any adjustable parameters. The initial part of the
master curve is well fitted 
by the expected form \eq{eq-cf-scaling-form} using $\alpha=2/3$,
$C_{L}(F)=1/(1+A(L/L_{c}(h))^{2(\alpha-1/3)})$ with $A\sim 2.0$.
Note that the decay is faster for $L/L_{c}(h) \gg 1$.

\subsection {Modified Sinai model}
\label{subsec:numerics-tilted-sinai}

In section \ref{sec.sinai} we proposed a modified 
Sinai model as an effective model for the free-ends of 
DPRM under uniform tilt field. 
Here we study numerically the properties of ``ground states'' 
of the modified Sinai model and study the mean squared displacement
corresponding to \eq{eq-b-numerical} and the correlation function of the
ground state energies corresponding to \eq{eq-c-f-numerical}.
The effective Hamiltonian given in \eq{eq-modified-sinai-1} 
and \eq{eq-modified-sinai-2} at a given length $L$
reads as \footnote{Note that here we do not use re-scaled variables 
such as in \eq{effh}.}\footnote{the same results are obtained if we 
consider that the two strings are tilted with $-h$ and $+h$ respectively.},
\begin{eqnarray}
H= H_A+H_B \qquad
H_A=\frac{1}{ 2L}x_A^2+{\cal V}(x_A) \qquad
H_B=\frac{1}{ 2L}x_B^2+{\cal V}(x_B)-h\;x_B
\end{eqnarray} 
where ${\cal V}(x)$ is the modified Sinai potential with correlations,
\begin{equation}
\overline{({\cal V}(x)-{\cal V}(y))^2}=
u+(u^{*}(L)-u)\theta(u-u^{*}(L)) \qquad \mbox{with}
\qquad u^{*}(L) = L^{2/3}
\end{equation}

First we prepared Sinai potential $V(x)$ on a 1 dimensional lattice 
$u=1,2,\ldots,R$ of size $R$ by generating random walks in 1 
dimensional space (regarding the 1 dimensional space coordinate 
as the ``time'' coordinate for the random walk).
We generated the {\it bounded} Sinai potential by a 1 dimensional
random walk confined in a box of size $u^{*}$.
Each step of the random walk has variance $1$.
The same  random potential is generated for two replicas A and B. 
For the B replica, we add an extra tilting potential $-hu$.
Then we numerically looked for the ``ground states'' of A and B replicas.
We examined various system sizes up to $R=10^{4}$ and used
$10^{4}$ samples for the disorder averages.

\begin{figure}[h]
\begin{center}
\includegraphics*[scale=0.45]{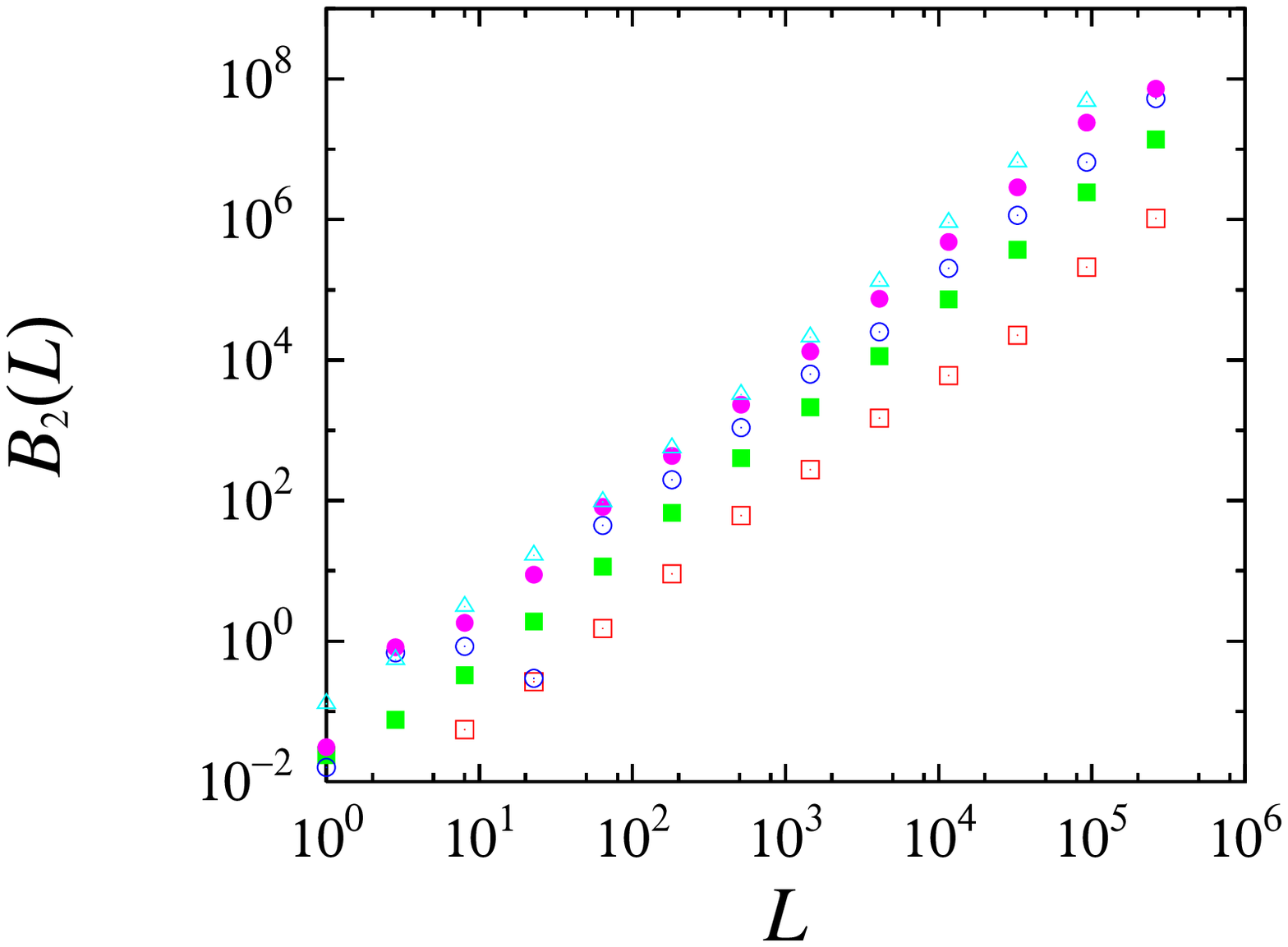}
\includegraphics*[scale=0.45]{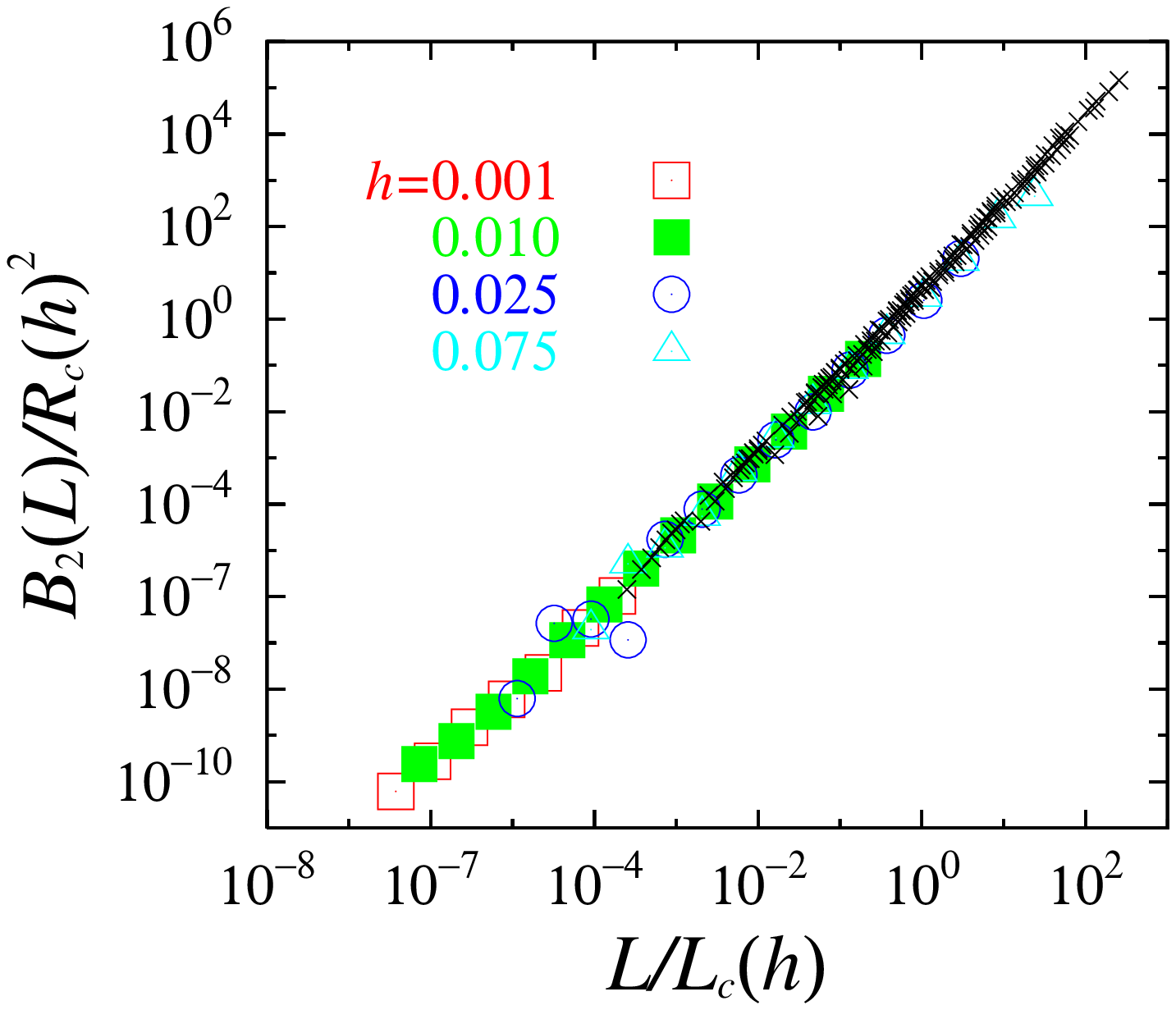}
\end{center}
\caption{$B_{2}(L)$ computed by the modified 
Sinai model and its scaling plot with $R_{c}(h)=(0.9 h)^{-2}$. 
In the scaling
plot, the master curve of DPRM under uniform tilt field 
plotted vs $L/L_{c}(h)$ as in Fig. \ref{corr-e-uni.fig}
is also included for comparison (black points). 
}
\label{b-sinai.fig}
\end{figure}

The 2nd moment of the distance between the minimum is computed for
various $L$ and $h$ as,
\begin{equation}
B_{2}(L)=\overline{[u^{\rm min}_A(L)-u^{\rm min}_B(L)]^{2}}.
\end{equation}
In Fig. \ref{b-sinai.fig}, the mean-squared displacement is shown
together with the scaling plot.
In the scaling plot, we included
the master curve of the equivalent DPRM problem 
shown in Fig \ref{b-uni.fig}.
We used the anticipated scaling factors
$L_{c}(h)= (0.9 h)^{-3}$ and $R_{c}(h)=1.2 h^{-4}$. 
The numerical prefactors are chosen
such  that the master curve of the modified Sinai model lies
on that of the DPRM problem.

\begin{figure}[h]
\begin{center}
\includegraphics*[scale=0.4]{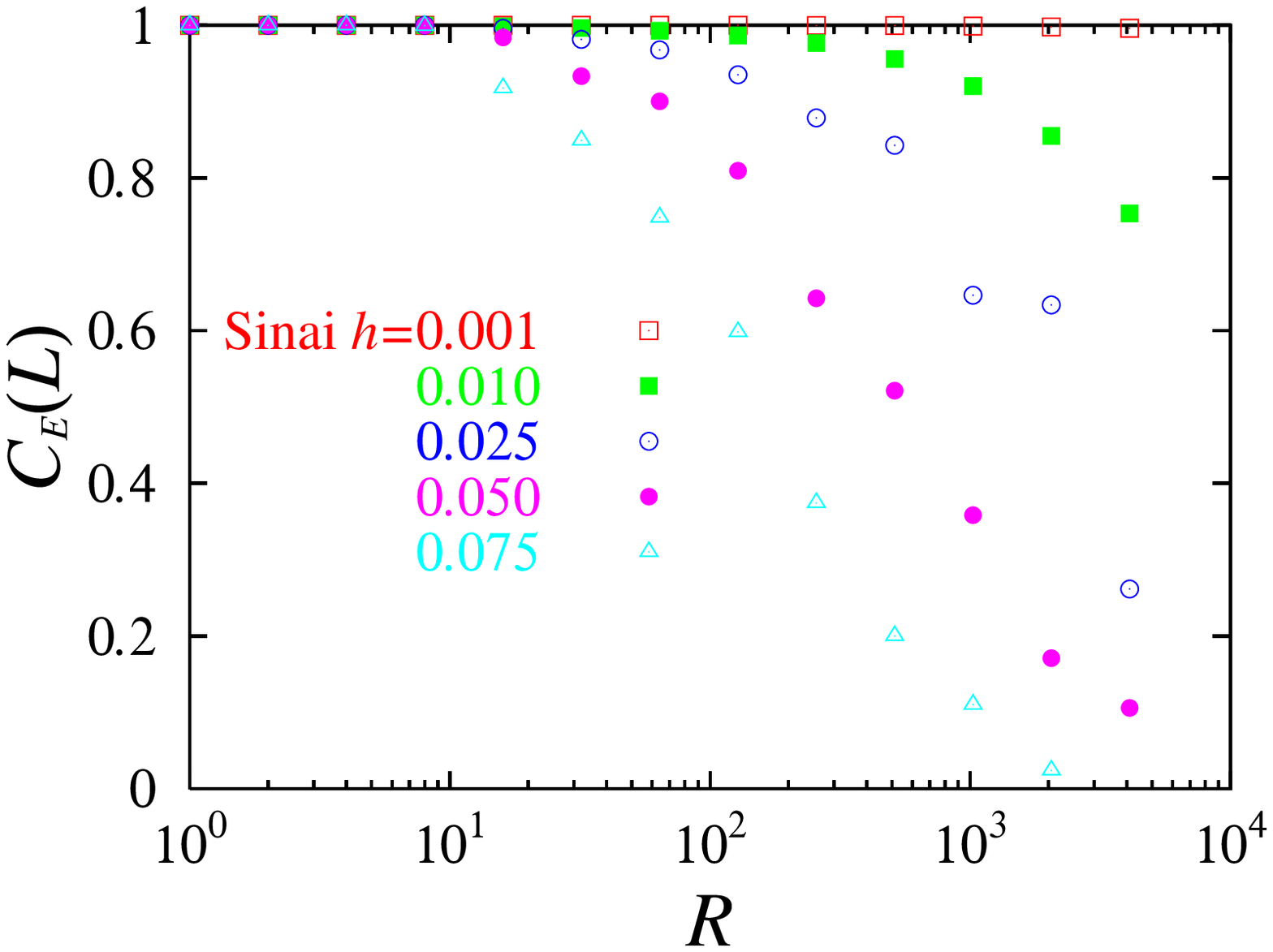}
\includegraphics*[scale=0.4]{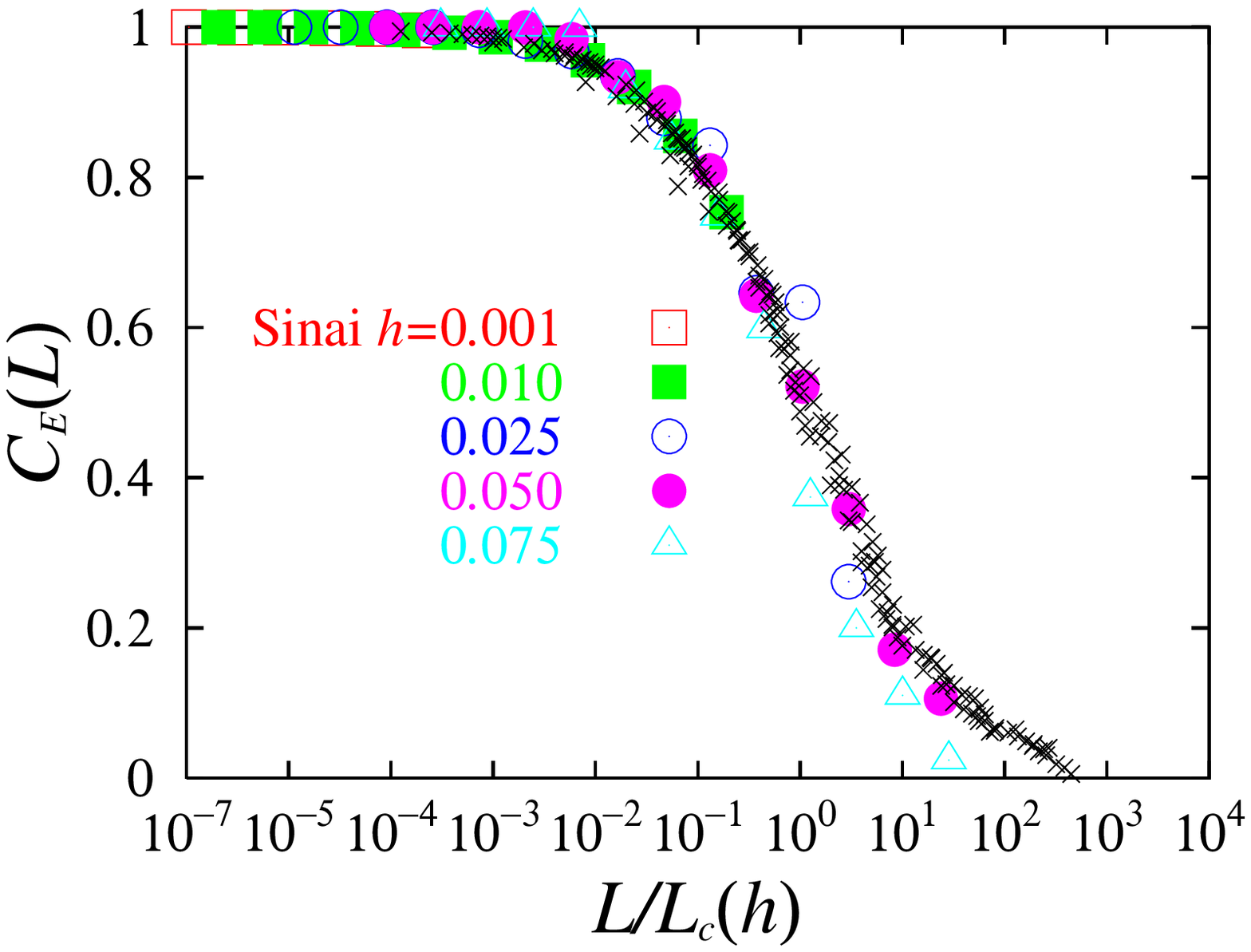}
\end{center}
\caption{$C_{E}(L)$ of Sinai model under  uniform tilt field 
and its scaling plot with $R_{c}(h)=(0.9 h)^{-2}$. In the scaling
plot, the master curve of DPRM under uniform tilt field 
plotted vs $L/L_{c}(h)$ as in Fig. \ref{corr-e-uni.fig}
is also included for comparison (black points).
}
\label{corr-e-sinai.fig}
\end{figure}

The correlation function of the fluctuation of ground state energies
is computed for various $L$ and $h$ as,
\begin{equation}
C_{E}(L)=\frac{\overline{\Delta E_A(L)\Delta E_B(L)}}{
\sqrt{\overline{\Delta E_A^{2}(L)}}\sqrt{\overline{\Delta E_B^{2}(L)}}}.
\end{equation}
where $\Delta E(L)$ is the deviation of a ground-state energy
from the mean ground-state energy.
In Fig. \ref{corr-e-sinai.fig} we show the correlation function 
of the fluctuation of the ground state energy as well as
its scaling plot using the anticipated scaling variable
$L/L_{c}(h)$.  In the plot, we have included
the master curve of the equivalent DPRM problem 
shown in Fig \ref{corr-e-uni.fig}.

It can be seen that the agreement between the modified Sinai
model and the original DPRM under uniform tilt field is good.
We checked that if the original unbounded Sinai potential is used,
the agreement becomes very bad for large lengthscales. Especially,
the correlation function $C_{E}(L)$ tends to saturate.
These results support the picture that RSB is needed to account
for the de-correlation of energy landscape of DPRM under uniform
tilt field.

\subsection{Explicit Repulsive Coupling}

Next we consider the case of explicit repulsive coupling.
The two replicas are at zero temperature, 
have exactly the same random potential
and are coupled by an explicit repulsive coupling $\epsilon$,
\begin{equation}
E[V,V,\epsilon,u_A,u_B] =  \sum_{z=1}^{L} \left ( V(u_A(z),z) + V(u_B(z),z)
+\epsilon \delta_{u_A(z),u_B(z)} \right)
\end{equation}
Here $\epsilon>0 $ is the strength of the perturbation.
M\'{e}zard \cite{M90} proposed a transfer matrix method
to deal with such a coupled system at $T>0$.
Here we used  a $T=0$ version of the method
and studied the ground states with different repulsive couplings
$\epsilon=0.05,0.07,0.1,0.2,0.3$.

\begin{figure}
\begin{center}
\includegraphics*[scale=0.425]{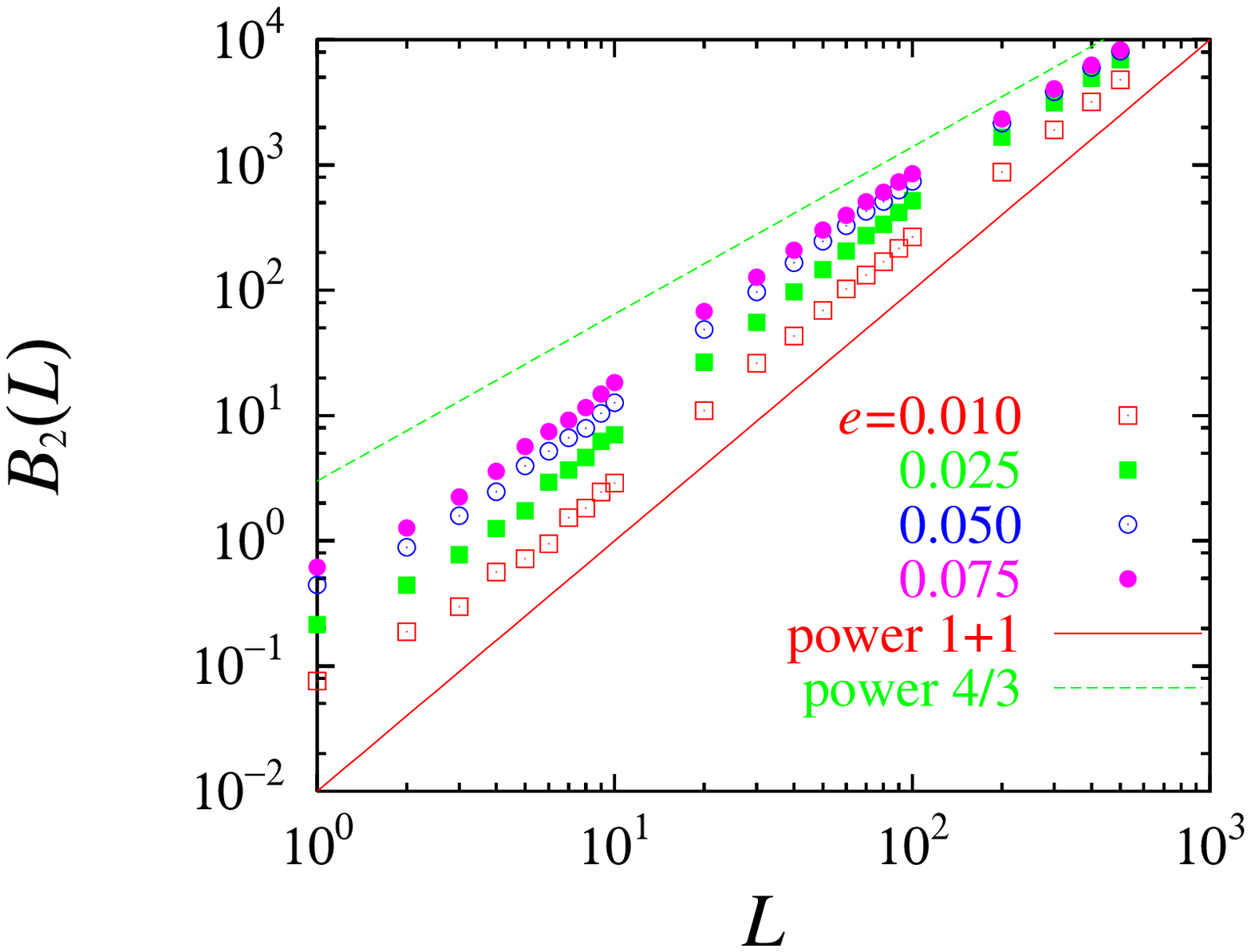}
\includegraphics*[scale=0.425]{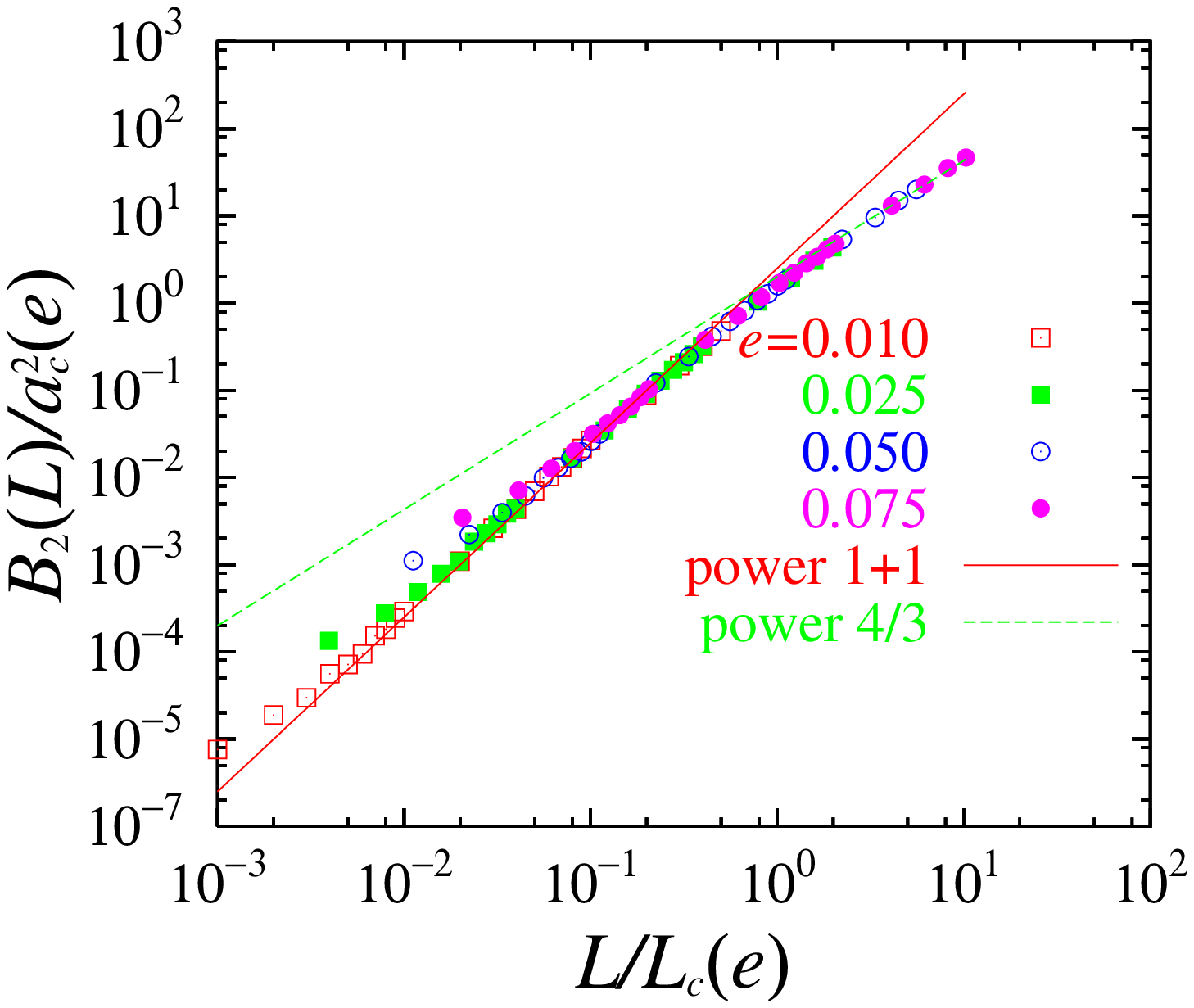}
\end{center}
\caption{$B_{2}(L)$ of the explicit repulsive coupling case
and its scaling plot.
Here the scaling parameters are $L_{c}(\epsilon)=\epsilon^{-3/2}$ and
$u_{c}(\epsilon)=L_{c}(\epsilon)^{\zeta=2/3}$.
}
\label{b-rep.fig}
\end{figure}

In Fig. \ref{b-rep.fig} the data of the mean-squared 
distance of the end points of the two replicas $B_{2}(L)$
is shown together with its scaling plot. 
From the discussion in section \ref{subsec:b},
it is expected to grow with increasing size $L$ as $L^{1+1=2}$ in the weakly 
perturbed regime and  $L^{4/3}$ in the strongly perturbed regime.
Here we have used the exponent of the perturbation
 corresponding to the explicit repulsive coupling perturbation $\alpha=1$ 
found in \ref{subsec:perturbation-rsb-1} (which is
related to the order of the perturbation 
$p=2$ in the replica analysis in section  \ref{sec.perturbation-rsb})) .
The crossover between both regimes is expected to take place at
the overlap length $L_{c} \sim \epsilon^{-2/3}$ given in \eq{lc-rep}.
These features are well confirmed by the data and the scaling plot.

\begin{figure}
\begin{center}
\includegraphics*[scale=0.4]{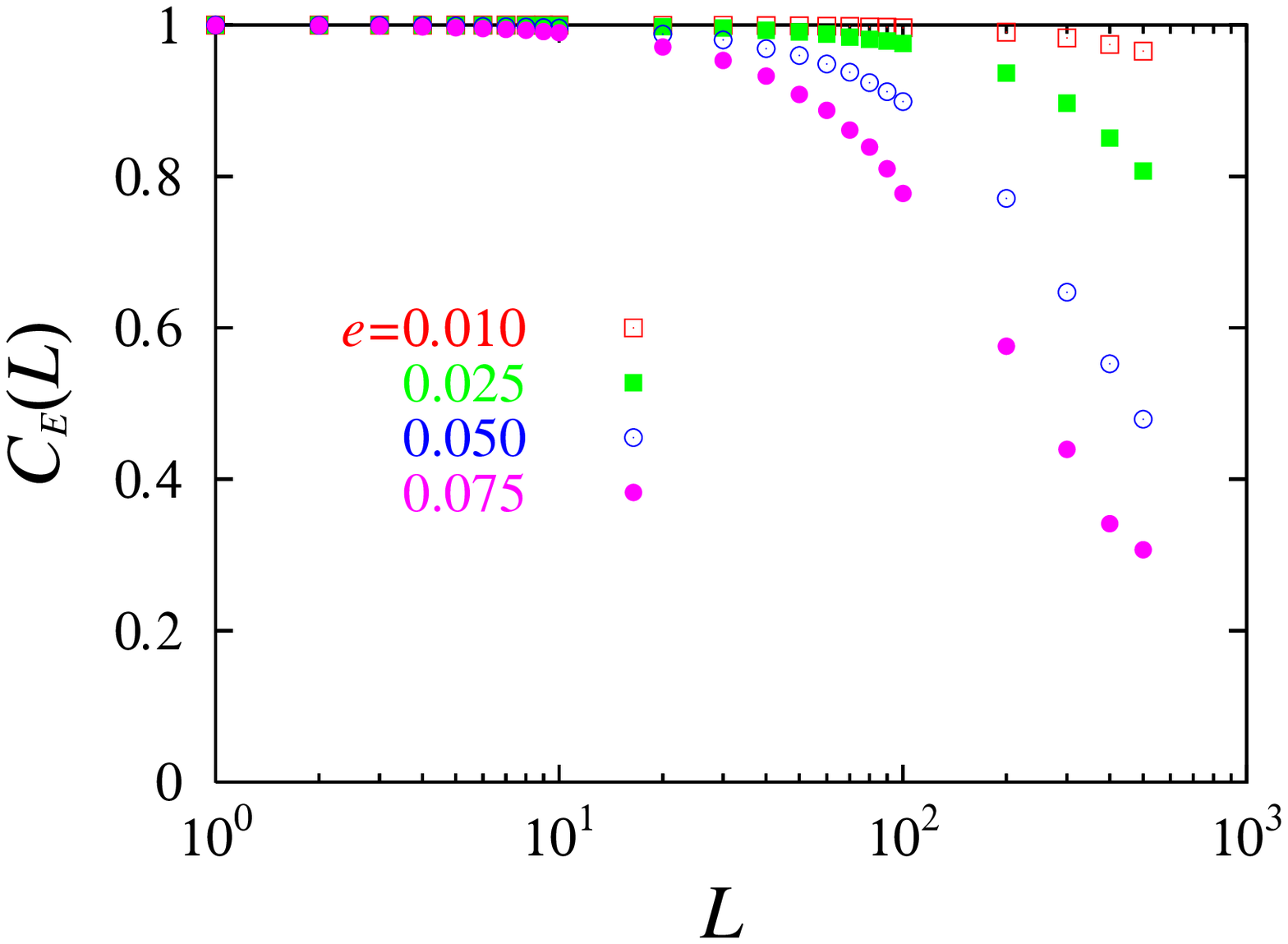}
\includegraphics*[scale=0.4]{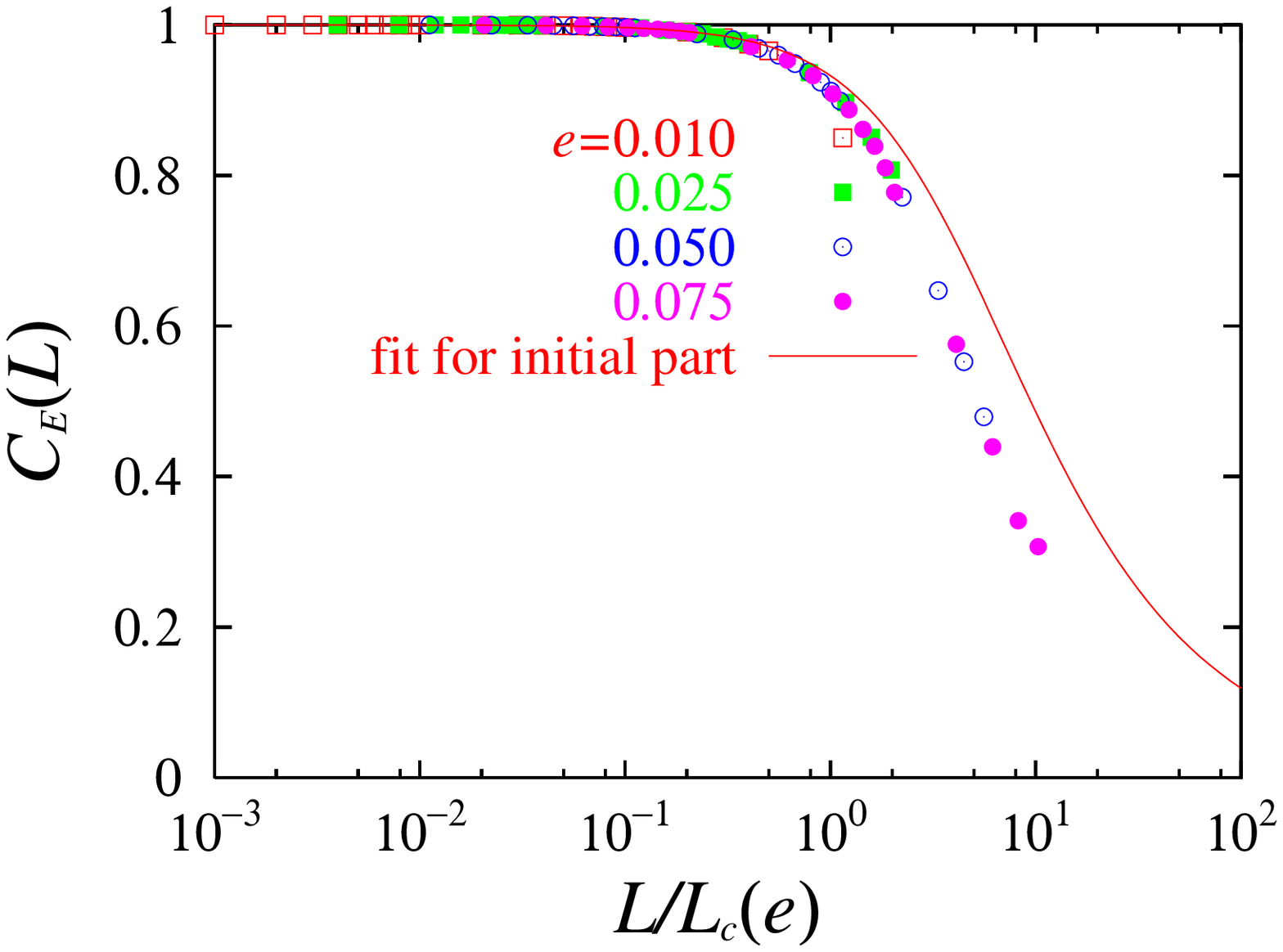}
\end{center}
\caption{$C_{E}(L)$ of the explicit repulsive coupling case
and its scaling plot.
Here the scaling parameter is $L_{c}(\epsilon)=\epsilon^{-3/2}$.
The fit is $C_{F}(L)=1/(1+A(L/L_{c}(h))^{2(1-1/3)})$ with $A\sim 2.0$.
}
\label{corr-e-rep.fig}
\end{figure}

In Fig. \ref{corr-e-rep.fig}, the correlation of 
the energies of the two replicas $E_{A}=\sum_{z=1}^{L}V(u_{A}(z),z)$
and $E_{B}=\sum_{z=1}^{L}V(u_{B}(z),z)$ is shown together with its scaling plot.
The initial part of the master curve matches properly 
with the expected form \eq{eq-cf-scaling-form} using $\alpha=1$,
$C_{L}(F)=1/(1+A(L/L_{c}(h))^{2(1-1/3)})$ with $A\sim 0.35$.
For $L/L_{c}(h) \gg 1$ the decay is much faster.

\subsection{
Perturbation on temperature, random potential and random tilt field}
\label{subsec-transfer-temp-potential-randomtilt}

Finally we examine the class of  perturbations which include
temperature-shift, potential change and random tilt field. 
These perturbations are characterized by the exponent
$\alpha=1/2$ found in section \ref{subsec:perturbation-rsb-2}
(which is related to the order of the perturbation 
$p=2$ in the replica analysis in section  \ref{sec.perturbation-rsb}).
Our primary interest here 
is to clarify whether these apparently different perturbations lead
indeed to the same universal scaling properties as anticipated by
the analytical arguments based on the replica-symmetry breaking ansatz.

\begin{enumerate}

\item{\bf Potential change}  

The Hamiltonian is given as,
\begin{equation}
E_{A}[V,u_A] =  \sum_{z=1}^{L} V(u_A(z),z)
\qquad E_{A}[V',u_B] =  \sum_{z=1}^{L}  V'(u_B(z),z)
\end{equation}
The temperature is set to zero $T=0$.
First we generate a random potential $V(u,z)$ with random 
numbers drawn  from a uniform distribution between $-1$ and $1$. This is the potential for replica A. 
In order to construct the perturbed random potential for replica B, we draw another independent
random number $U(u,z)$ from the same distribution and define,
\begin{equation}
V'(u,z)=\frac{V(u,z)+\delta U(u,z)}{\sqrt{1+\delta^{2}}}
\end{equation}
where $\delta$ is the strength of the perturbation.
We have used $T=0$ transfer matrix method and examined the ground states
for different strengths of the perturbation $\delta=0.1,0.2,0.3,0.4,0.5,0.6,0.8,1.0,1.2$.

\item{\bf Random Tilt Field} 

The Hamiltonian is given by,
\begin{eqnarray}
E_{A}[V,u_A,h_{A}=0] &=&  \sum_{z=1}^{L}  V(u_A(z),z)  \nonumber \\
E_{B}[V,u_B,h_{B}] &=&  \sum_{z=1}^{L} \left ( V(u_B(z),z) \right)
-\delta \sum_{z=1}^{L-1}h_{B}(z)(u_{B}(z+1)-u_{B}(z))\}.\qquad
\end{eqnarray}
The temperature is set to zero $T=0$. 
The two replicas have the same random potential $V(u,z)$.
The difference is that replica B is subjected to a random tilt field
$h_{B}(z)$ which for each $z$ takes a different random value which is drawn from a 
uniform distribution between $-1$ and $1$.
We have used $T=0$ transfer matrix method to examine the ground states
with different random tilt intensities
$\delta=0.1,0.2,0.3,0.4,0.5$ for each realization of random potential.

\item{\bf Temperature-shift} 

In the case of temperature perturbation, the Hamiltonian of A and
B replicas are exactly the same,  
\begin{equation}
E_{B}[V,u_A] =  \sum_{z=1}^{L} V(u_A(z),z) 
\qquad E_{B}[V,u_B] =  \sum_{z=1}^{L} V(u_B(z),z) 
\end{equation}
We have used the finite temperature version 
of the transfer matrix method. The temperature of replica A is set to
$T_{A}=0.1$.  The temperature of replica B is varied as 
 $T_{B}=T_{A}+\delta T$ with  different 
temperature shifts $\delta T=0.1,0.2,0.3,0.4,0.5,0.6,1.2$.

\end{enumerate}

\subsubsection{Transverse Jumps}

Let us first examine the mean-squared transverse displacement 
of the end point  $B_{2}(L)$ due to this class of perturbations.
By substituting $\alpha=1/2$ in \eq{eq:b-weak} we see that $B_{2}(L)$   
 is expected to grow with increasing size $L$ as $L^{1+(1/2)=3/2}$
in the weakly perturbed regime. 
In the strongly perturbed regime, it should grow as  $L^{4/3}$
, as discussed in section \ref{subsubsec:b-strong},
which is slightly slower than the growth in the weakly perturbed regime.
The difference between exponents is of only $1/6$.
The crossover between both regimes is expected to take place at
the overlap length $L_{c} \sim L_{0}\delta^{-6}$ as in
\eq{lc-potential},\eq{lc-temperature},\eq{lc-random-tilt} with  $\delta$
being the strength of the perturbation.

\begin{figure}[h]
\begin{center}
\includegraphics*[scale=0.45]{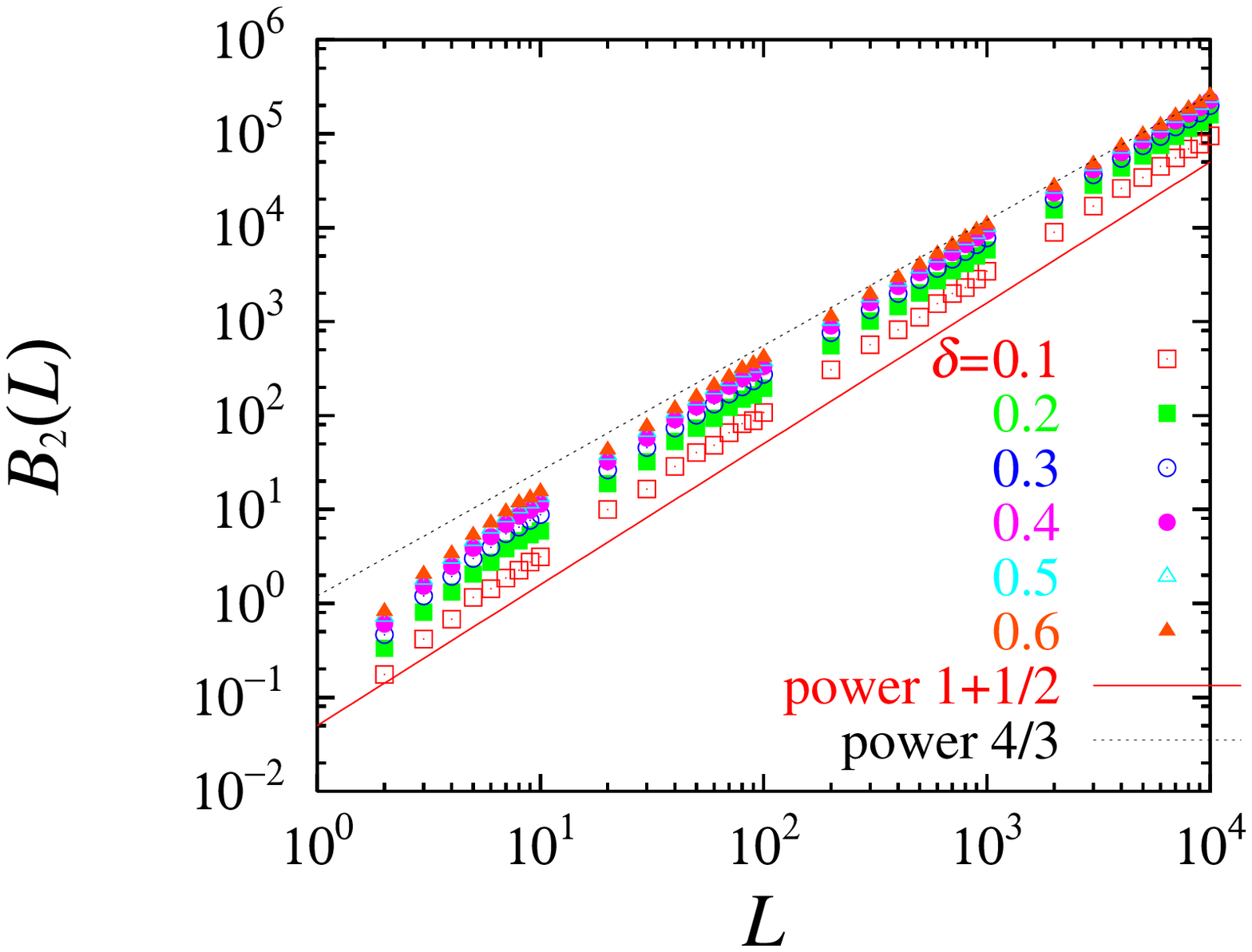}
\includegraphics*[scale=0.45]{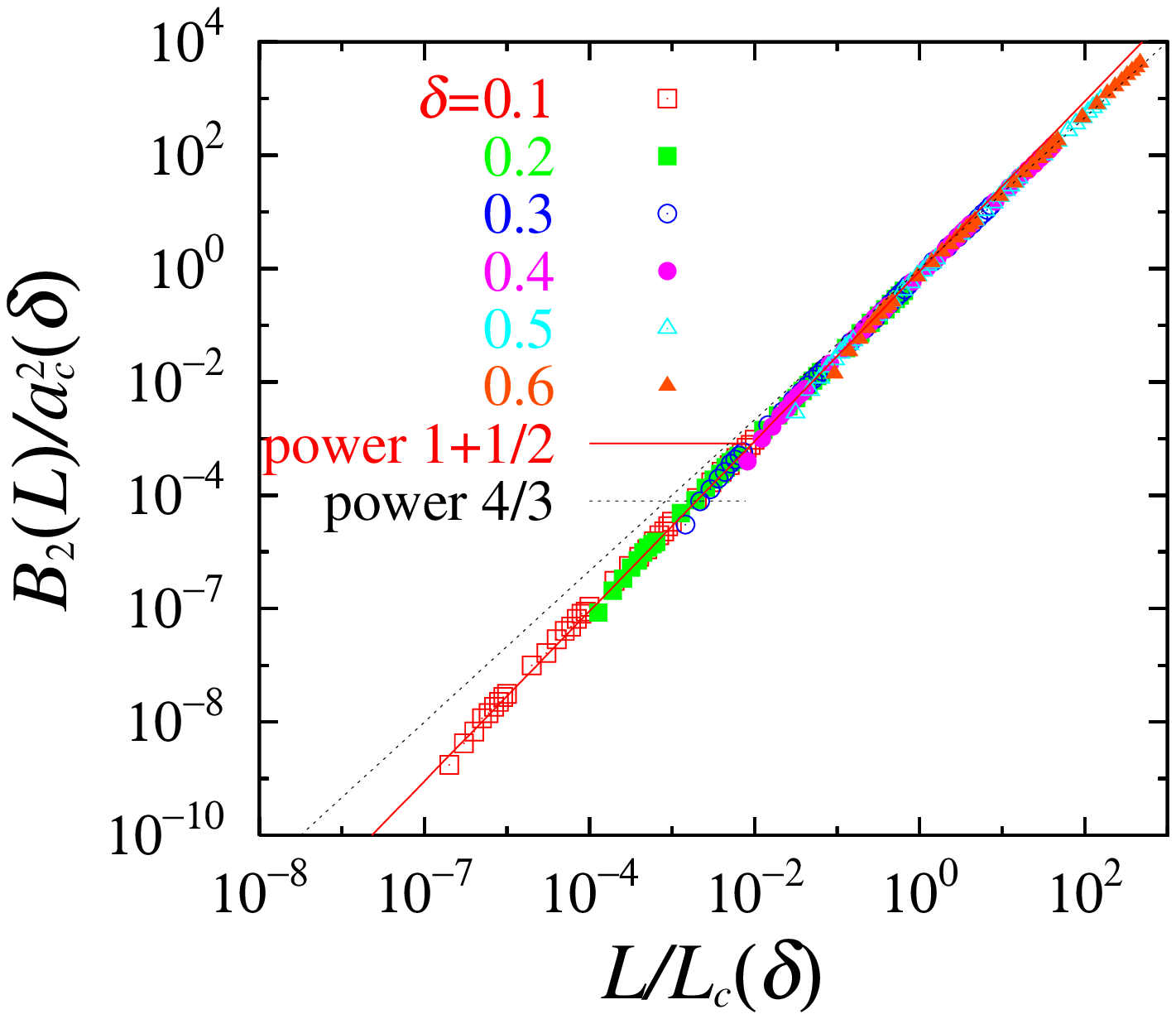}
\end{center}
\caption{$B_{2}(L)$ of the potential perturbation case
and its scaling plot with $L_{c}(\delta)= \delta^{-6}$ and
$u_{c}(\delta)=L_{c}(\delta)^{\zeta=2/3}$.}
\label{b-potential.fig}
\end{figure}

\begin{figure}[h]
\begin{center}
\includegraphics*[scale=0.45]{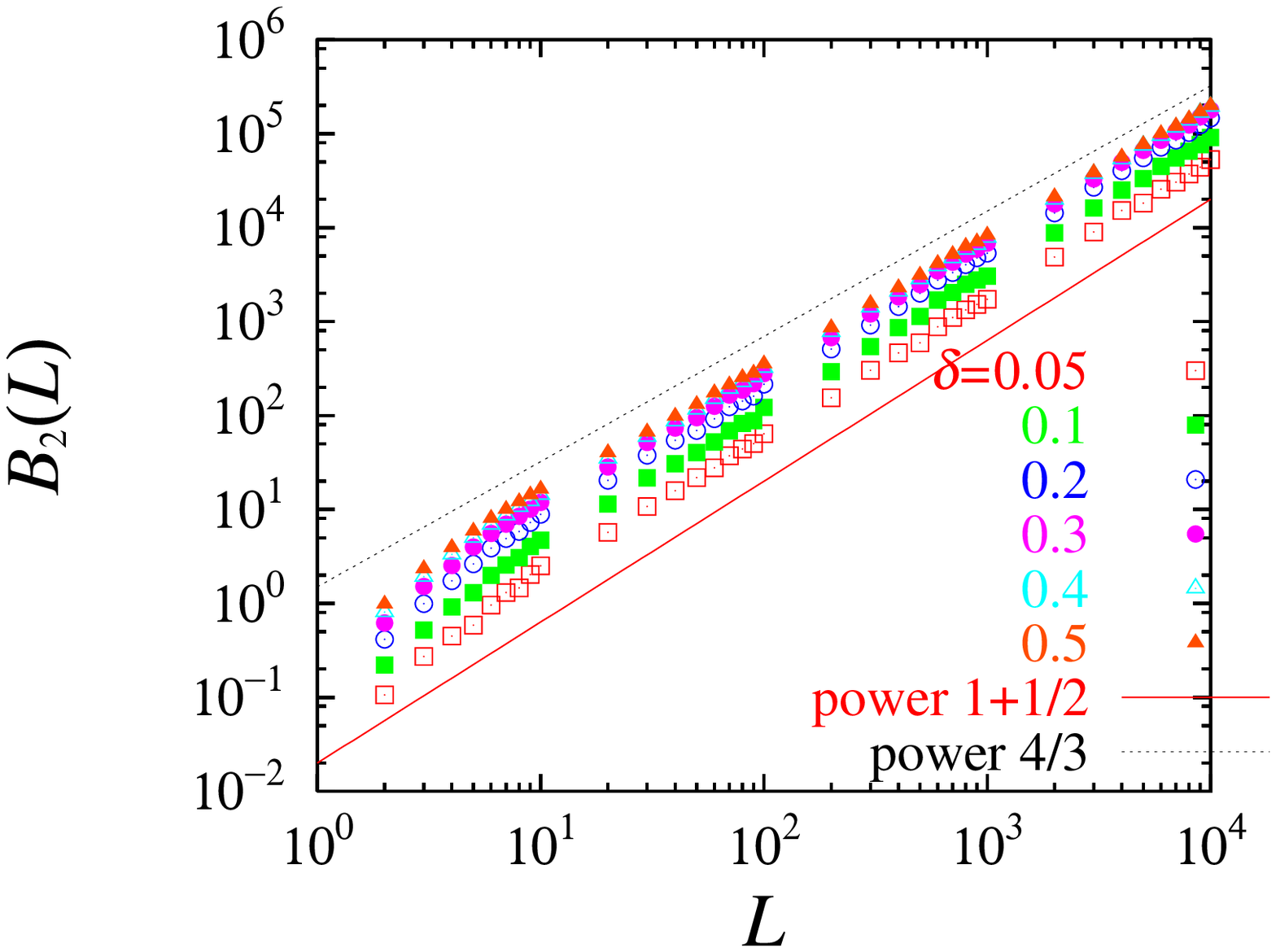}
\includegraphics*[scale=0.45]{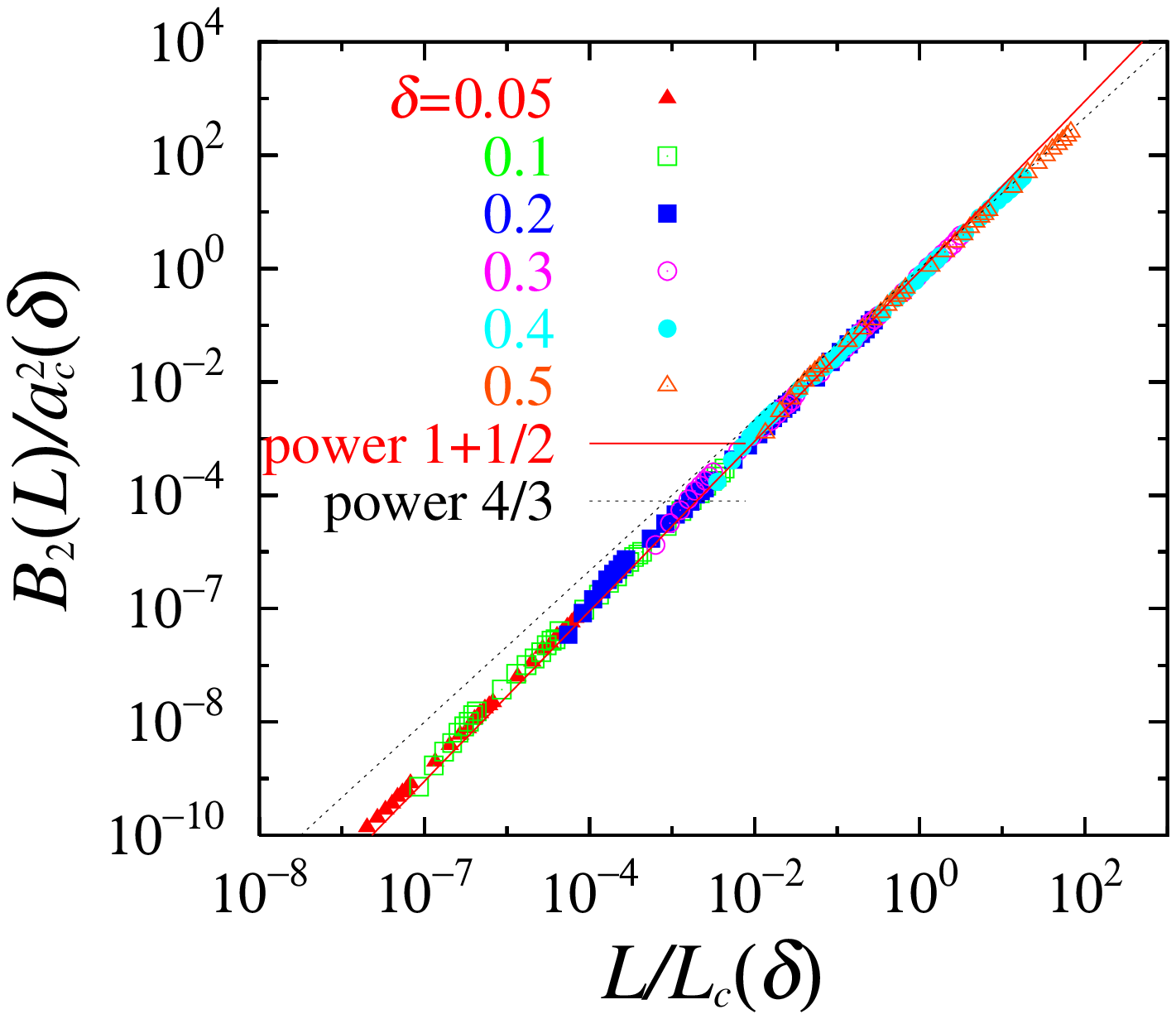}
\end{center}
\caption{$B_{2}(L)$ of the random tilt field perturbation case
and its scaling plot with $L_{c}(\delta)=(0.87 \delta)^{-6}$ and
$u_{c}(\delta)=L_{c}(\delta)^{2/3}$.}
\label{b-random-tilt.fig}
\end{figure}

\begin{figure}[h]
\begin{center}
\includegraphics*[scale=0.45]{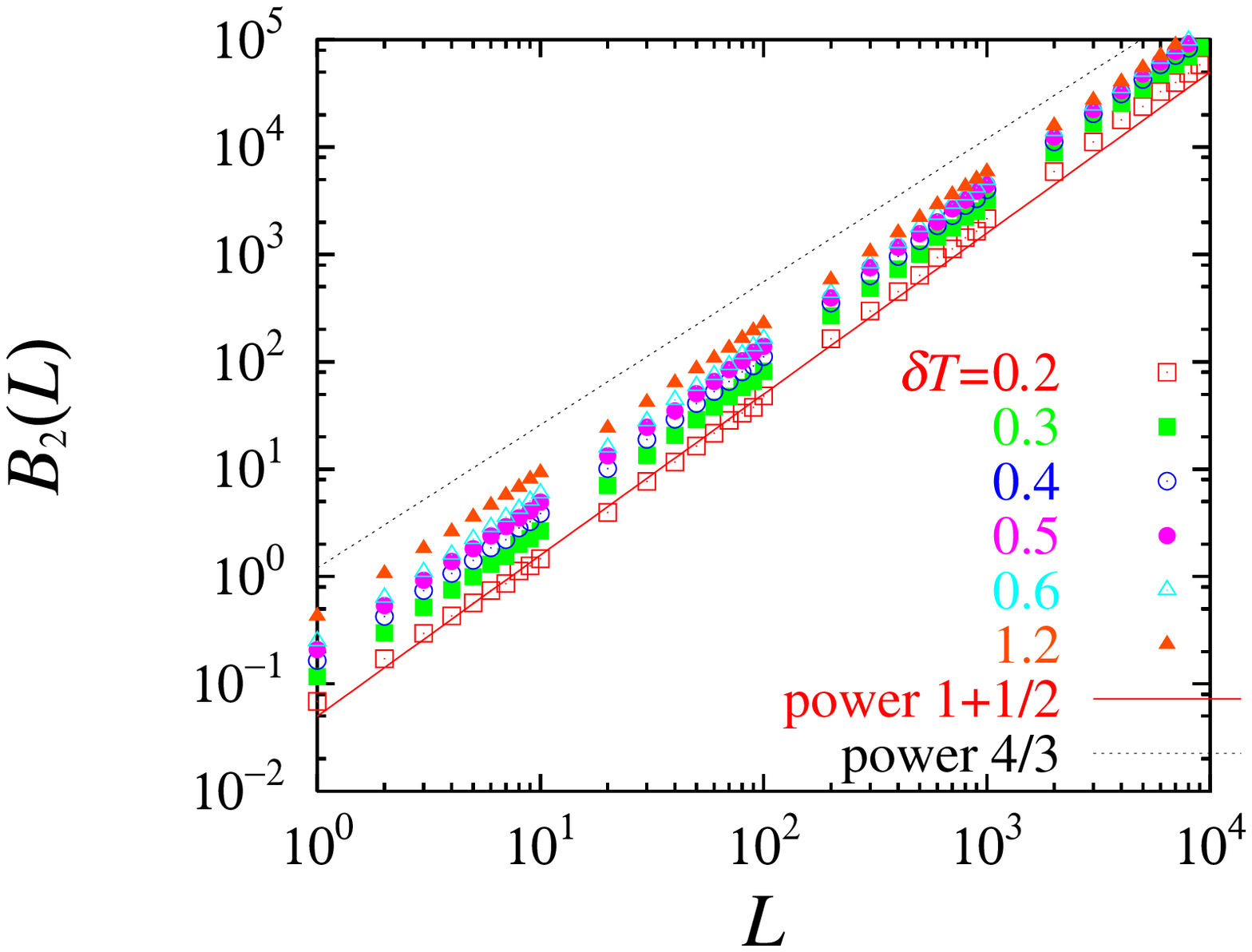}
\includegraphics*[scale=0.45]{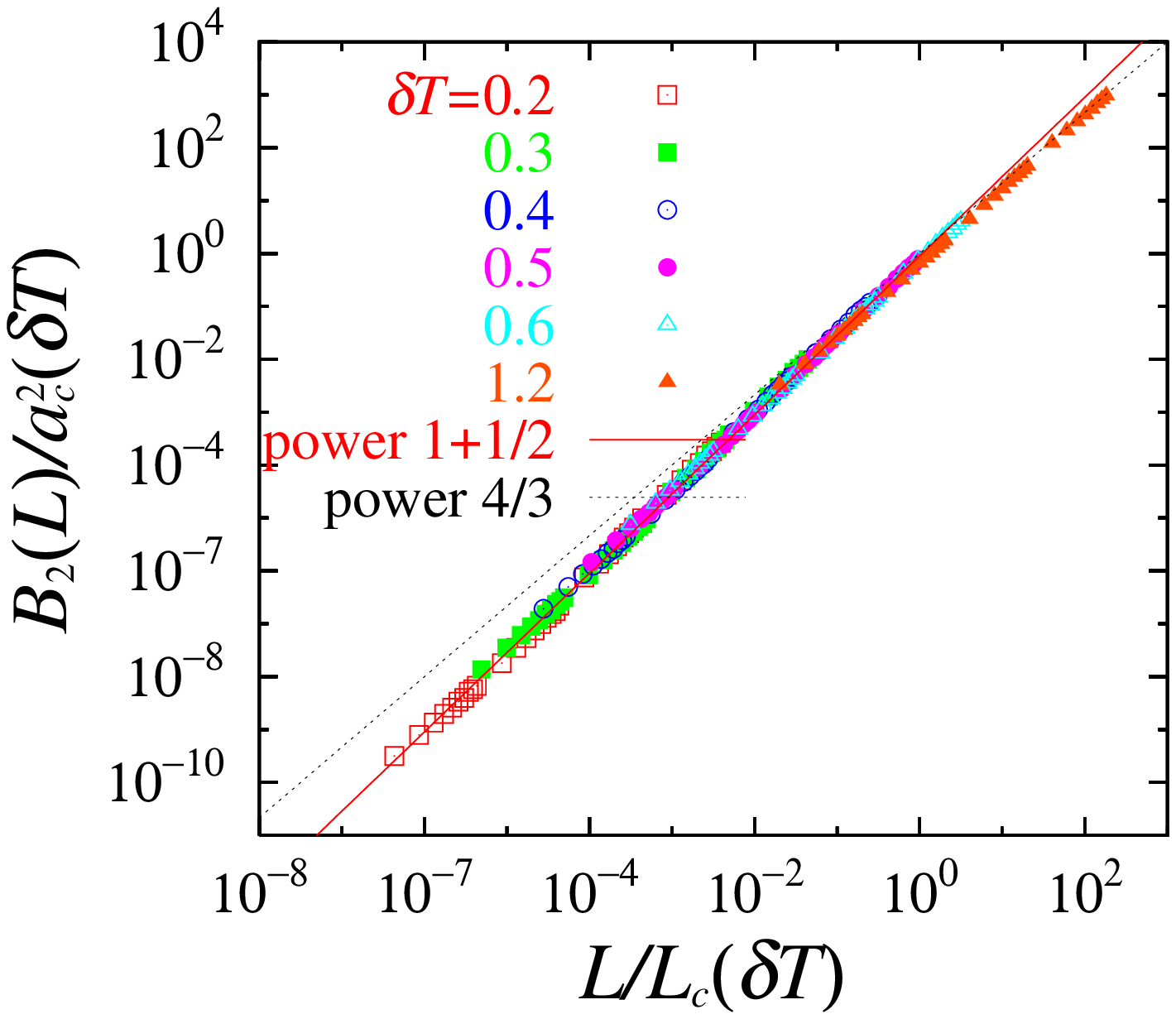}
\end{center}
\caption{$B_{2}(L)$ of the temperature-shift perturbation case
and its scaling plot with
$L_{c}(\delta T)=(0.43 \delta T)^{-6}$ and
$u_{c}(\delta T)=L_{c}^{2/3}(\delta T)$.}
\label{b-temperature.fig}
\end{figure}

In Fig. \ref{b-potential.fig} -Fig. \ref{b-temperature.fig} 
the data for $B_{2}(L)$ corresponding to the three perturbations
are shown together with their scaling plots. 
In the scaling plots we have chosen an adequate numerical prefactor $c$ in
$L_{c}(\delta)=(c\delta)^{-6}$ in order that the master curves
corresponding to the three perturbations lay on the same curve.
The resultant master-curves become indistinguishable:
the expected crossover behavior between weak and strong 
perturbation regimes is indeed the {\it same}
for the three kinds of apparently different looking perturbations.

\begin{figure}
\begin{center}
\includegraphics*[scale=0.45]{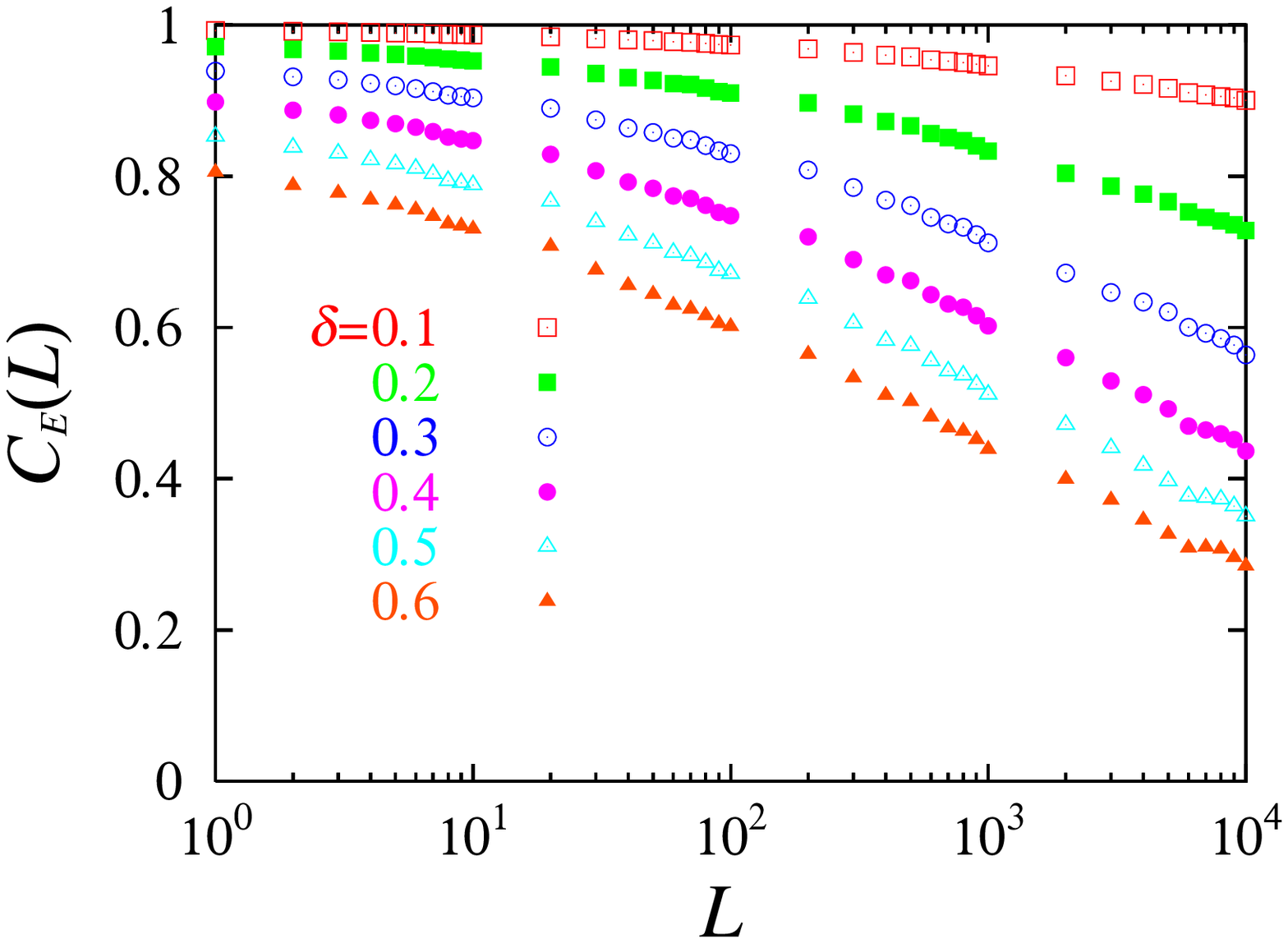}
\includegraphics*[scale=0.45]{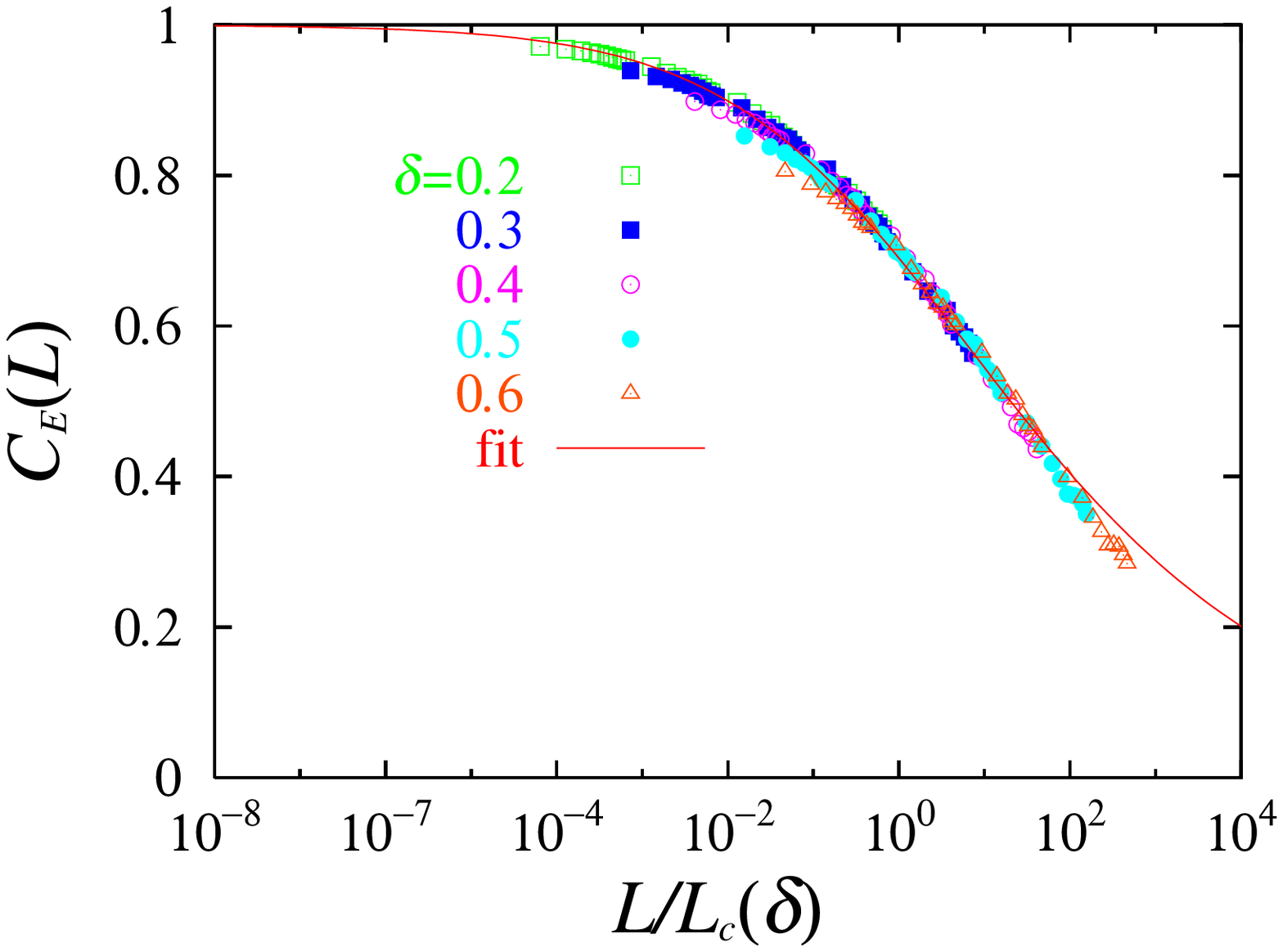}
\end{center}
\caption{$C_{E}(L)$ of the potential perturbation case
and its scaling plot with with $L_{c}(\delta)=(\delta)^{-6}$.
}
\label{corr-e-potential.fig}
\end{figure}

\begin{figure}
\begin{center}
\includegraphics*[scale=0.45]{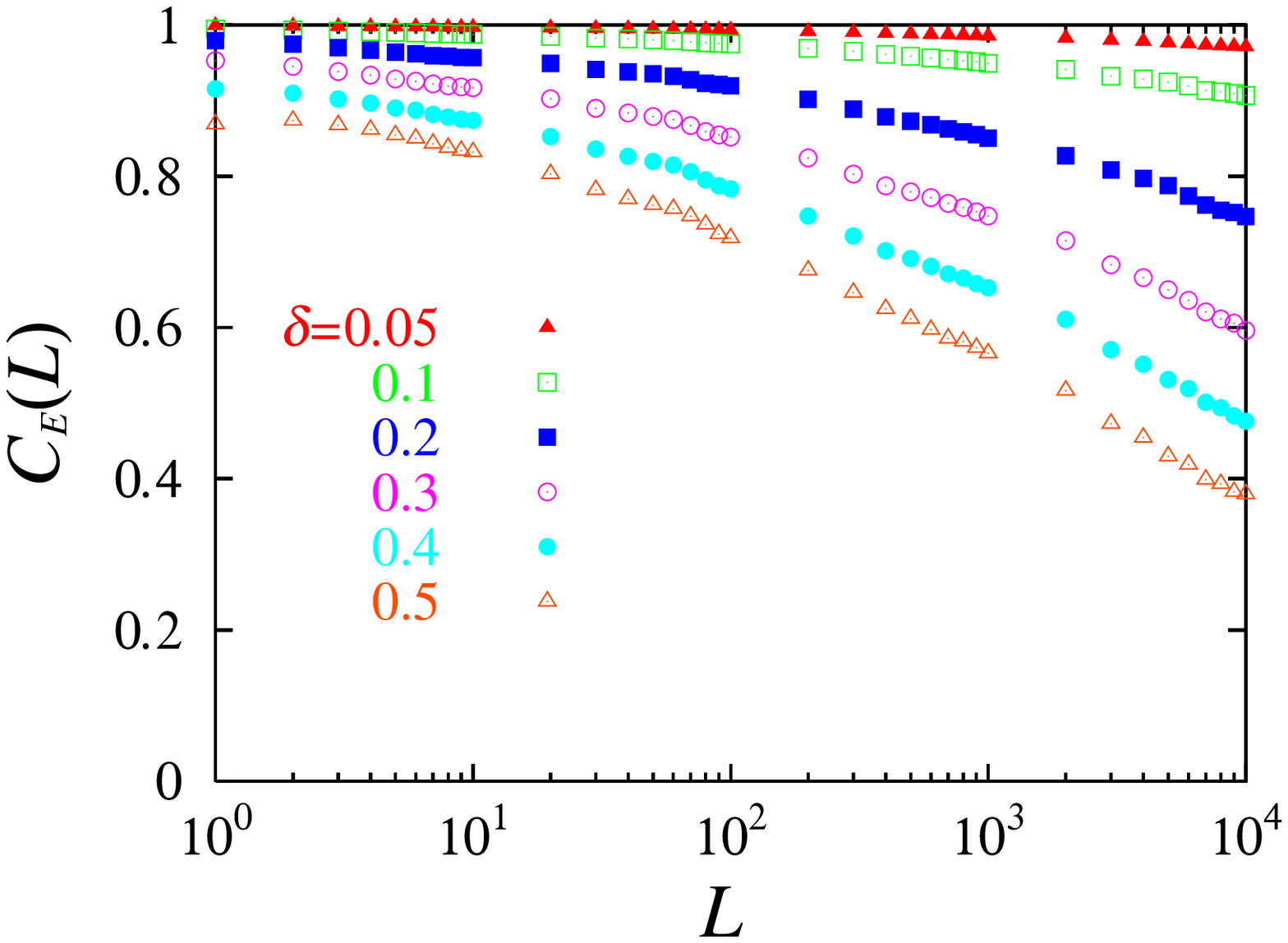}
\includegraphics*[scale=0.45]{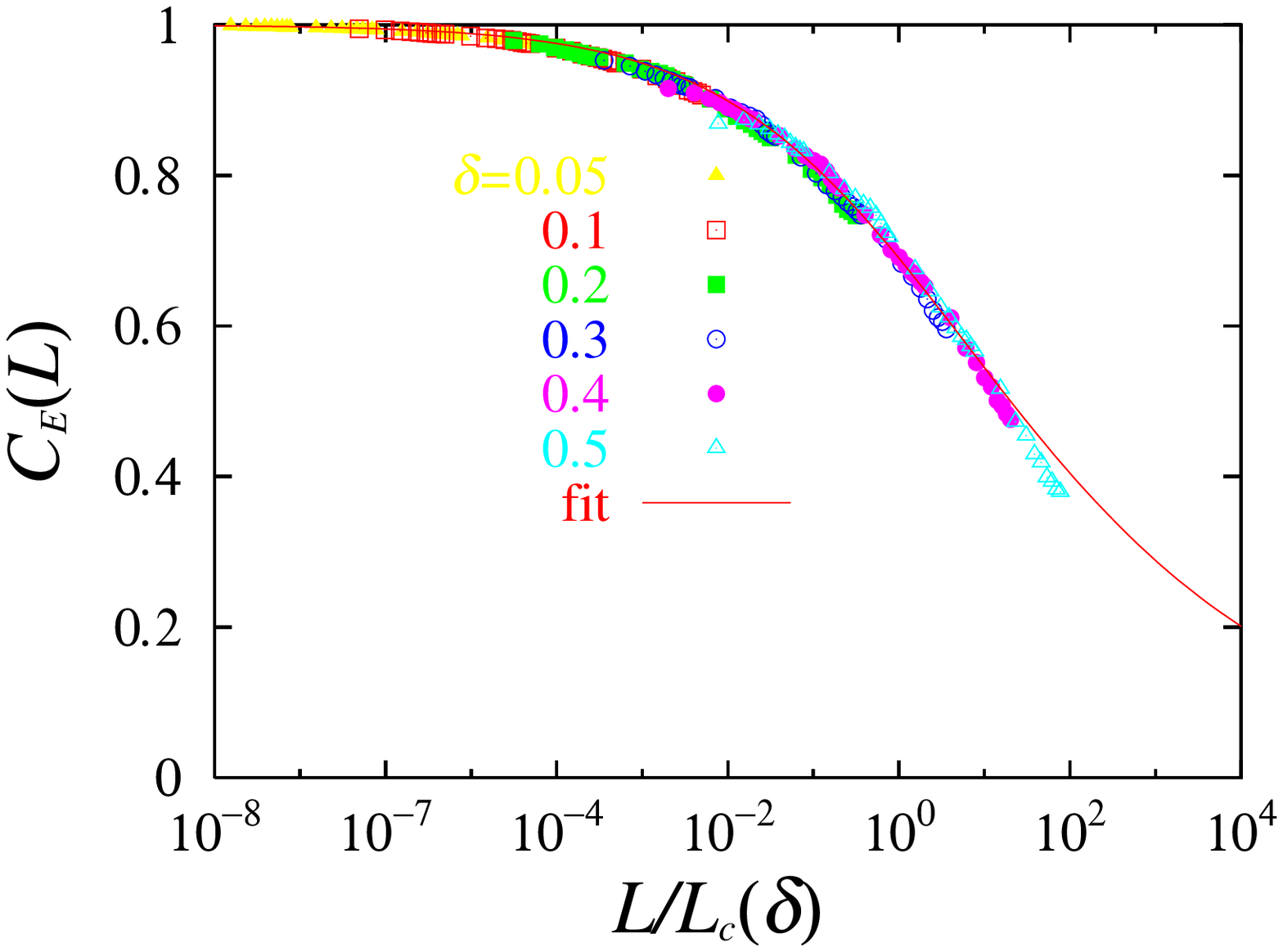}
\end{center}
\caption{$C_{F}(L)$ of the random tilt field perturbation case
and its scaling plot with $L_{c}(\delta)=(0.87 \delta)^{-6}$.
}
\label{corr-e-randmtilt.fig}
\end{figure}

\begin{figure}
\begin{center}
\includegraphics*[scale=0.45]{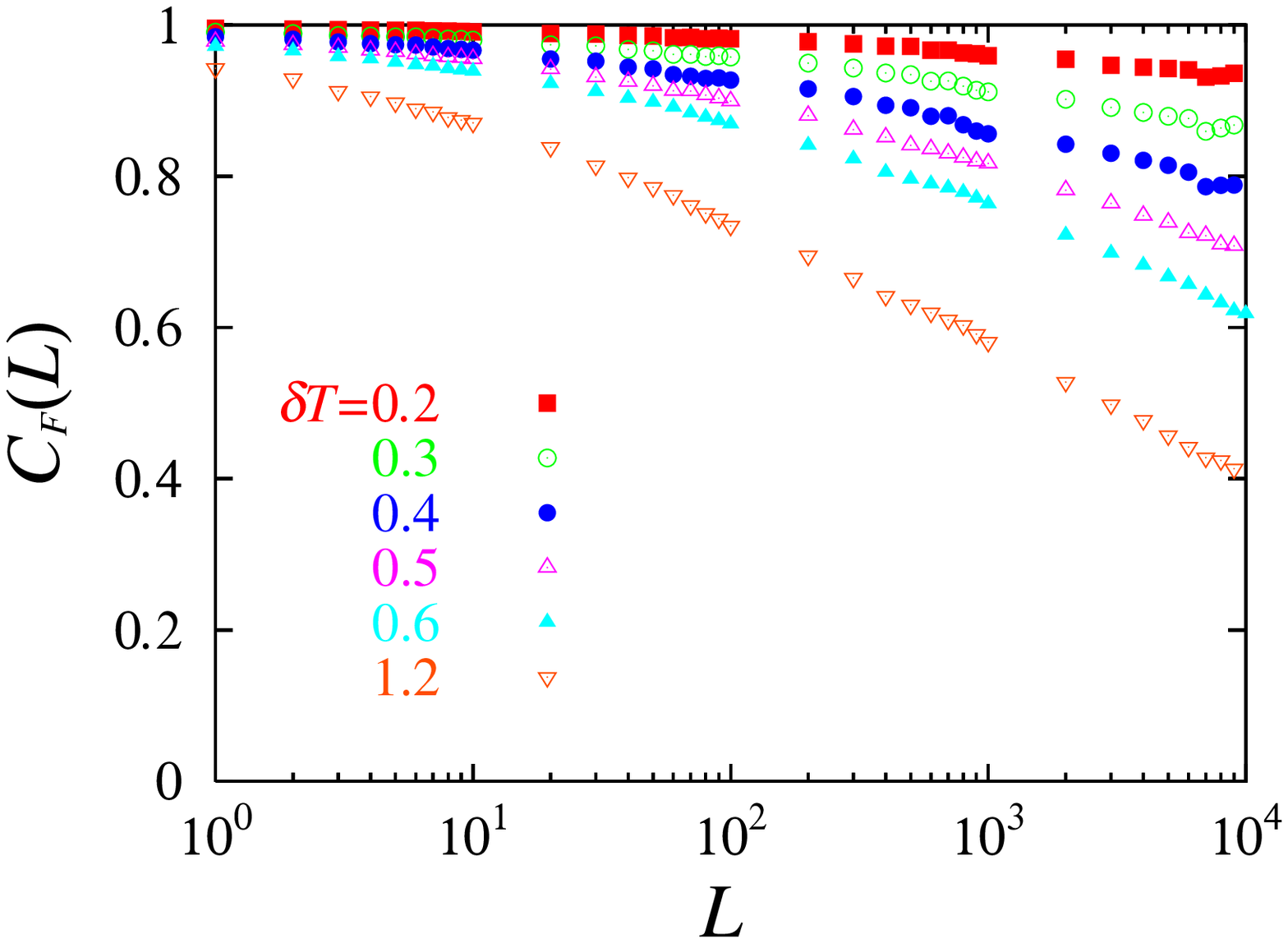}
\includegraphics*[scale=0.45]{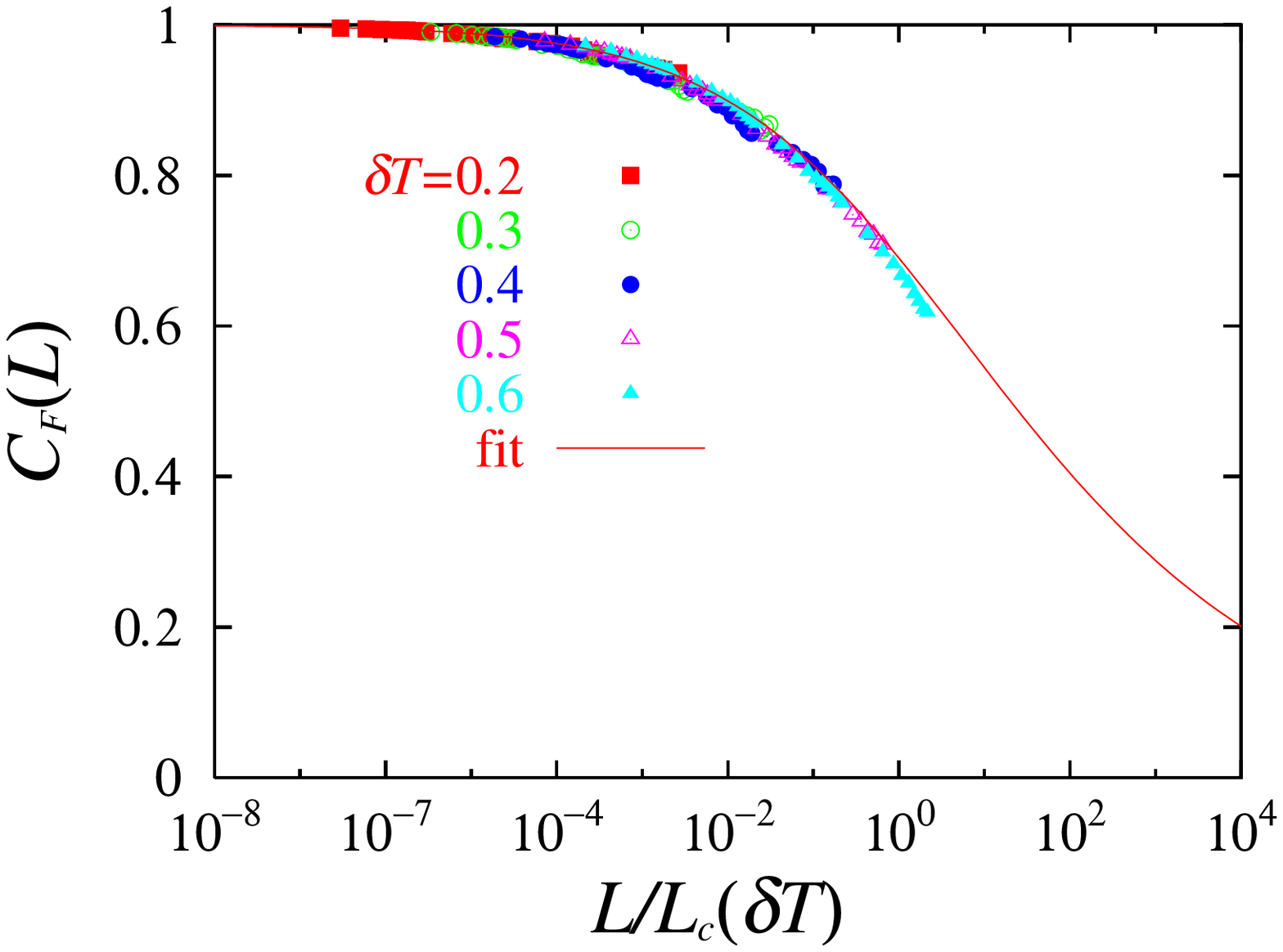}
\end{center}
\caption{$C_{F}(L)$ of the temperature-shift perturbation case
and its scaling plot with
$L_{c}(\delta T)=(0.43 \delta T)^{-6}$.
}
\label{corr-f-temperature.fig}
\end{figure}

\subsubsection{De-correlation of fluctuations of free-energies 
and ground state energies}

In Fig. \ref{corr-e-potential.fig}-\ref{corr-f-temperature.fig}, 
the correlation of the correlation of
the ground-state energies or free-energies 
of the perturbed and unperturbed systems
are shown together with its scaling plot.
In the scaling plots we have used the same numerical prefactor $c$ in
$L_{c}(\delta)=(c\delta)^{-6}$ used in the scaling plot
of $B_{2}(L)$. As one can see, the master curves
 for the three perturbations merge.
The initial part of the master curve fits nicely into
the expected form \eq{eq-cf-scaling-form} using $\alpha=1/2$,
$C_{L}(F)=1/(1+A(L/L_{c}(h))^{2(1-1/3)})$ with $A\sim 1.5$.
One can see that the decay is faster for $L/L_{c}(h) \gg 1$.
To sum up, the expected crossover behavior from 
weakly perturbed regime and strongly perturbed regime is indeed the same
for the three kinds of apparently different perturbations.

\section{Conclusion}
\label{sec.conclusion}

In this work we have studied the sensitivity of the glassy phase 
of DPRM against various types of thermal and non-thermal perturbations.
For the first time, we have obtained  very 
coherent results
which strongly support the picture anticipated by the 
phenomenological scaling arguments. As we increase the length 
scale $L$ at which  observations are made, there is a crossover from the weakly 
perturbed regime dominated by rare events (i.e. jumps between 
neighboring free-energy valleys) $L \ll L_{c}(\delta)$ to the strongly perturbed 
regime where these events become typical $L \gg L_{c}(\delta)$. 
This means that  perturbations become strong at large length 
scales $L/L_{c}(\delta) \to \infty$ such that the configuration 
can easily jump from one valley to another, i.e. it becomes ``chaotic'' 
in the sense that the visited landscape is totally different 
from that before. 

In replica space we proposed a new definition of chaos \eq{eq-dec}
in terms of the global partition function (A$+$B) rather than 
the correlation function itself. There is chaos if in the adequate 
limits the partition function factorizes, so that we have two non-interacting systems. 
The decorrelation of systems A and B 
when introducing a perturbation can be understood as a concrete example 
of explicit replica symmetry breaking as proposed by Parisi and Virasoro \cite{PV89}. 
Concerning the mapping to the Sinai model, it means that the free-energy 
landscape of the perturbed DPRM cannot be described anymore by a single
Sinai potential. Instead, RSB requires the coexistence of statistically 
independent Sinai potentials. 

The key point in our DPRM case is the fact that the RS bound 
state of the quantum problem is marginally stable with respect 
to RSB as noticed by Parisi~\cite{P90}.
Infinitesimally weak perturbations $\Delta \ll 1$ 
induce small replica symmetry breaking terms and induce a symmetry
breaking transition from a RS to a RSB state which takes place 
in the $n/n^{*}(\Delta) \to 0$ limit for any small but non-zero
strength of the perturbation.  It turns out that we can read 
off the overlap length $L_{c}(\delta)$ from $n^{*}(\Delta)$. 
Within the replica space, the perturbations are 
naturally classified according to their order of perturbation $p$ and
the symmetries which are left conserved. 
For each class we have numerically verified that, indeed, there are
universal scaling functions of correlation functions in terms 
of $L/L_{c}(\delta)$ describing the crossover from the weakly 
to strongly perturbed regimes.
It is notable that the decay of the free-energy fluctuation $C_{F}(L)$
is very slow in all cases we studied. 
It will not be surprising that one cannot have an impression of  ``chaos'' 
by only making observations within some limited length scales. 

In mean-field models, RSB is always associated with the existence of many pure
states \cite{MFT}, which is not the case in DPRM in a strict sense.
In the DPRM, the mapping to Quantum Mechanics is always possible for
an arbitrarily large number of dimensions of the transverse space 
(usually denoted as $N$, being $N=1$ in our case) in which
 the ground state must contain the bosonic symmetry of the 
Schr\"{o}dinger operator \cite{MP92}. In this sense, RSB has 
to be {\it weak} in the DPRM problem\cite{M90,P90}, i.e. 
it is a latent feature which only manifests under certain circumstances. In 
the present $1+1$ model the existence of a 'hidden' RSB excited state
with vanishingly small gap with respect to the RS ground state 
in the $n \to 0$ limit is extremely important.
Loosely speaking, the situation is not very far from having many pure states.
In a really stable  RS phase like (like ferromagnetic phases), 
this phenomena cannot happen. It is tempting to speculate
that some lessons obtained in the present $1+1$ dimensional 
DPRM case may turn out to be more general.

In the present paper, the temperature-chaos is confirmed. 
Thus the present model serves
as a suitable testing ground to examine the possible connection between the 
temperature-chaos and the restart of aging (rejuvenation) observed experimentally 
\cite{JVHBN,sg-experiment-review,uppsala-reentrantsg,saclay-reentrantsg}.
Experimentally, almost complete restart of {\it aging} or relaxation 
take place only by slight temperature-changes. 
Whether this restart of aging can be associated with
the temperature-chaos remains an interesting open question.
Interestingly enough, recent experiments 
\cite{uppsala-reentrantsg,saclay-reentrantsg} 
suggest rejuvenation (chaos) effect in random ferromagnetic systems.
A candidate to account for the mechanism of the phenomena may be the 
temperature-chaos of pinned domain walls of the ferromagnets, which is
directly related to the present study.

Another surprise revealed by the temperature-cycling experiments is that
initial aging is resumed when the temperature is cycled back to
the initial temperature, giving rise to the so called memory effect.
It appears contradictory to the temperature chaos effect at first sight.
Recently a coarsening model under cycling of target equilibrium states was
studied \cite{YLB00}.
There a hidden {\it dynamical memory by  ghost domains}
was found and a scenario was proposed to explain the intriguing
coexistence of the rejuvenation and memory effect. In the present
context of pinned elastic manifold rejuvenation and memory  can be
easily explained by considering Fourier components of the temporal
configuration.  When the temperature is shifted, Fourier components
at wave length larger than the overlap length will be subjected to
rejuvenation. At time $t$ after the temperature-shift,
Fourier components at wave length shorter than $L(t)$ will be adopted
to the new temperature. Here $L(t)$ is a dynamical length scale
over which the system can be equilibrated within a given time $t$.
However, Fourier components of even larger wavelength $ > L(t)$
remain the same as before the temperature-shift. Thus {\it dynamical
memory} exists at the coarse-grained level of $L(t)$.                   

Recalling that relaxational dynamics is extremely slow in glassy
systems because of the dominance of the activated processes,
one has to consider seriously how large time scale is needed to go
beyond the overlap length. If it is  too large, even experimental time scales 
(typically $10^{14}-10^{17}$ $\tau_{0}$ where $\tau_{0} \sim 10^{-13}$ (sec)) 
may not be sufficient and one must look for other 
mechanisms \cite{JP,BDHV01,BB02} to explain the rejuvenation phenomena 
observed experimentally. Previous numerical studies of the relaxational 
dynamics of the present DPRM model \cite{Y98} implies that the needed time 
lies within the time window of experiments and numerical simulations
for some realistic parameters. More work in 
this direction would certainly be interesting.

{\bf Acknowledgements}

We thank the Service de Physique de l'\'Etat Condens\'e, CEA Saclay
for financial support and kind hospitality where this work was
started. M. S. acknowledges the MEC of the Spanish 
government for grant AP98-36523875. 
We especially thank M. M\'{e}zard, J. P. Bouchaud and F. Ritort for kind 
suggestions and many useful discussions.


\begin{thebibliography}{99}
\bibitem{note-energy-entropy} We observed the following
 numerically within the lattice model we used in section 
\ref{subsec-transfer-temp-potential-randomtilt}.
We computed root-mean-square (RMS) average 
of the sample-to-sample fluctuations of 
energy, entropy and free-energy. Fluctuation of energy
with respect to the mean value depends on temperature $T$ as 
$\sqrt{[\Delta E(L)]^{2}}=U_{0}(L/L_{0})^{\theta}+c \kb T(L/L_{0})^{1/2}$
where $c$ is a numerical constant. Note that the $L^{1/2}$ behavior
shows up only at sufficiently large length scales (as found in \cite{FH91}).
The temperature dependence is consistent with
the observation that fluctuation of energy at $T=0$ (ground state)
scales as $\lim_{T \to 0}\sqrt{[\Delta E(L)]^{2}}=U_{0}(L/L_{0})^{\theta}$,
which is often assumed in the scaling arguments.
On the other hand the entropic fluctuation scales as 
$\sqrt{[S(L)]^{2}}=c (L/L_{0})^{1/2}$ which is consistent with the
observation that
the fluctuation of free-energy $\Delta F =\Delta E - \kb T \Delta S$
scales as $\sqrt{[\Delta F(L)]^{2}}=U_{0}(L/L_{0})^{\theta}$.
\end{thebibliography}
\end{document}